\documentclass[11pt]{article}
\usepackage{amsfonts}
\usepackage{dsfont}
\usepackage{amssymb}
\usepackage{amsmath}
\usepackage{amsxtra}
\usepackage{graphicx}
\usepackage{psfrag}
\usepackage{mathrsfs}
\usepackage{multirow}
\usepackage[sort, numbers]{natbib}
\usepackage{stmaryrd}
\usepackage{subfigure}
\usepackage{tikz}
\usepackage{empheq}
\usepackage[normalem]{ulem}
\usepackage{pdfpages}
\usepackage{color}
\usepackage{longtable}
\usepackage{footmisc}
\usepackage{hyperref}

\usepackage{lineno}

\begin{document}
\def\BGamma{\mbox{\boldmath$\Gamma$}}
\def\BDelta{\mbox{\boldmath$\Delta$}}
\def\BTheta{\mbox{\boldmath$\Theta$}}
\def\BLambda{\mbox{\boldmath$\Lambda$}}
\def\BXi{\mbox{\boldmath$\Xi$}}
\def\BPi{\mbox{\boldmath$\Pi$}}
\def\BSigma{\mbox{\boldmath$\Sigma$}}
\def\BUpsilon{\mbox{\boldmath$\Upsilon$}}
\def\BPhi{\mbox{\boldmath$\Phi$}}
\def\BPsi{\mbox{\boldmath$\Psi$}}
\def\BOmega{\mbox{\boldmath$\Omega$}}
\def\Balpha{\mbox{\boldmath$\alpha$}}
\def\Bbeta{\mbox{\boldmath$\beta$}}
\def\Bgamma{\mbox{\boldmath$\gamma$}}
\def\Bdelta{\mbox{\boldmath$\delta$}}
\def\Bepsilon{\mbox{\boldmath$\epsilon$}}
\def\Bzeta{\mbox{\boldmath$\zeta$}}
\def\Beta{\mbox{\boldmath$\eta$}}
\def\Btheta{\mbox{\boldmath$\theta$}}
\def\Biota{\mbox{\boldmath$\iota$}}
\def\Bkappa{\mbox{\boldmath$\kappa$}}
\def\Blambda{\mbox{\boldmath$\lambda$}}
\def\Bmu{\mbox{\boldmath$\mu$}}
\def\Bnu{\mbox{\boldmath$\nu$}}
\def\Bxi{\mbox{\boldmath$\xi$}}
\def\Bpi{\mbox{\boldmath$\pi$}}
\def\Brho{\mbox{\boldmath$\rho$}}
\def\Bsigma{\mbox{\boldmath$\sigma$}}
\def\Btau{\mbox{\boldmath$\tau$}}
\def\Bupsilon{\mbox{\boldmath$\upsilon$}}
\def\Bphi{\mbox{\boldmath$\phi$}}
\def\Bchi{\mbox{\boldmath$\chi$}}
\def\Bpsi{\mbox{\boldmath$\psi$}}
\def\Bomega{\mbox{\boldmath$\omega$}}
\def\Bvarepsilon{\mbox{\boldmath$\varepsilon$}}
\def\Bvartheta{\mbox{\boldmath$\vartheta$}}
\def\Bvarpi{\mbox{\boldmath$\varpi$}}
\def\Bvarrho{\mbox{\boldmath$\varrho$}}
\def\Bvarsigma{\mbox{\boldmath$\varsigma$}}
\def\Bvarphi{\mbox{\boldmath$\varphi$}}
\def\bone{\mbox{\boldmath$1$}}
\def\bzero{\mbox{\boldmath$0$}}
\def\bnabla{\mbox{\boldmath$\nabla$}}
\def\bvarepsilon{\mbox{\boldmath$\varepsilon$}}
\def\bA{\mbox{\boldmath$ A$}}
\def\bB{\mbox{\boldmath$ B$}}
\def\bC{\mbox{\boldmath$ C$}}
\def\bD{\mbox{\boldmath$ D$}}
\def\bE{\mbox{\boldmath$ E$}}
\def\bF{\mbox{\boldmath$ F$}}
\def\bG{\mbox{\boldmath$ G$}}
\def\bH{\mbox{\boldmath$ H$}}
\def\bI{\mbox{\boldmath$ I$}}
\def\bJ{\mbox{\boldmath$ J$}}
\def\bK{\mbox{\boldmath$ K$}}
\def\bL{\mbox{\boldmath$ L$}}
\def\bM{\mbox{\boldmath$ M$}}
\def\bN{\mbox{\boldmath$ N$}}
\def\bO{\mbox{\boldmath$ O$}}
\def\bP{\mbox{\boldmath$ P$}}
\def\bQ{\mbox{\boldmath$ Q$}}
\def\bR{\mbox{\boldmath$ R$}}
\def\bS{\mbox{\boldmath$ S$}}
\def\bT{\mbox{\boldmath$ T$}}
\def\bU{\mbox{\boldmath$ U$}}
\def\bV{\mbox{\boldmath$ V$}}
\def\bW{\mbox{\boldmath$ W$}}
\def\bX{\mbox{\boldmath$ X$}}
\def\bY{\mbox{\boldmath$ Y$}}
\def\bZ{\mbox{\boldmath$ Z$}}
\def\ba{\mbox{\boldmath$ a$}}
\def\bb{\mbox{\boldmath$ b$}}
\def\bc{\mbox{\boldmath$ c$}}
\def\bd{\mbox{\boldmath$ d$}}
\def\be{\mbox{\boldmath$ e$}}
\def\bff{\mbox{\boldmath$ f$}}
\def\bg{\mbox{\boldmath$ g$}}
\def\bh{\mbox{\boldmath$ h$}}
\def\bi{\mbox{\boldmath$ i$}}
\def\bj{\mbox{\boldmath$ j$}}
\def\bk{\mbox{\boldmath$ k$}}
\def\bl{\mbox{\boldmath$ l$}}
\def\bm{\mbox{\boldmath$ m$}}
\def\bn{\mbox{\boldmath$ n$}}
\def\bo{\mbox{\boldmath$ o$}}
\def\bp{\mbox{\boldmath$ p$}}
\def\bq{\mbox{\boldmath$ q$}}
\def\br{\mbox{\boldmath$ r$}}
\def\bs{\mbox{\boldmath$ s$}}
\def\bt{\mbox{\boldmath$ t$}}
\def\bu{\mbox{\boldmath$ u$}}
\def\bv{\mbox{\boldmath$ v$}}
\def\bw{\mbox{\boldmath$ w$}}
\def\bx{\mbox{\boldmath$ x$}}
\def\by{\mbox{\boldmath$ y$}}
\def\bz{\mbox{\boldmath$ z$}}
\newcommand*\mycirc[1]{%
  \begin{tikzpicture}
    \node[draw,circle,inner sep=1pt] {#1};
  \end{tikzpicture}
}
\newcommand{\upcite}[1]{\textsuperscript{\textsuperscript{\cite{#1}}}}
\newcommand{\kg}[1]{\textcolor{black}{\textbf{(kg)} #1}}
\newcommand{\xh}[1]{\textcolor{blue}{XH: #1}}          \makeatletter
\def\@biblabel#1{#1.}
\makeatother
\title{Variational system identification of the partial differential equations governing the physics of pattern-formation: Inference under varying fidelity and noise}
\author{Z. Wang\thanks{Department of Mechanical Engineering, University of Michigan}, X. Huan\thanks{Department of Mechanical Engineering, Michigan Institute for Computational Discovery \& Engineering, University of Michigan, University of Michigan} and K. Garikipati\thanks{Departments of Mechanical Engineering, and Mathematics, Michigan Institute for Computational Discovery \& Engineering, University of Michigan, corresponding author, {\tt krishna@umich.edu}}}
\maketitle
\begin{abstract}
We present a contribution to the field of system identification of partial differential equations (PDEs), with emphasis on discerning between competing mathematical models of pattern-forming physics. The motivation comes from developmental biology, where pattern formation is central to the development of any multicellular organism, and from materials physics, where phase transitions similarly lead to microstructure. In both these fields there is a collection of nonlinear, parabolic PDEs that, over suitable parameter intervals and regimes of physics, can resolve the patterns or microstructures with comparable fidelity. This observation frames the question of which PDE best describes the data at hand. This question is particularly compelling because identification of the closest representation to the true PDE, while constrained by the functional spaces considered relative to the data at hand, immediately delivers insights to the physics underlying the systems. While building on recent work that uses stepwise regression, we present advances that leverage the variational framework and statistical tests. We also address the influences of variable fidelity and noise in the data.
\end{abstract}

\section{Introduction}
The widespread use of sensors, high throughput experiments and simulations, as well high performance computing have made ``big data'' available for a range of engineering physics systems. This has sparked an explosion of interest in data-driven modeling for these systems. An extreme manifestation of this still-developing field is seen in ``model-free'' approaches that do not rely on physics-based knowledge. Such a path to modelling holds the promise of very high computational efficiency by circumventing solutions to potentially complex physics. However, it  draws criticism for its ``black box'' nature, and often for lack of interpretability. More seriously, these approaches offer scant openings for analysis when the model fails. Conversely, the availability of abundant data also presents opportunities to discover mathematical frameworks, with well-understood physical meaning, which govern the underlying behavior. This has fueled the field of \emph{system identification}, which is particularly compelling for determining the governing partial differential equations (PDEs). This is so, because knowledge of the PDE directly translates to deep insights to the physics, guided by differential and integral calculus. 

%
Approaches for data-assisted and data-driven discovery of PDEs have proceeded along several fronts.
(a) Early work in parameter identification can be traced to nonlinear regression approaches \cite{VossPLA1997, VossOPRL1999}. (b) If the governing equation is unknown, an approximate model---whether physical, empirical, or mixed---could be learned from data. For example, the approximate model could be a neural network \cite{GarciaCCE1998}, {\color{black} a reduced-order model \cite{AttarJFS2005, KhalilMSSP2007}, a phenomenological effective model formed by a linear combinations of basis functions in finite dimensions \cite{GuoIJC2010, DanielsCC2015}}, or a linear mapping from a current to a future state where the dynamic characteristics of the mapping could be learned by Dynamic Mode Decomposition \cite{MezicJFM2009, SchmidJFM2010}.  (c) Lastly, the complete underlying governing equations could be extracted from data by combining symbolic regression and genetic programming to infer algebraic expressions along with their coefficients \cite{SchmidtSCI2009, SchmidtPB2011}. In this context, while genetic programming balances model accuracy and complexity, it proves very expensive when searching for a few relevant terms from a large library of candidates. 
In a recent approach to solving inverse problems \cite{Raissi2019}, the strong form of a specified PDE was directly embedded in the loss function while training deep neural network representations of the solution variable. This approach results in the minimum-residual solution subject to a deep neural network global basis. The work contained examples where up to two unknown PDE coefficients were estimated, and did not explore system identification from a larger set of possible operators. A somewhat more ambitious and extensive approach is based on the recognition that most physical systems have only a few relevant terms, thus creating an opportunity to develop sparse regression techniques for system identification. This class of approaches has recently been applied with success \cite{KutzPNAS2015, KutzIEEE2016, KutzSCIADV2017} to determining the operators in PDEs from a comprehensive library of candidates, thus efficiently discovering governing equations from data. More recent extensions have been applied to model recovery using sparse data after abrupt changes in the system \cite{KutzChaos2018}, hybrid dynamical systems \cite{KutzHybrid2018} and multiscale systems \cite{KutzSIAM2019}. The work presented here is based on this approach, inspired by its direct treatment of operator forms in PDEs, and their connections with physical mechanisms, while presenting several novel advances. These include the adoption of the variational setting, via the weak forms of PDEs; step-wise regression with the statistical F-test; a treatment of noise and its amelioration by varying fidelity.

For context, we briefly discuss the role of pattern forming systems of equations in developmental biology and materials physics. Following Alan Turing's seminal work on reaction-diffusion systems \cite{Turing1952}, a robust literature has developed on the application of nonlinear versions of this class of PDEs to model pattern formation in developmental biology \cite{Gierer1972,Murray1981,Dillon1994,Barrio1999,Barrio2009,MainiByrne2012,Spill2015,Korvasova2015,GarikipatiJMPS2017}. The Cahn-Hilliard phase field equation \cite{CahnHilliard1958} has been applied to model other biological processes with evolving fronts, such as tumor growth and angiogenesis \cite{Wise2008,Cristini2009,Lowengrub2010,Lowengrub2009,Vilanova2013,Vilanova2014,Oden2010,Xu2016}. Pattern formation during phase transformations in materials physics can happen as the result of instability-induced bifurcations from a uniform composition \cite{Jiangetal2016,Rudrarajuetal2016,Teichertetal2017}, which was the original setting of the Cahn-Hilliard treatment. Another phase field treatment, the Allen-Cahn equation \cite{Allen1979} models nucleation and growth of precipitates and also has seen wide use \cite{Vaithyanathan2002,Hu2001,Zhu2002,Su1996,Kim1999,Gao2012,Liu2013,Ji2014,Liu2017,Jou1997,Teichert2018a}. All these pattern forming systems fall into the class of nonlinear, parabolic PDEs, and have spawned vast literature in mathematical physics. They can be written as systems of first-order dynamics driven by a number of time-independent terms of algebraic and differential form. The spatio-temporal, differentio-algebraic operators act on either a composition (normalized concentration) or an order parameter. It also is common for the algebraic and differential terms to be coupled across multiple species. 

It is compelling to attempt to discover the physics governing these pattern forming systems by identifying their PDE forms from data. However proper evaluation of the numerical derivatives is very challenging. The commonly used finite-difference approximations require data to be available in a very finely discretized space. Polynomial interpolation has proven to be reliable and robust \cite{KnowlesEJD}, and recently has found use in data-driven discovery of PDEs \cite{KutzSCIADV2017}. However, this approach encounters substantial difficulties in estimating the coefficients of high-order derivatives, because of the significant error associated with numerical differentiation. Even still, to the best of our knowledge, the published literature in discovering governing equations has remained focused on trying to identify the strong form of the PDEs. Every strong form, however, has equivalent weak form(s), which seem to have not been exploited for system identification. As will be seen in this communication, the advantages of using weak forms include: the natural occurrence of the loss function for the regression problem, allowing identification of the boundary conditions, which has proven to be a challenge, and successfully identifying higher-order derivatives with the aid of smooth basis functions (such as non-uniform rational B-splines or NURBS in this work). 

Here, after briefly reviewing how the Galerkin weak form may be employed for identification of a general dynamical system (Section \ref{sec:weak_form}), we focus on introducing our methods for identifying PDEs in this variational setting, using stepwise regression (Section \ref{sec:IDSWF}). In this first communication, we focus on examples to identify reaction-diffusion equations evolving under Schnakenberg kinetics \cite{Schnakenberg1976}, and the Cahn-Hilliard and Allen-Cahn equations driven by non-convex energy density functions (Section \ref{sec:example}). This choice is driven by the fact that, as explained above, these are widely used models in biological patterning and morphogenesis, and have well-established traditions of use in materials physics, also. The datasets we use are all obtained by high fidelity direct numerical simulations. The more challenging problem of datasets from physical experiments will be addressed in a subsequent communication. We do call attention to our treatment of both high and low fidelity data, with and without noise, as a precursor to using data from physical experiments. Concluding remarks appear under Discussion and Conclusions (Section \ref{sec:conclusions}). 

We note in passing that, very recently, there has been growing interest in combining data-driven methods with PDEs in a more direct manner: to find solutions to the canonical initial and boundary value problems (IBVPs) of mathematical physics. These approaches include Gaussian processes with kernels defined by specific PDEs in linearized form for one- and two-dimensions \cite{RaissiJCP2018}, and deep neural networks for solution of high-dimensional IBVPs on semi-linear parabolic PDEs \cite{Han2018} or with strong forms of the target one- and two-dimensional PDEs embedded in the loss functions \cite{Raissi2019}.

\subsection{The Galerkin weak form}
\label{sec:weak_form}

We first provide a brief discussion of how the weak form of PDEs will be used in this work.
We start with the general strong form for first-order dynamics written as
\begin{align}
   \frac{\partial C}{\partial t}-\Bchi\cdot\Bomega=0, 
   \label{eq:general_strongForm}
\end{align}
where $\Bchi$ is a vector containing all possible independent terms expressed as algebraic and differential operators on the scalar solution $C$:
\begin{align}
\Bchi=[1, C, C^2,...,\nabla^2 C,...],
\end{align}
and $\Bomega$ is the vector of pre-factors for each term. Thus, the one-field diffusion reaction equation
\begin{align}
\frac{\partial C}{\partial t}-D\nabla^2 C-f=0
\end{align}
with constant diffusivity $D$ and reaction rate $f$ has
\begin{align}
\Bchi=[1, C, C^2,\nabla^2 C]\hspace{2em} \mathrm{and} \hspace{2em}
\Bomega=[f,0,0,D].
\end{align}
Note that the time derivative $\partial C/\partial t$ is treated separately from the other terms in order to highlight the first-order dynamics of all problems we target in this work. 

Next, we recall the weak form that is equivalent to a general strong form written as above with vector of operators $\boldsymbol{\chi}$ and vector of pre-factors $\boldsymbol{\omega}$. For infinite-dimensional problems with Dirichlet boundary conditions on $\Gamma^c$ the weak form can be stated as: $\forall ~w \in \mathscr{V}$ where $\mathscr{V}= \{w\vert ~w = ~0 \;\mathrm{on}\;  \Gamma^c\}$, find $c \in \mathscr{S}$ where $\mathscr{S} = \{C\vert C = \bar{C} \text{ on } \Gamma^c\}$ such that
\begin{align}
\int_{\Omega}w\left(\frac{\partial C}{\partial t}-\Bchi\cdot\Bomega\right) \text{d}v=0.
\label{eq:weak_form}
\end{align}
For finite-dimensional fields $C^h, w^h$, the weak form is as follows: find $C^h\in \mathscr{S}^h \subset \mathscr{S}$ where $\mathscr{S}^h= \{ C^h \in \mathscr{H}^2(\Omega) ~\vert  ~C^h = ~\bar{u}\; \mathrm{on}\;  \Gamma^u\}$,  such that $\forall ~w^h \in \mathscr{V}^h \subset \mathscr{V}$ where $\mathscr{V}^h= \{ w^h \in\mathscr{H}^2(\Omega)~\vert  ~w^h = ~0 \;\mathrm{on}\;  \Gamma^u\}$, the finite-dimensional (Galerkin) weak form of the problem is satisfied. The choice of $\mathscr{H}^2(\Omega)$ as the Sobolev space is motivated by the differential operators, which reach the highest order of two in the weak forms we consider (four in strong form). The variations $w^h$ and trial solutions $C^h$ are defined component-wise using a finite number of basis functions,
\begin{align}
w^h = \sum_{a=1}^{n_\mathrm{b}} d^a N^a \label{eq:basisw} \\
C^h = \sum_{a=1}^{n_\mathrm{b}} c^a N^a,
\label{eq:basisu}
\end{align}
\noindent where $n_\mathrm{b}$ is the dimensionality of the function spaces $\mathscr{S}^h$ and $\mathscr{V}^h$, and $N^a$ represents the basis functions. Substituting $w^h$ and $C^h$ in Equation (\ref{eq:weak_form}) by Equations (\ref{eq:basisw}) and (\ref{eq:basisu}) and decomposing the integration over $\Omega$ by a sum over subdomains $\Omega^e$ leads to:
\begin{align}
\sum_e\int_{\Omega^e}\sum_{a=1}^{n_\mathrm{b}} d^a N^a\left(\frac{\partial C^h}{\partial t}-\Bchi(C^h)\cdot\Bomega\right) \text{d}v=0.
\end{align}
Upon integration by parts, application of appropriate boundary conditions, and accounting for the arbitrariness of $w^h$ in $\mathscr{V}^h$, the finite-dimensionality leads to a system of residual equations for each degree of freedom (DOF):
\begin{align}
\mathscr{R}_i=\mathscr{F}_i\left(\frac{\partial C^h}{\partial t}, C^h, \nabla C^h,...,N,\nabla N ...\right),
\label{eq:residual}
\end{align}
where $\mathscr{R}_i[C^h]$ is the $i^\text{th}$ component of the residual vector, a functional of $C^h$ satisfying $\mathscr{R}_i[C^h] = 0$ for the Galerkin solution of the problem, $C^h$.

\section{Identification of governing parabolic PDEs in weak form}
\label{sec:IDSWF}

The PDE identification problem in strong form is to find the correct pre-factor $\Bomega$ in Equation (\ref{eq:general_strongForm}), which chooses among many possible candidate basis operators acting on $C$ collected in $\Bchi$, given the solution/data of the dynamical system. Alternately, we could identify the PDE in finite-dimensional weak form by adopting finite-dimensional basis functions and also finding $\Bomega$ for many possible basis operators, but now written in weak form, acting on $C^h$. In preparation, we first recall several basis operators, including Neumman boundary condition operators, all written in weak form. We next introduce NURBS basis functions, which have been popularized for isogeometric analysis \cite{CottrellHughesBazilevs2009}. The algorithm of stepwise regression for model selection is introduced following these developments.

\subsection{Candidate basis operators in weak form} 
\label{sec:candidateBasis}

Suppose we have data for the field $C$ on certain 
degrees of freedom (DOFs) of a discretization for a series of time steps. These DOFs are the control variables if NURBS basis functions are used. Selecting all or a subset of these DOFs, a Galerkin representation of the field could be constructed over the domain. The use of fewer DOFs represents lower fidelity data, which is discussed separately in Section \ref{sec:Low_fidelity_noisy_data}. We can then obtain the system of residual equations in \eqref{eq:residual}, with the independent candidate basis terms in weak form populating $\Bchi$.  Below we illustrate the procedure with four example terms.

\noindent \textbf{Constant term $1$:} 
\begin{align}
\int_{\Omega}w1\text{d}v=\int_{\Omega} \sum_{i=1}^{n_\mathrm{b}} d^{i} N^{i} \sum_{a=1}^{n_\mathrm{b}} 1 N^{a}\text{d}v
\end{align} 
Since the $d^{i}$ are arbitrary, we obtain a vector of size $N_\text{DOF}$, where the $i$th component is 
\begin{align}
\Xi^\text{cons}_i=\int_{\Omega} N^i \sum_{a=1}^{n_\mathrm{b}}  N^a\text{d}v.
\end{align}

\noindent \textbf{Time derivative $\partial C/\partial t$ at time $t_n$:} We write a backward difference approximation of the time derivative at $t_n$ as
\begin{align}
\Xi^{\dot{C}}_i|_n=\int_{\Omega} N^i \sum_{a=1}^{n_\mathrm{b}} \frac{c_{n}^{a}-c_{n-1}^{a}}{\Delta t} N^a\text{d}v  \label{eq:basis_Ct}
\end{align}
where $\Delta t = t_n-t_{n-1}$ is the time step, and $c_n$ and $c_{n-1}$ are the control variable values at times $t_n$ and $t_{n-1}$. 

\noindent \textbf{Laplace operator $\nabla^2 C$ at time $t_n$:} Multiplying by the weighting function and integrating by parts:
\begin{align}
\int_{\Omega}w\nabla^2 C^h_{n}\text{d}v=&-\int_{\Omega} \sum_{i=1}^{n_\mathrm{b}} d^{i} \nabla N^{i}  \cdot\sum_{a=1}^{n_\mathrm{b}}  c_{n}^{a}\nabla N^a\text{d}v\\
&+\int_{\Gamma}\sum_{i=1}^{n_\mathrm{b}} d^{i} N^{i}  \sum_{a=1}^{n_\mathrm{b}} c_{n}^{a}  \nabla N^a\cdot\bn\text{d}s.
\end{align} 
Note that the second term on the right is the Neumann boundary term. For a flux boundary condition on $\Gamma$, such that $-\nabla C^h_{n}\cdot \bn=1$. We thus have two basis functions after accounting for the arbitrariness of $d^{i}$:
\begin{align}
\Xi^{\nabla^2 C}_i|_n=-\int_{\Omega} \nabla N^i \cdot\sum_{a=1}^{n_\mathrm{b}}  c_{n}^{a}\nabla N^a\text{d}v
\label{eq:basisLaplaceVol}
\end{align}
is the basis for $\nabla^2 C^h_{n}$ (with zero Neumann boundary condition), and
\begin{align}
 \Xi^{\nabla^2 C_{\text{BC}}}_i|_n=\int_{\Gamma}N^i \sum_{a=1}^{n_\mathrm{b}} 1\text{d}s   
 \label{eq:basis_BC}
 \end{align}
is the basis for the Neumann boundary term arising from $\nabla^2 C$ at the surface $\Gamma$. As is well known, in the variational setting the weak form of the divergence operator yields volume \eqref{eq:basisLaplaceVol} and surface \eqref{eq:basis_BC} terms, with the latter corresponding to the Neumann boundary condition. In this work we treat them as distinct bases, thus allowing us to separately identify the Neumann boundary term. 

We also call attention to the fact that for the unsteady diffusion problem, equating \eqref{eq:basis_Ct} to the sum of \eqref{eq:basisLaplaceVol} and \eqref{eq:basis_BC} amounts to the Backward Euler time integration algorithm, which is first-order accurate. The mid-point rule, on the other hand is second-order in time.

\noindent \textbf{Biharmonic operator $\nabla^4 C_n$ at time $t_n$:} Multiplying by the weighting function and integrating by parts:
\begin{align}
\int_{\Omega}w\nabla^4 C^h_{n}\text{d}v&=\int_{\Omega} \sum_{i=1}^{n_\mathrm{b}} d^{i} \nabla^2N^{i} \sum_{a=1}^{n_\mathrm{b}}  c_{n}^{a}\nabla^2 N^a\text{d}v\nonumber\\
&-\int_{\Gamma}\sum_{i=1}^{n_\mathrm{b}} d^{i} \nabla N^{i}\cdot\bn \sum_{a=1}^{n_\mathrm{b}} c_{n}^{a} \nabla^2 N^a\text{d}s \nonumber\\
& +\int_{\Gamma}\sum_{i=1}^{n_\mathrm{b}} d^{i} N^{i}  \sum_{a=1}^{n_\mathrm{b}}  c_{n}^{a}  \nabla(\nabla^2 N^a)\cdot\bn\text{d}s.  \label{eq:weakBiharmonic3}
\end{align}
Since the $d^{i}$ are arbitrary, we define the first term
\begin{align}
\Xi^{\nabla^4 C}_i|_n=-\int_{\Omega}  \nabla^2N^{i} \sum_{a=1}^{n_\mathrm{b}}  c_{n}^{a}\nabla^2 N\text{d}v
\label{eq:basis_Biharmonic}
\end{align}
as the basis for the biharmonic operator $\nabla^4 C$ (with zero Neumann boundary condition).  The last two terms in \eqref{eq:weakBiharmonic3} are, respectively, a higher-order Dirichlet boundary condition (as emerges from variational calculus applied in the context of the full equation \cite{Rudrarajuetal2016}) and the Neumman boundary condition, from which we define:
\begin{align}
    \Xi^{\nabla^4 C_{\text{DBC}}}_i|_n &= -\int_{\Gamma} \nabla N^{i}\cdot\bn \sum_{a=1}^{n_\mathrm{b}} c_{n}^{a} \nabla^2 N^a\text{d}s\\
    \Xi^{\nabla^4 C_{\text{NBC}}}_i|_n &= \int_{\Gamma} N^{i}  \sum_{a=1}^{n_\mathrm{b}}  c_{n}^{a}  \nabla(\nabla^2 N^a)\cdot\bn\text{d}s.
\end{align}
The second-order gradients in the Equation (\ref{eq:basis_Biharmonic}) require the solutions and basis functions to lie in $\mathscr{H}^2(\Omega)$, while the Lagrange polynomial basis functions traditionally used in finite element analysis only lie in $\mathscr{H}^1(\Omega)$. We therefore draw the basis functions, $N$, from the family of Non-Uniform Rational B-Splines (NURBS), and adopt Isogeomeric Analysis (IGA) in our simulations to find the solutions in  $\mathscr{H}^2(\Omega)$. A discussion of the NURBS basis and IGA is beyond the scope of this paper; interested readers are directed to the original works on this topic \cite{CottrellHughesBazilevs2009}. We briefly present the construction of the NURBS basis functions in the following subsection. 

\subsection{NURBS basis functions}
\label{sec:iga}
Similar to Lagrange polynomial basis functions traditionally used in the Finite Element Method (FEM), NURBS basis functions are partitions of unity with compact support, satisfy affine covariance (i.e., an affine transformation of the basis is obtained by the affine transformation of its nodes/control points), and support an isoparametric formulation, thereby making them suitable for a Galerkin framework. They enjoy advantages over Lagrange polynomial basis functions in being able to ensure $C^n$-continuity,  possessing the positive basis and convex hull properties, and being variation diminishing.

The building blocks of the NURBS basis functions are univariate B-spline functions defined as follows. Consider two positive integers $p$ and $n$, and a non-decreasing sequence of values $\chi=[\xi_1, \xi_2,...., \xi_{n+p+1}]$, where $p$ is the polynomial order, $n$ is the number of basis functions, $\xi_i$ are coordinates in the parametric space referred to as knots (equivalent to nodes in FEM), and $\chi$ is the knot vector. The B-spline basis functions $B_{i,p}(\xi)$ are defined starting with the zeroth order basis
\begin{align}
B^i_0(\xi) &= \left\{\begin{array}{ll}
1 &\mathrm{if}\;\xi_i \le \xi < \xi_{i+1},\\
0 &\mathrm{otherwise},
\end{array}\right.
\end{align}
and higher orders using the Cox-de Boor recursive formula for $p \geq 1$ \cite{Piegl1997}
\begin{align}
 B^i_p (\xi) &=
  \frac{\xi-\xi_i}{\xi_{i+p}-\xi_i} B^i_{p-1} (\xi) + \frac{\xi_{i+p+1}-\xi}{\xi_{i+p+1}-\xi_{i+1}} B^{i+1}_{p-1} (\xi).
\end{align}

\begin{figure}[hbtp]
  \centering
\includegraphics[width=0.6\textwidth]{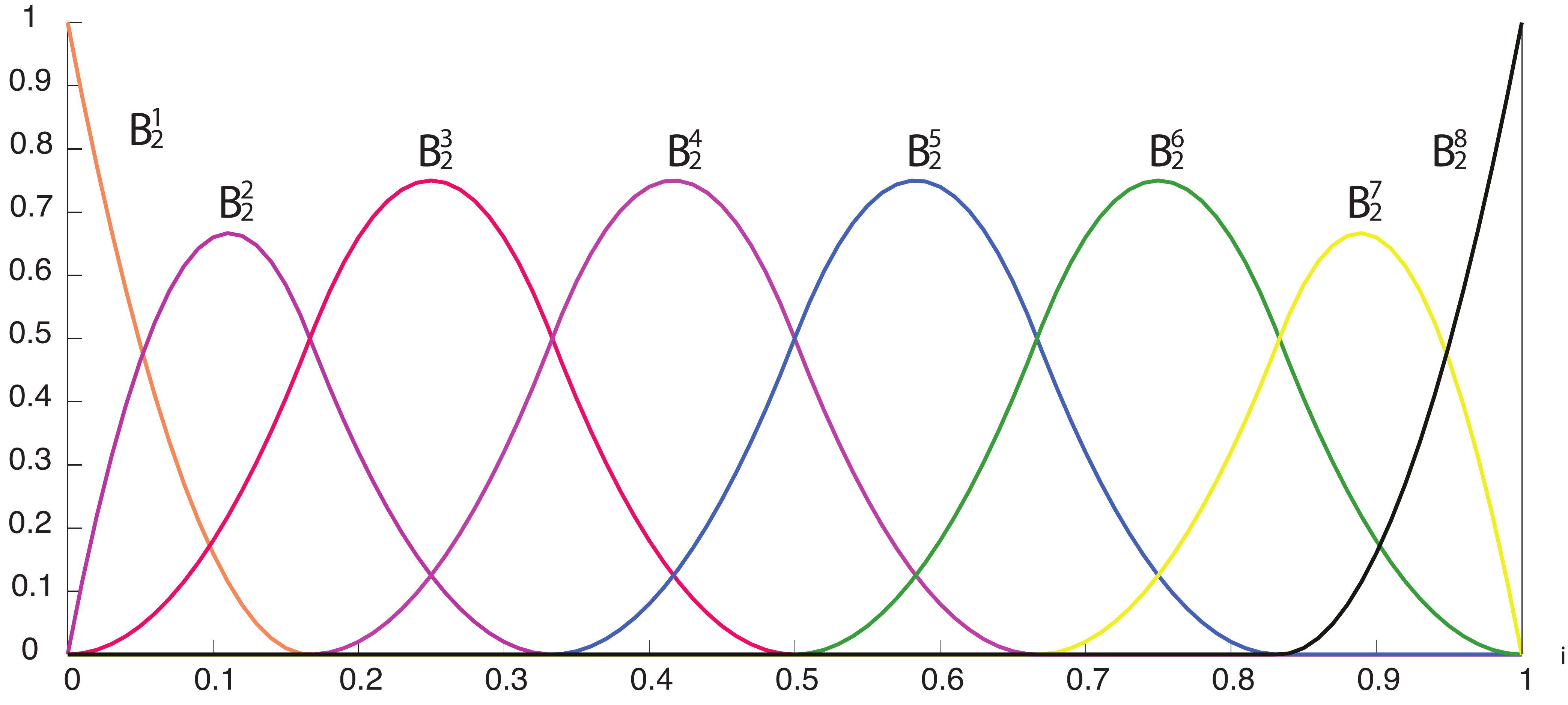}
\caption{A one-dimensional quadratic B-spline basis constructed from the knot vector $\chi = [0, 0, 0, 1/6, 1/3, 1/2, 2/3, 5/6, 1, 1, 1]$.}
\label{fig:bsplines}
\end{figure}

The knot vector divides the parametric space into intervals referred to as knot spans (equivalent to elements in FEM). A B-spline basis function is $C^{\infty}$-continuous inside knot spans and $C^{p-1}$-continuous at the knots. If an interior knot value repeats, it is referred to as a multiple knot. At a knot of multiplicity $k$, the continuity is $C^{p-k}$. Using a quadratic B-spline basis (Figure (\ref{fig:bsplines})), a $C^1$-continuous one dimensional NURBS basis can now be constructed: \begin{align}
N^{i}_p (\xi) =
  \frac{B^i_p (\xi) \textit{w}_{i}}{\sum_{i'=1}^{n_b} B^{i'}_p (\xi) \textit{w}_{i'}}
  \label{eq:onednurbs}
\end{align}
where $w_i$ are the weights associated with each of the B-spline functions. In higher dimensions, NURBS basis functions are constructed as a tensor product of their one-dimensional counterparts:
\begin{align}
 N^{ij}_p (\xi,\eta) &=
  \frac{B^i_p (\xi) B^j_p (\eta) \textit{w}_{ij}}{\sum_{i'=1}^{n_{b1}} \sum_{j'=1}^{n_{b2}} B^{i'}_p(\xi) B^{j'}_p (\eta) \textit{w}_{i'j'}} &\mathrm{(2D)} \label{eq:higherordernurbs2D} \\
 N^{ijk}_p (\xi,\eta, \zeta) &=
  \frac{B^i_p (\xi) B^j_p (\eta) B^k_p (\zeta) \textit{w}_{ijk}}{\sum_{i'=1}^{n_{b1}} \sum_{j'=1}^{n_{b2}} \sum_{k'=1}^{n_{b3}} B^{i'}_p(\xi) B^{j'}_p (\eta) B^{k'}_2 (\zeta) \textit{w}_{i'j'k'}}. &\mathrm{(3D)}
\label{eq:higherordernurbs3D}
\end{align}

Using NURBS basis functions, we are able to smoothly evaluate the higher order derivatives of the data by interpolating them from the DOF control variable values. This is crucial in enabling the identification of higher order derivatives (e.g., the second order derivatives that arise in the weak form of the Cahn-Hilliard equations) using the stepwise regression methods described next.

\subsection{Identification of basis operators via stepwise regression}
To identify a dynamical system, we need to generate all possible basis operators acting on the solution, compute the non-zero pre-factor for each basis operator that is in the model (relevant bases), while also attaining pre-factors of zero for the the basis operators that are not in the model (irrelevant bases). {\color{black} For situations where the true model is not contained within the set of candidate basis operators, our method then finds the optimal model (optimal in the sense of the loss function to be defined shortly) out of the set of all possible models. 
The resolution of such situations requires investigations into model inadequacy/misspecification (e.g., discussed in~\cite{Kennedy2001}), which is a challenging problem that we do not attempt to tackle here; instead, we limit our scope to work within a given set of candidate operators. } We begin by putting together the time derivative basis operators at each time $\{ \dots, t_{n-1}, t_n, t_{n+1},\dots\}$. Let
\begin{align}
\by=\left[
\begin{array}{c}
\vdots\\
\BXi^{\dot{C}}|_{n-1}\\
\BXi^{\dot{C}}|_{n}\\
\BXi^{\dot{C}}|_{n+1}\\
\vdots\\
\end{array}
\right]
\label{eq:targety}
\end{align}
be our target vector. Note that each element $\BXi^{\dot{C}}|_n \in \mathbb{R}^{N_{DOF}}$ 
in (\ref{eq:targety}) is a vector formed in Equation (\ref{eq:basis_Ct}):
\begin{align}
\BXi^{\dot{C}}|_{n}=\left[
\begin{array}{c}
\vdots\\
\Xi^{\dot{C}}_{i-1}|_{n}\\
\Xi^{\dot{C}}_i|_{n}\\
\Xi^{\dot{C}}_{i+1}|_{n}\\
\vdots\\
\Xi^{\dot{C}}_{N_\text{DOF}}|_{n}
\end{array}
\right]
\label{eq:targety2}
\end{align}
Likewise we can form the matrix, $\BXi$, containing all possible terms:
\begin{align}
\BXi=\left[
\begin{array}{cccccccc}
\vdots&\vdots&\vdots&\vdots&\vdots&\vdots&\vdots&\vdots\\
\BXi^\text{cons}|_{n-1}& \BXi^{C}|_n& \BXi^{{C}^2}|_{n-1}&...& \BXi^{\nabla^2 C}|_{n-1}&\BXi^{\nabla^4 C}|_{n-1}&\Xi^{\nabla^2 C_{\text{BC}}}|_{n-1}& ...\\
\BXi^\text{cons}|_{n}& \BXi^{C}|_n& \BXi^{{C}^2}|_{n}&...& \BXi^{\nabla^2 C}|_{n}&\BXi^{\nabla^4 C}|_{n}&\Xi^{\nabla^2 C_{\text{BC}}}|_n& ...\\
\BXi^\text{cons}|_{n+1}& \BXi^{C}|_{n+1}& \BXi^{{C}^2}|_{n+1}&...& \BXi^{\nabla^2 C}|_{n+1}&\BXi^{\nabla^4 C}|_{n+1}&\Xi^{\nabla^2 C_{\text{BC}}}|_{n+1}& ...\\
\vdots&\vdots&\vdots&\vdots&\vdots&\vdots&\vdots
\end{array}
\right]
\label{eq:xi}
\end{align}
The residual accounting for all DOFs and time steps can then be expressed as
\begin{align}
\mathscr{R}=\by-\BXi\Bomega.
\end{align}
The solution $\mathscr{R} = 0$ gives
\begin{align}
\by=\BXi\Bomega
\label{eq:least-square}
\end{align}
Note that the target vector in \eqref{eq:targety2} and  matrix $\BXi$ in \eqref{eq:xi} are currently structured to contain all DOFs. However we can always exclude some DOFs by deleting their corresponding rows. Additional attention is needed if one wishes to eliminate an entire basis operator. For instance, doing so to the Neumann boundary basis requires deletion of the corresponding column. The reduced Equation \eqref{eq:least-square} would then represent data with the homogeneous Neumann boundary condition.

If Equation \eqref{eq:least-square} were directly solved for $\boldsymbol{\omega}$ as an ordinary least squares regression problem, the analytic solution is
\begin{align}
\Bomega=(\BXi^T\BXi)^{-1}\BXi^T\by,
\label{eq:sol_least-square}
\end{align}
where a nontrivial $\by$ leads to a non-trivial solution to $\Bomega$. The least squares formulation can also be interpreted as an optimization problem with the loss function
\begin{align}
l=(\by-\BXi\Bomega)^T(\by-\BXi\Bomega),
\label{eq:loss_least-square}
\end{align}
which is simply the squared Euclidean norm of the residual vector obtained from the weak form. The system identification problem can then be viewed as finding $\Bomega$ that minimizes the loss:
\begin{align}
\Bomega =\text{arg }\underset{\widetilde{\Bomega}}\min\; l(\widetilde{\Bomega}).
\label{eq:OP1}
\end{align}
If the linear system in Equation \eqref{eq:least-square} is free of error, the loss function $l$ and the prefactors for irrelevant terms can all be driven down to machine zero. For a linear system containing error terms (further discussed in Section \ref{sec:Low_fidelity_noisy_data}), however, performing a standard regression will result in a solution of $\Bomega$ with nonzero contributions in all component of this vector. 
Such a solution of the system identification will not sharply delineate the relevant bases. Additionally, we could apply a separate least squares regression for each possible combination of all bases and choose the best one \cite{ISL}.\footnote{A technique called best subset selection.} However the problem of selecting the best model from among the $2^m$ possibilities grows exponentially in $m$, the number of all candidate basis. A viable alternative is to select the relevant bases by shrinking the pre-factors towards zero for the irrelevant bases. Ridge regression and $\ell_1$ regression (also known as least absolute shrinkage and selection operator (LASSO), a popular problem in compressive sensing research) are two well-known techniques to induce regularization. The optimization problem Equation \eqref{eq:OP1} is updated with additional penalty terms:
\begin{align}
\Bomega =&\text{arg }\underset{\Bomega}{\min} \text{ } \left\{l(\Bomega) +\lambda_2 \sum_j\omega_j^2\right\} \quad \text{Ridge Regression} \label{eq:ridge}\\
\Bomega =&\text{arg }\underset{\Bomega}{\min} \text{ } \left\{l(\Bomega)+ \lambda_1\sum_j|\omega_j|\right\}   \quad \text{LASSO},
\label{eq:OP_lasso}
\end{align}
where $\lambda_2$ and $\lambda_1$ are regularization parameters.  
For both methods, however, selecting a good regularization parameter is critical, and may need to leverage prior knowledge about the magnitude of pre-factors\cite{ISL}.

\subsubsection{Stepwise regression}
In this work, we use backward model selection by stepwise regression \cite{ISL}. The algorithm is summarized below.
\\

\noindent\fbox{%
\parbox{\textwidth}{%
\textbf{Algorithm for Model selection by Stepwise regression:}
\\
\\
\texttt{Step 0: Establish target vector $\by$ and matrix of bases, $\BXi$.}
\vspace{0.25cm}

\texttt{Step 1: Solve $\Bomega^i$ in the linear regression problem, Equation (\ref{eq:least-square}) using ordinary least squares regression. Calculate the loss function at this iteration, $l^i$.}
\vspace{0.25cm}

\texttt{Step 2: 
Eliminate basis terms in matrix $\BXi$  by deleting their columns, using the criterion to be introduced below. Set to zero the corrpesonding components of $\Bomega^i$. GOTO Step 1. Note that in this case the value of loss function remains small (\text{$l^{i}\sim l^{i-1}$}), and the solution may be overfitted. }
\vspace{0.25cm}

\texttt{Step 3: The algorithm stops if the pre-specified criterion does not allow us to eliminate any more basis terms. Beyond this, the loss function increases dramatically for any further reduction. }
}
}
\\
\\
There are several choices for the criterion for eliminating basis terms. Kutz and co-workers applied a hard threshold on the pre-factors in each iteration \cite{KutzPNAS2015, KutzIEEE2016, KutzSCIADV2017}. Since the results of sparse regression and sparsity profiles of $\Bomega$ vary with different values of the threshold, a different method is needed for the final solution corresponding to the optimal tolerance. In the work mentioned above, Kutz and co-workers used Pareto front analysis by cross-validation to find the optimal tolerance. The drawback of this approach, in our experience, is that its performance degrades if the pre-factors of the relevant basis operators differ significantly in magnitude.
For the approach to perform well, a carefully chosen rescaling is needed for the basis operators along with cross-validation.

Here, we explore, as an alternative, a widely used statistical criterion called the $F$-test \cite{ISL}, which can be motivated as follows: Since the model at iteration $i$ contains fewer bases than at iteration $i-1$, and is therefore more restrictive, the loss function, in general, is higher at iteration $i$ than at iteration $i-1$. In this sense, model always becomes ``worse'' after eliminating any basis. We want to determine whether the model at iteration $i$ is significantly worse than at iteration $i-1$. If not, the basis could be eliminated. We can evaluate the significance of the change by the $F$-test:
\begin{align}
F=\frac{\left( \frac{l^i-l^{i-1}}{p^{i-1}-p^{i}}\right)}{\frac{l^{i-1}}{n-p^{i-1}}}
\end{align}
where $p^i$ is the number of bases at iteration $i$ and $n$ is the total number of bases. The $F$ value essentially evaluates the increasing loss function under the consideration of model complexity. If $F$ does not exceed a pre-defined tolerance, $\alpha$, meaning that the model at iteration $i$ is not significantly worse than at iteration $i-1$, we eliminate the considered basis. Using the $F$-test in Step 2 of the above algorithm, this step can be re-stated as following: 
\\

\begin{figure}[hbtp]
\vspace{-5cm}
  \centering
\includegraphics[width=0.8\textwidth]{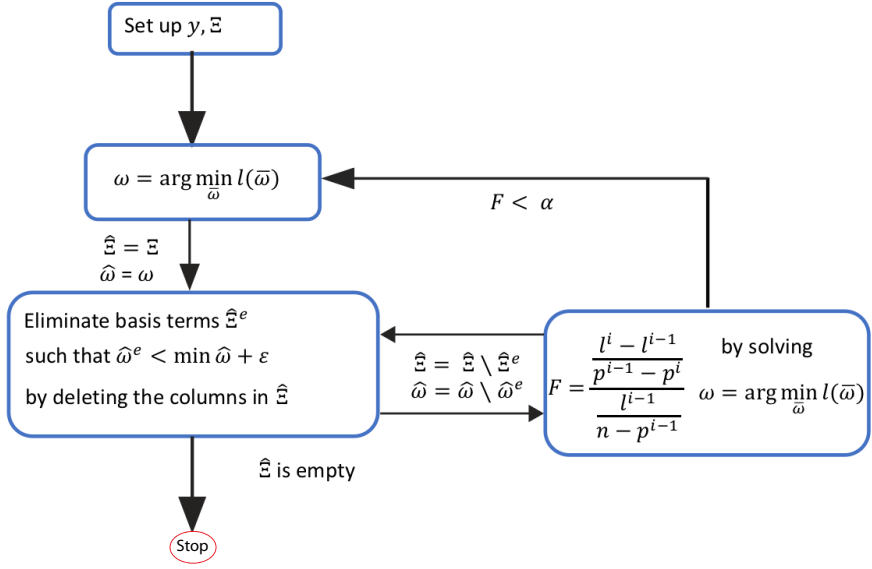}
\vspace{-5cm}
\caption{Schematic of the complete algorithm for stepwise regression.}
\label{fig:system_algorithm_flowchart}
\end{figure}

\noindent\fbox{%
\parbox{\textwidth}{%
\textbf{Step 2 with $F$-test:}
\\
\\
\texttt{Step 2.1:}\\
\texttt{Tentatively eliminate the basis corresponding to pre-factors in $\Bomega^i$ which are smaller than the pre-defined threshold, evaluate the $F$ value followed by ordinary least squares regression on the reduced bases set.}
\vspace{0.25cm}

\texttt{Step 2.2:}\\
\texttt{IF $F<\alpha$}\\
\texttt{THEN formally eliminate these bases in matrix $\BXi$, by deleting the corresponding columns. GOTO Step 1.}\\
\texttt{ELSE GOTO Step 2.1, and choose another basis.}
}
}
\\
\\
In this work we defined the threshold to be:
\begin{align}
\text{threshold}^i=\omega^i_\text{smallest}+\epsilon
\label{eq:threshold}
\end{align}
where $\omega^i_\text{smallest}$ is the smallest value in $\Bomega^i$, and $\epsilon$ is a small tolerance, and chosen to be $10^{-5}$. This threshold will eliminate the basis with the smallest pre-factor, as well as all bases with pre-factors within $\epsilon$ of the smallest one. Note that $\alpha$ in the $F$-test is the only hyperparameter in the algorithm. 
System identification is significantly more challenging if greater numbers of candidate bases are retained in the first few iterations. A small value of $\alpha$ may not eliminate all irrelevant bases at once, but can serve as a good initial value as it will filter some irrelevant bases, thus mollifying the problem to a degree. In this work, we choose $\alpha=1$ initially, and increase it to 10 when no more basis can be eliminated for $\alpha=1$.\footnote{For a different problem, the value can be rigorously chosen by cross-validation.}. We have summarized the step-wise algorithm in Figure \ref{fig:system_algorithm_flowchart}.
As mentioned previously, our approach builds on recent work in sparse regression \cite{KutzPNAS2015, KutzIEEE2016, KutzSCIADV2017}. However, it differs prominently from those papers in adopting a variational setting, using the $F$-test for determining bases, and, as will be seen in Section \ref{sec:Low_fidelity_noisy_data}, in considering the influence of data fidelity and noise.

{\color{black} More broadly, the purpose of the $F$-test is to perform model selection: a criterion to judge whether the new candidate model with fewer operators is more suitable than the previous one. At the core, the comparison is made to balance the desires of (a) fitting the data and (b) reducing the model complexity. Indeed, model selection is a rich and active area of research, encompassing methods such as cross-validation~\cite{Hastie2009,Picard2010}, Akaike information criterion (AIC)~\cite{Akaike1974}, Bayesian information criterion (BIC)~\cite{Schwarz1978}, and Bayes factor~\cite{Kass1995,Wasserman2000}. We will explore additional model selection techniques in the future, especially those that require a Bayesian version of our system identification framework.}


\subsection{Low fidelity and noisy data}
\label{sec:Low_fidelity_noisy_data}
\begin{figure}[hbtp]
  \centering
\includegraphics[width=0.6\textwidth]{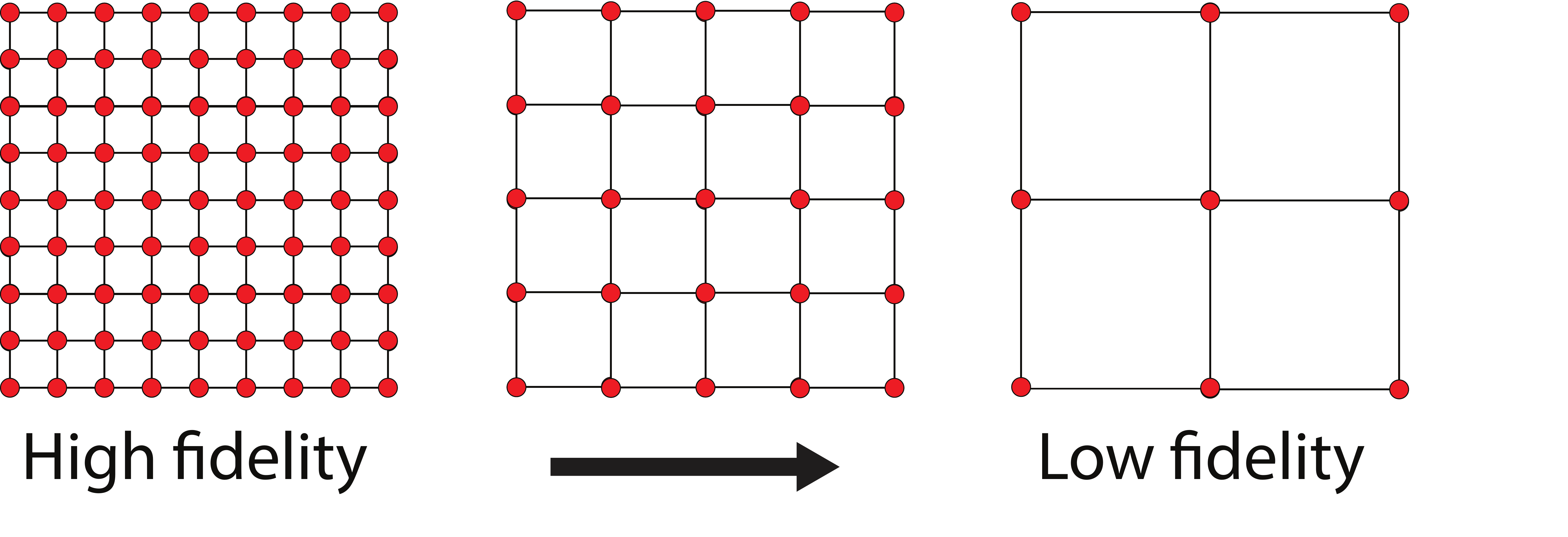}
\caption{Different fidelity of data on different mesh sizes.  }
\label{fig:mesh}
\end{figure}

In practice, we may not be able to collect high fidelity data in space. As illustrated in Figure \ref{fig:mesh}, we consider the situation in which data is gathered at fewer locations in space, or equivalently from fewer instances in time. The low fidelity data yields larger element sizes when constructing the finite dimensional spatial bases, and larger time steps, for use in the weak form. 
Furthermore, the data could be contaminated by background noise as well as data collection error, such that the observed data $\widehat{C}$ may be noisy. We model these imperfections using an additive form
 \begin{align}
    \widehat{C}^h=C^h+\epsilon.
    \label{eq:C_noise}
 \end{align}
 Consequently the target vector $\widehat{\by}$ and every basis in $\widehat{\BXi}$ constructed from $\widehat{C}^h$ will contain error terms.  
 For $\widehat{\by}$, we have 
 \begin{align}
\Xi^{\dot{\widehat{C} }}_i|_n=\int_{\Omega} N^i \sum_{a=1}^{n_\mathrm{b}} \frac{c_{n}^{a}+\epsilon_{n}-(c_{n-1})^{a}-\epsilon_{n-1} }{\Delta t} N^a\text{d}v  
\label{eq:basis_Ct_noisy}
\end{align}
We further assume that $\epsilon$, the noise on $C^h$, follows an independent and identically distributed (i.i.d.) Gaussian distribution\footnote{While the observation noise may have systematic bias and correlation, we do not explore these possibilities in this paper, and instead take an initial step to assume an i.i.d. setting.} with zero mean and standard deviation $\sigma$ :
 \begin{align}
    \epsilon\sim \mathcal{N}(0,\sigma^2)
    \label{eq:dis_noise}
 \end{align}
Since linear combinations of mutually independent Gaussian random variables are also Gaussian, we have: 
 \begin{align}
\Xi^{\dot{\widehat{C} }}_i|_n = \int_{\Omega} N^i \sum_{a=1}^{n_\mathrm{b}} \frac{c_{n}^{a}-(c_{n-1})^{a} }{\Delta t} N^a\text{d}v + \int_{\Omega} N^i \sum_{a=1}^{n_\mathrm{b}} \frac{\epsilon_{\dot{C}} }{\Delta t} N^a\text{d}v
\label{eq:basis_C_dot_noise}
 \end{align}
 where $\epsilon_{\dot{C}}$ is a Gaussian distribution with zero mean and standard deviation $\sqrt{2}\sigma$. Since the integral is computed using quadrature, we have:
 \begin{align}
\Xi^{\dot{\widehat{C} }}_i|_n = \sum_{q=1}^Q N^i \sum_{a=1}^{n_\mathrm{b}} \frac{c_{n}^{a}-(c_{n-1})^{a} }{\Delta t} N^a|J|w_q + \sum_{q=1}^Q N^i \sum_{a=1}^{n_\mathrm{b}} \frac{\epsilon_{\dot{C}} }{\Delta t} N^a|J|w_q,
\label{eq:basis_C_dot_noise_q}
 \end{align}
where $|J|$ is the determinant of the Jacobin matrix of the element mapping from an isoparametric domain, and $w_q$ is the weight for the quadrature point $q$. The first term on the right hand side reflects the noiseless (true) component of $\Xi^{\dot{\widehat{C} }}_i|_t$, and the second term is the error. 
the true term converges to the time derivative:
 \begin{align}
     \lim\limits_{\Delta t \rightarrow 0}\frac{c^a_{n}-(c^a_{n-1}) }{\Delta t} = \frac{\partial C^h}{\partial t},
     \label{eq:back_euler}
 \end{align}
however the error term 
scaled by the time step diverges as $\Delta t\longrightarrow 0$. It also indicates that the error due to noise decreases if larger time steps are used:
\begin{align}
    \frac{\text{error term}}{\text{true term}}\sim\frac{1}{\Delta t},
    \label{eq:basis_back_euler_error}
\end{align}
For gradient operators, such as the Laplacian and the Biharmonic operator, we have
\begin{align}
\Xi^{\nabla^2 \widehat{C}}_i|_{t}=\sum_{q=1}^Q \nabla N^i \cdot\sum_{a=1}^{n_\mathrm{b}}  c_{n}^{a}\nabla N^a|J|w_q +\sum_{q=1}^Q \nabla N^i \cdot\sum_{a=1}^{n_\mathrm{b}}  \epsilon\nabla N^a|J|w_q
\label{eq:basis_laplace_noise}
\end{align}
\begin{align}
\Xi^{\nabla^4 \widehat{C}}_i|_{n}=\sum_{q=1}^Q \nabla^2 N^{i} \sum_{a=1}^{n_\mathrm{b}}  c_{n}^{a}\nabla^2 N|J|w_q+\sum_{q=1}^Q  \nabla^2 N^{i}\sum_{a=1}^{n_\mathrm{b}}  \epsilon\nabla^2 N|J|w_q,
\label{eq:basis_Biharmonic_noise}
\end{align}
from which, since the forward solution converges with decreasing mesh size, $h$, we have:
\begin{align}
\lim\limits_{h\rightarrow 0}\nabla N^i\sum_{a=1}^{n_\mathrm{b}}  c_{n}^{a}\nabla N^a = &\nabla^2 C^h  \label{eq:nabla2_basis}\\
\lim\limits_{h\rightarrow 0}\nabla^2 N^i\sum_{a=1}^{n_\mathrm{b}}  c_{n}^{a}\nabla^2 N^a =&  \nabla^4 C^h.
\label{eq:nabla4_basis}
\end{align}
However the error terms are scaled by $h^{-2}$ in the noisy Laplacian operator and $h^{-4}$ in the noisy Biharmonic operator, where the mesh size $h$ is introduced by the gradients of the basis functions. The error on $\widehat{C}^h$ is amplified by the mesh size acting through the spatial gradient operators and diverges as $h\longrightarrow 0$. Consequently, we have
\begin{align}
    \frac{\text{error term}}{\text{true term}}\sim h^{-2}& \quad \text{for the Laplacian operator} \label{eq:basis_laplace_error}\\
        \frac{\text{error term}}{\text{true term}}\sim h^{-4}& \quad \text{for the Biharmonic operator.} \label{eq:basis_Biharmonic_error}
\end{align}
The above analysis illustrate that the final error induced by noise on $C^h$  will decrease with larger time steps and element sizes. In Section \ref{sec:example} we show that for noisy $C^h$, using low fidelity data improves the results of system identification. For purely algebraic bases, i.e., without differential operators from time derivatives or spatial gradients, time step and element size do not affect the ratio of error to true value. Note also that the approximations in Equations (\ref{eq:back_euler}), (\ref{eq:nabla2_basis}) and (\ref{eq:nabla4_basis}) degrades with increasing time step and element size, bringing in discretization errors.

\section{Examples}
\label{sec:example}
We now turn to using the framework detailed in the preceding section to identify the parabolic PDEs that govern patterning. Instead of directly applying our algorithms to data from physical experiments, in this first communication, we test our methods on identifying PDEs from data obtained through high-fidelity, direct numerical simulations (DNS).

We consider test cases with the following four data-generating models, with their true coefficients summarized in Table \ref{ta:parameters}: 

\noindent \textbf{Model 1:}
\begin{align}
\frac{\partial C_1}{\partial t}&=D_1\nabla^2C_1+R_{10}+R_{11}C_1+R_{13}C_1^2C_2\\    
\frac{\partial C_2}{\partial t}&=D_2\nabla^2C_2+R_{20}+R_{21}C_1^2C_2\\
\text{with}& \quad \nabla C_1\cdot\bn=0; \quad \nabla C_2\cdot\bn=0 \text{ on }\Gamma 
\end{align}
where $\Gamma$ is the domain boundary.

\noindent \textbf{Model 2:}
\begin{align}
\frac{\partial C_1}{\partial t}&=D_1\nabla^2C_1+R_{10}+R_{11}C_1+R_{11}C_1^2C_2\\    
\frac{\partial C_2}{\partial t}&=D_2\nabla^2C_2+R_{20}+R_{21}C_1^2C_2\\
\text{with} \quad &\nabla C_1\cdot\bn=j_n \text{ on }\Gamma_2;\quad
 \quad \nabla C_1\cdot\bn=0 \text{ on }\Gamma \backslash \Gamma_2; \nonumber\\
 &\nabla C_2\cdot\bn=0 \text{ on }\Gamma 
\end{align}
where $\Gamma_2$ and $\Gamma_4$ are shown in Figure \ref{fig:ini_BC}.
Models 1 and 2 represent coupled diffusion-reaction equations for two species following Schnakenberg kinetics \cite{Schnakenberg1976}, but with different boundary conditions. For an activator-inhibitor species pair, these equations use auto-inhibition with cross-activation of a short range species, and auto-activation with cross-inhibition of a long range species to form so-called Turing patterns \cite{Turing1952}.

\noindent \textbf{Model 3:}
\begin{align}
  \frac{\partial C_1}{\partial t}&=\nabla \cdot (M_1\nabla\mu_1)\label{eq:Cahn-C1}\\
    \frac{\partial C_2}{\partial t}&=\nabla \cdot (M_2\nabla\mu_2)\label{eq:Cahn-C2}\\
\mu_1&=\frac{\partial g}{\partial C_1}-k_1\nabla^2C_1 \label{eq:Cahn-mu_1}\\
\mu_2&=\frac{\partial g}{\partial C_2}-k_2\nabla^2C_2 \label{eq:Cahn-mu_2}\\
 \text{with}& \quad \nabla \mu_1\cdot\bn=0; \nabla C_1\cdot\bn = 0 \text{ on }\Gamma  \label{eq:CahnHillDirBC}\\
 & \quad \nabla \mu_2\cdot\bn=0; \nabla C_2\cdot\bn = 0  \text{ on }\Gamma
\end{align}
where $g$ is a non-convex, ``homogeneous'' free energy density function, whose form has been chosen from \cite{GarikipatiJMPS2017}:
\begin{align}
   g(C_1,C_2)&=\frac{3d}{2s^4}\left((2C_1-1)^2+(2C_2-1)^2\right)^2+\frac{d}{s^3}(2C_2-1)\left((2C_2-1)^2-3(2C_1-1)^2\right)\nonumber\\
      &-\frac{3d}{2s^2}\left((2C_1-1)^2+(2C_2-1)^2\right).
   \label{eq:freeEnergy_g}
 \end{align}
 
\noindent \textbf{Model 4:}
\begin{align}
  \frac{\partial C_1}{\partial t}&=\nabla \cdot (M_1\nabla\mu_1)\label{eq:Alen-C1}\\
    \frac{\partial C_2}{\partial t}&=-M_2\mu_2\label{eq:ALen-C2}\\
\mu_1&=\frac{\partial g}{\partial C_1}-\nabla\cdot k_1\nabla C_1 \label{eq:Alen-mu-eta}\\
\mu_2&=\frac{\partial g}{\partial C_2}-\nabla\cdot k_2\nabla C_2 \label{eq:Alen-J}\\
 \text{with}& \quad \nabla \mu_1\cdot\bn=0; \nabla C_1\cdot\bn = 0 \text{ on }\Gamma  \label{eq:AllenCahnDirBC}\\
 & \quad \nabla C_2\cdot\bn=0 \text{ on }\Gamma 
\end{align}

\begin{figure}[hbtp]
\centering
\includegraphics[scale=0.2]{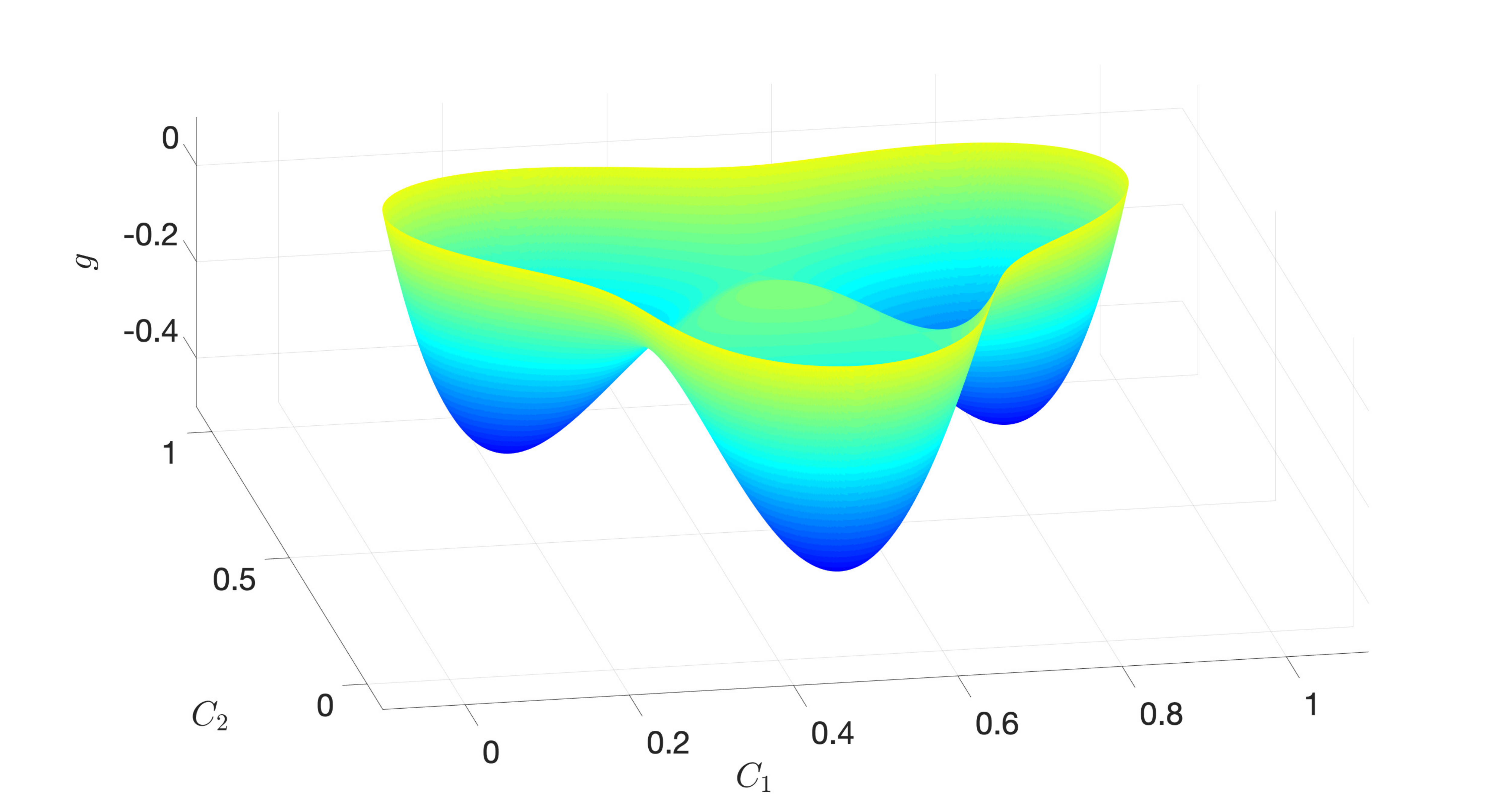}
\caption{The three wells non-convex tissue energy density function. }
\label{fig:free_energy}
\end{figure}

Model 3 is a two field Cahn-Hilliard system with the well-known fourth-order term in the concentration, $C_1$ and $C_2$. The three-well non-convex free energy density function (See Figure \ref{fig:free_energy}), $g(C_1,C_2)$, drives segregation of the system into two distinct types. We have previously used this system to make connections with cell segregation in developmental biology \cite{GarikipatiJMPS2017}.
The diffusion-reaction and Cahn-Hilliard equations are widely used in biological pattern generation, as discussed in the Introduction. The Cahn-Hilliad equation \cite{CahnHilliard1958} also occupies a central role in the materials physics literature for modelling phase transformations developing from a uniform concentration field in the presence of an instability. 

Model 4 is the coupled Cahn-Hilliard/Allen-Cahn equations system \cite{Allen1979} where the field $C_2$ describes growth of alloy precipitates from nuclei created by a process of spinodal decomposition that controls the field $C_1$. The free energy density $g(C_1, C_2)$ couples these processes through $\mu_1$ and $\mu_2$ in both Models 3 and 4. The Introduction also cites some of the relevant literature on the use of Cahn-Hilliard and Allen-Cahn equations in materials physics. An important difference between Models 3 and 4 is that $C_2$ in Model 4 is a non-conserved order parameter, which defines the identities of the precipitate, described by composition field $C_1$, and matrix phases. Both Cahn-Hilliard equations in Model 3 are conservative. Note that solving the Cahn-Hilliard and Allen-Cahn equations in the isogeometric analytic framework requires imposition of the higher-order Dirichlet boundary conditions \eqref{eq:CahnHillDirBC} and \eqref{eq:AllenCahnDirBC}. We use Nitsche's method  \cite{ShivaNPJCM2016} to impose these boundary conditions weakly.

\begin{table}[h]
\centering
 \begin{tabular}{|c|c|c|c|c|c|c|c|c|c|c|c|c|c|}
 \hline
$D_1$&$D_2$ & $R_{10}$ & $R_{11}$ & $R_{13}$ & $R_{20}$ &$R_{21}$& $j_n$ & $M_1$& $M_2$& $k_1$&$k_2$ &$d$ &$s$\\\hline
1&40&0.1&-1&1&0.9&-1&0.1& 0.1& 0.1& 10&10& 0.4& 0.7\\ \hline
\end{tabular}
\caption{parameters used in the simulations.}
\label{ta:parameters}
\end{table}
Substituting the parameter values from Table \ref{ta:parameters}, we present the weak form of each model:

\noindent \textbf{Model 1:}
\begin{align}
\int_{\Omega}w_1\frac{\partial C_1}{\partial t}\text{d}v&=\int_{\Omega}-1\nabla w_1 \cdot\nabla C_1\text{d}v
+\int_{\Omega}w_1 (0.1-C_1+1C_1^2C_2)\text{d}v 
\label{eq:weak_value_model1-1}\\    
\int_{\Omega}w_2\frac{\partial C_2}{\partial t}\text{d}v&=\int_{\Omega}-40\nabla w_2\cdot\nabla C_2+\int_{\Omega}w_2(0.9-1C_1^2C_2)\text{d}v
\label{eq:weak_value_model1-2}
\end{align}

\noindent \textbf{Model 2:}
\begin{align}
\int_{\Omega}w_1\frac{\partial C_1}{\partial t}\text{d}v&=\int_{\Omega}-1\nabla w_1 \cdot\nabla C_1\text{d}v
+\int_{\Omega}w_1 (0.1-C_1+1C_1^2C_2)\text{d}v +\int_{\Gamma_2}w_10.1\text{d}s
\label{eq:weak_value_model2-1}\\    
\int_{\Omega}w_2\frac{\partial C_2}{\partial t}\text{d}v&=\int_{\Omega}-40\nabla w_2\cdot\nabla C_2+\int_{\Omega}w_2(0.9-1C_1^2C_2)\text{d}v
\label{eq:weak_value_model2-2}
\end{align}

\noindent \textbf{Model 3:}
\begin{align}
\int_{\Omega}w_1\frac{\partial C_1}{\partial t}\text{d}v=&\int_{\Omega}\nabla w_1\cdot\left(-17.8126+47.98C_1+21.591C_2-47.98C_1^2-15.9933C_2^2 \right)\nabla C_1\text{d}v\nonumber\\
&+ \int_{\Omega}\nabla w_1\nabla\cdot\left(-10.7955+21.591C_1+15.9933C_2-31.9867C_1C_2 \right)\nabla C_2\text{d}v\nonumber\\
&+\int_{\Omega}-1\nabla^2w_1\nabla^2C_1
\label{eq:weak_value_model3-1}
\\
\int_{\Omega}w_2\frac{\partial C_2}{\partial t}\text{d}v=&\int_{\Omega}\nabla w_2\cdot\left(-10.7955+21.591C_1+15.9933C_2-31.9867C_1C_2 \right)\nabla C_1\text{d}v\nonumber\\
&+ \int_{\Omega}\nabla w_2\nabla\cdot\left(-12.2149+15.9933C_1+42.3823C_2-15.9933C_1^2-47.98C_2^2 \right)\nabla^2C_2\text{d}v\nonumber\\
&+\int_{\Omega}-1\nabla^2w_2\nabla^2C_2
\label{eq:weak_value_model3-2}
\end{align}

\noindent \textbf{Model 4:}
\begin{align}
\int_{\Omega}w_1\frac{\partial C_1}{\partial t}\text{d}v=&\int_{\Omega}\nabla w_1\cdot\left(-17.8126+47.98C_1+21.591C_2-47.98C_1^2-15.9933C_2^2)\nabla C_1 \right)\text{d}v\nonumber\\
&+ \int_{\Omega}\nabla w_1\nabla\cdot\left(-10.7955+21.591C_1+15.9933C_2-31.9867C_1C_2 \right)\nabla C_2\text{d}v\nonumber\\
&+\int_{\Omega}-1\nabla^2w_1\nabla^2C_1
\label{eq:weak_value_model4-1}\\
\int_{\Omega}w_2\frac{\partial C_2}{\partial t}\text{d}v&=\int_{\Omega}-1\nabla w_2\cdot\nabla C_2+\int_{\Omega}+w_2(3.5085-10.7955C_1-12.2149C_2 \nonumber\\
&+10.7955C_1^2+15.9933C_1C_2+21.1912C_2^2-15.9933C_1^2C_2-15.9933C_2^3)\text{d}v
\label{eq:weak_value_model4-2}
\end{align}
The higher order boundary conditions and stabilization terms in Nitsche's methods are not shown here. We do not infer these terms since they are of a numerical rather than physical nature.
\\
\\

\subsection{Data preparation}
\begin{figure}[hbtp]
\centering
\includegraphics[scale=0.2]{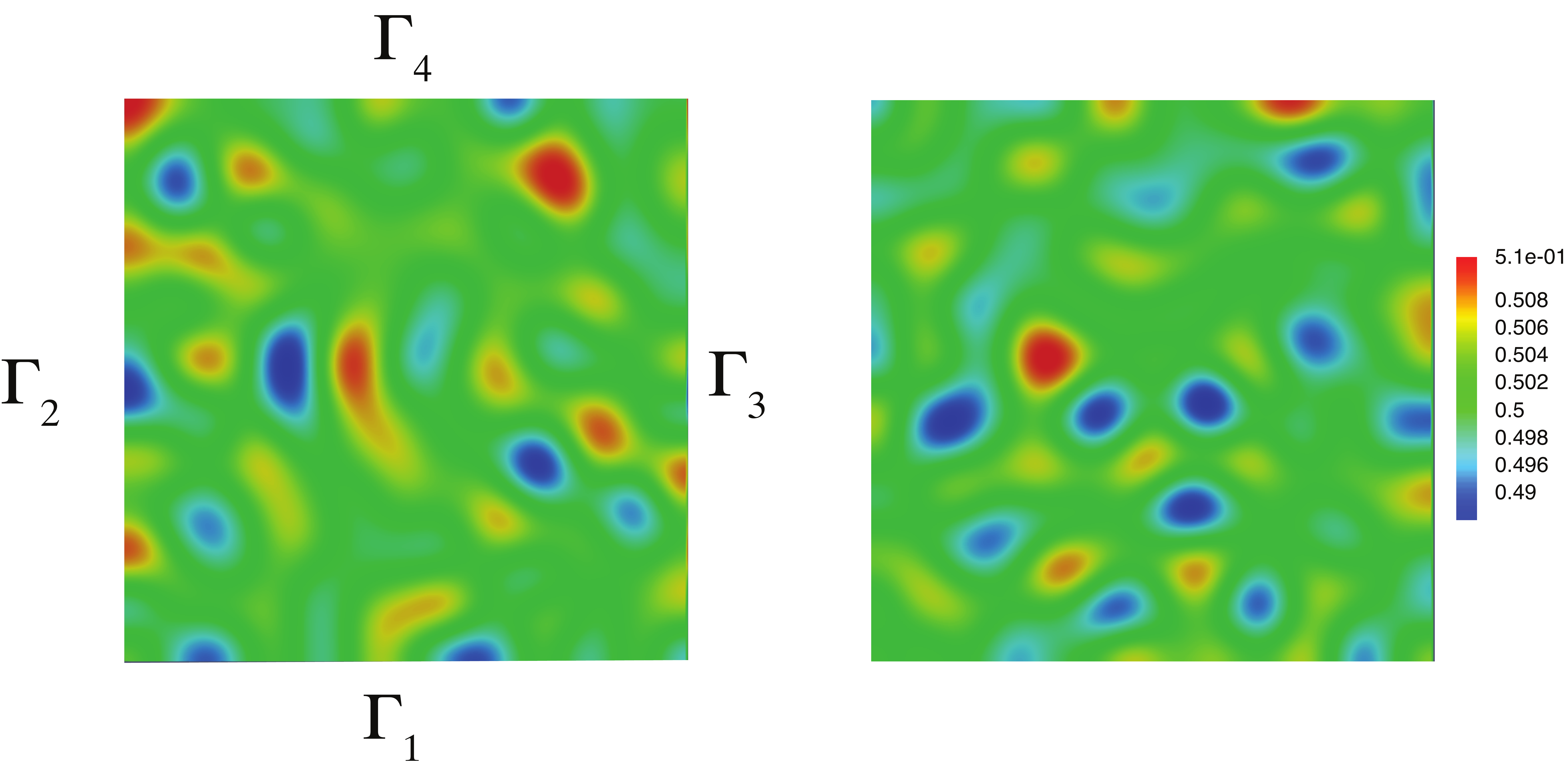}
\caption{Concentration of $C_1$ (left) and $C_2$ (right) in all four models at $t=0$.}
\label{fig:ini_BC}
\end{figure}

\begin{figure}[hbtp]
\centering
$C_1$\\
\includegraphics[scale=0.12]{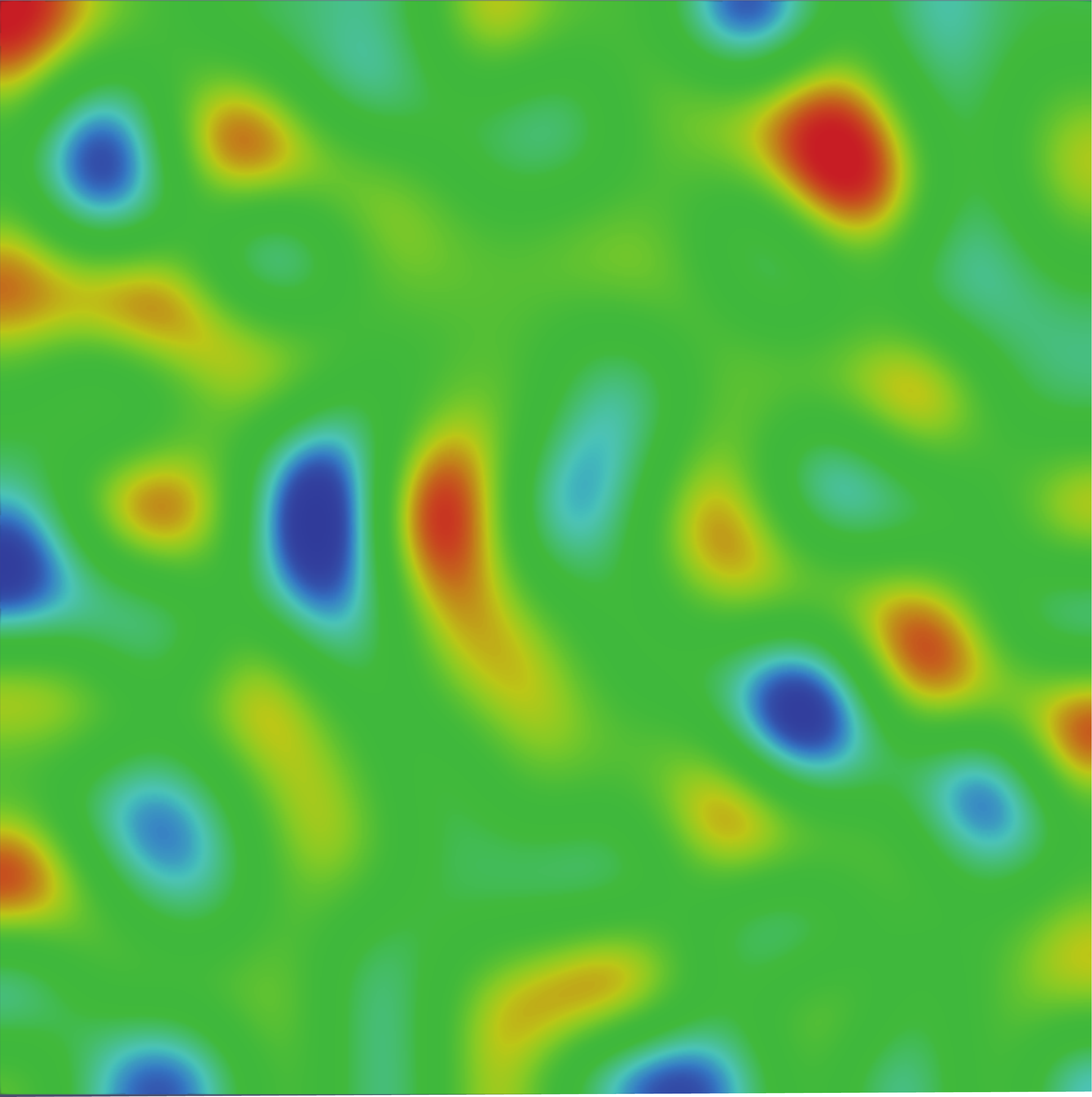}
\includegraphics[scale=0.12]{42_C1.pdf}
\includegraphics[scale=0.12]{42_C1.pdf}
\includegraphics[scale=0.12]{42_C1.pdf}\\
$C_2$\\
\includegraphics[scale=0.12]{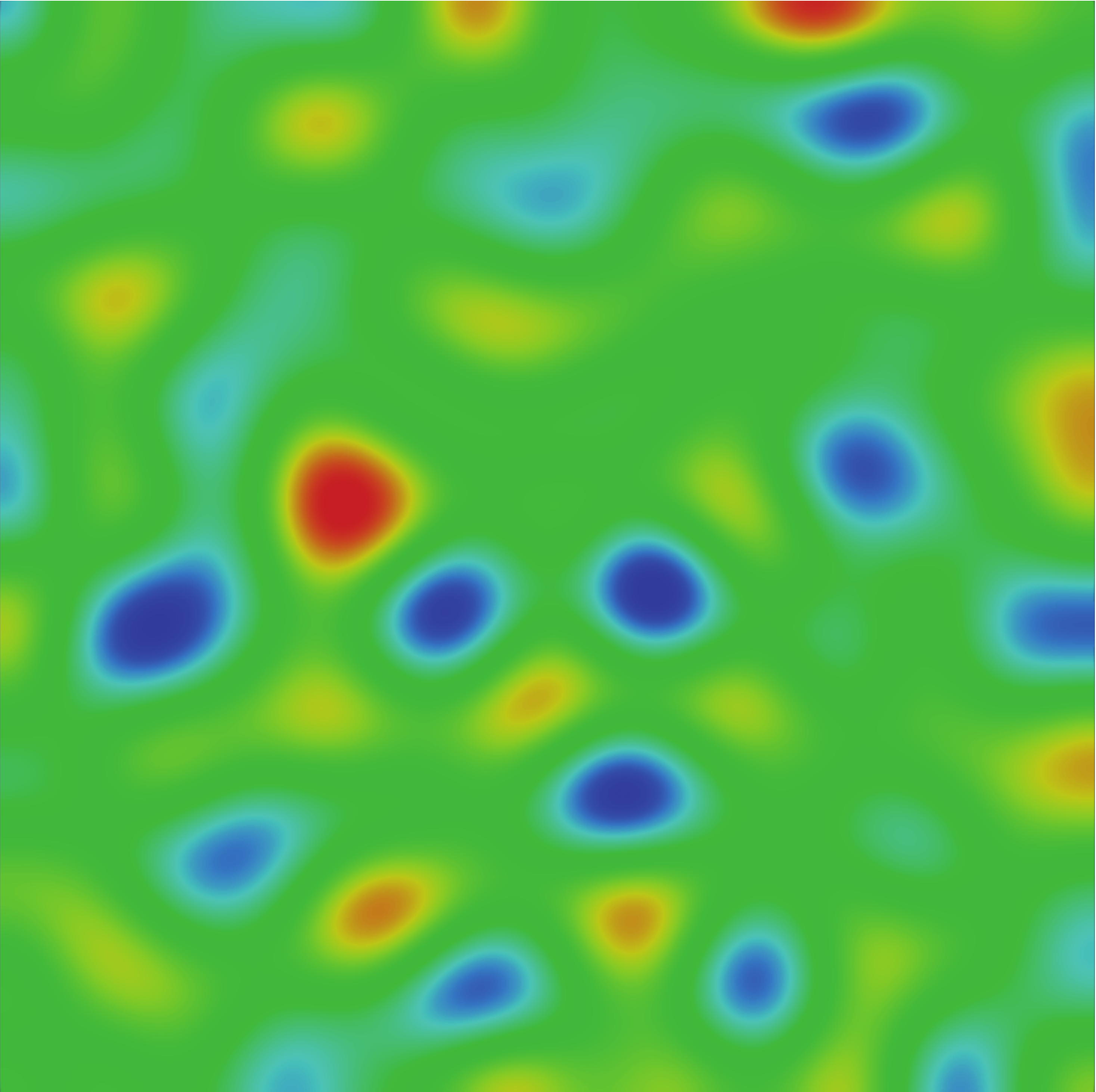}
\includegraphics[scale=0.12]{42_C2.pdf}
\includegraphics[scale=0.12]{42_C2.pdf}
\includegraphics[scale=0.12]{42_C2.pdf}
\caption{$C_1$ and $C_2$ for Model 1 to 4 ( left to right) at $t=0.01$ are almost identical.}
\label{fig:C_001}
\end{figure}

\begin{figure}[hbtp]
\centering
$C_1$\\
\includegraphics[scale=0.12]{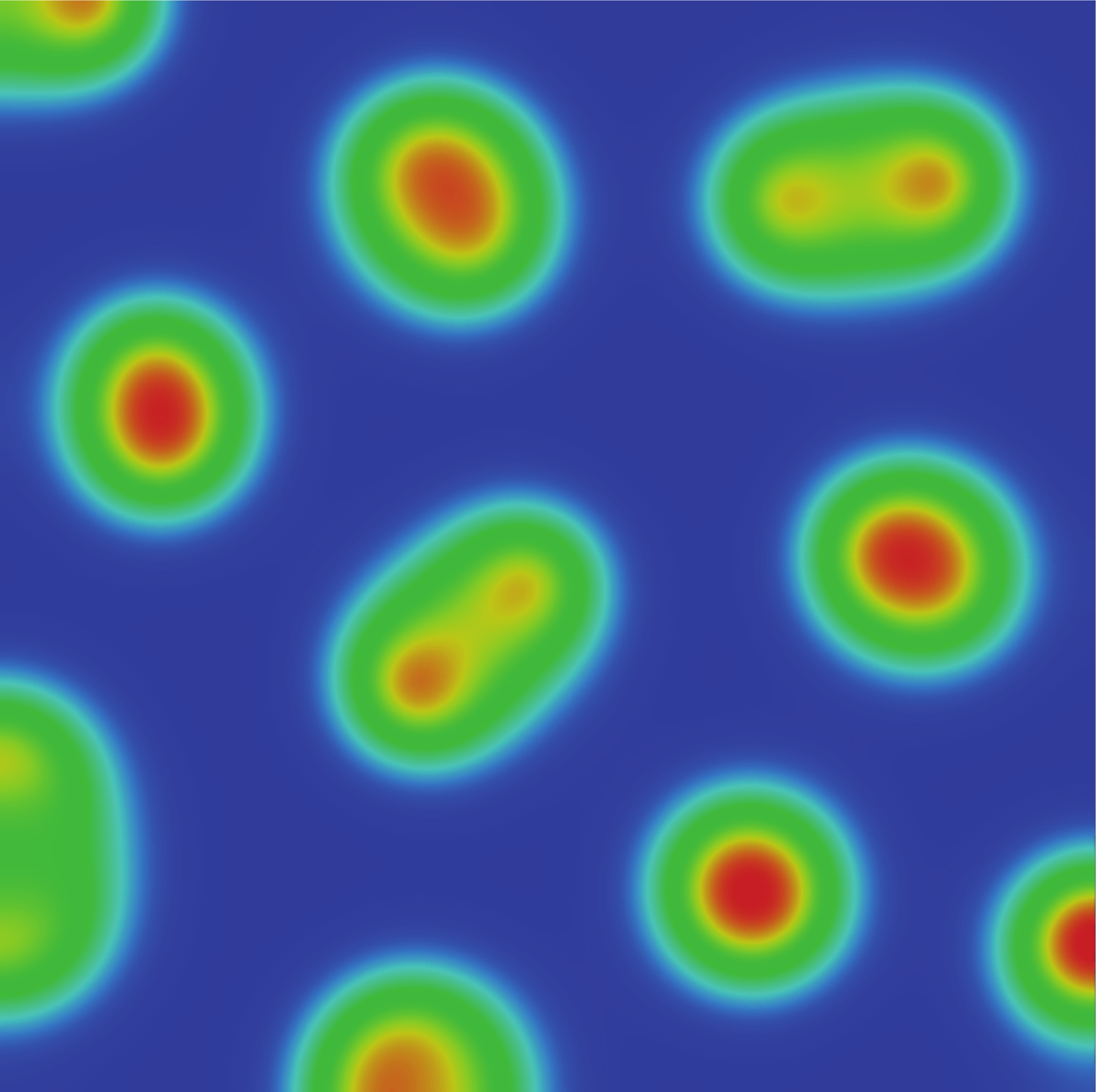}
\includegraphics[scale=0.12]{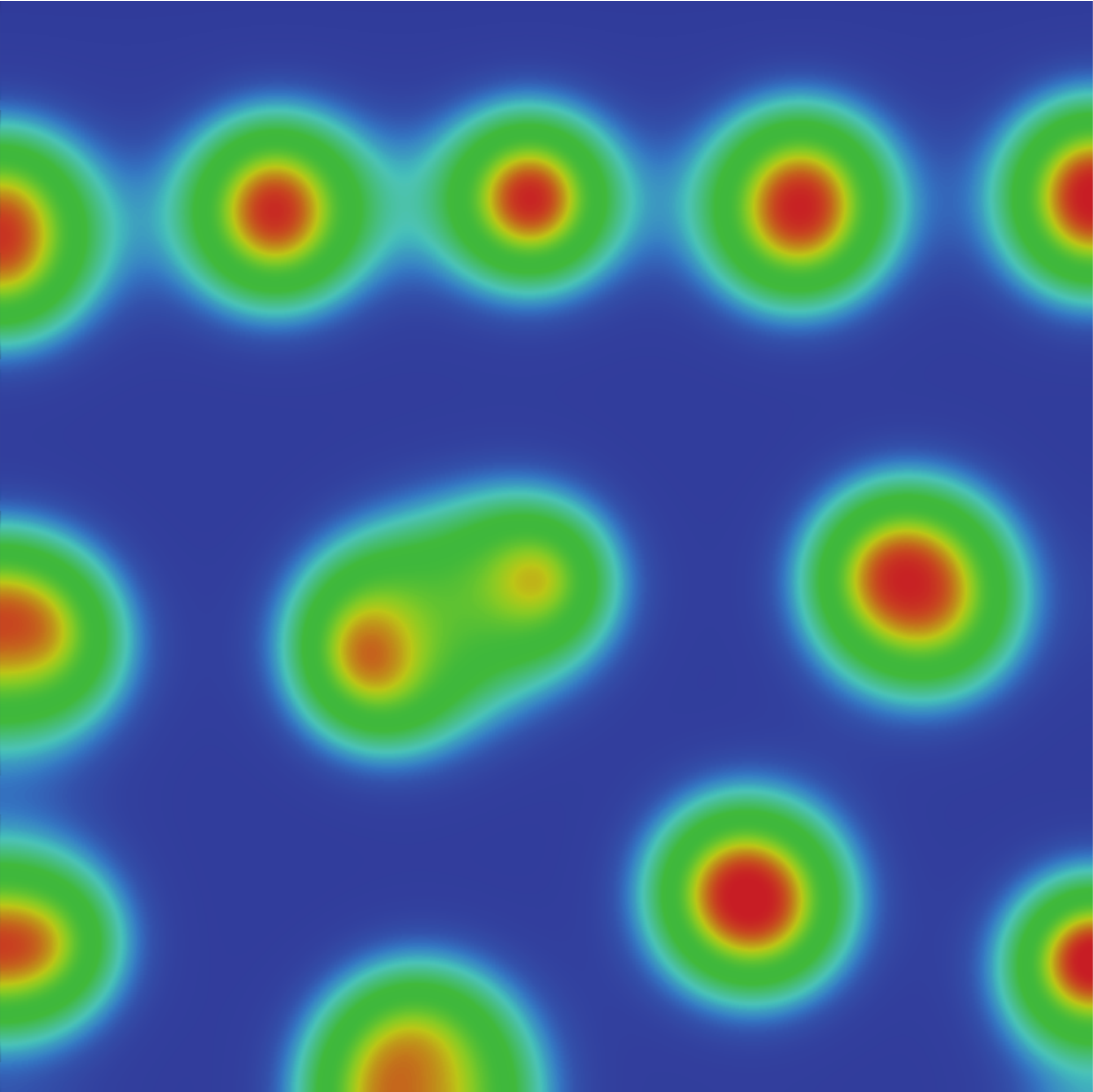}
\includegraphics[scale=0.12]{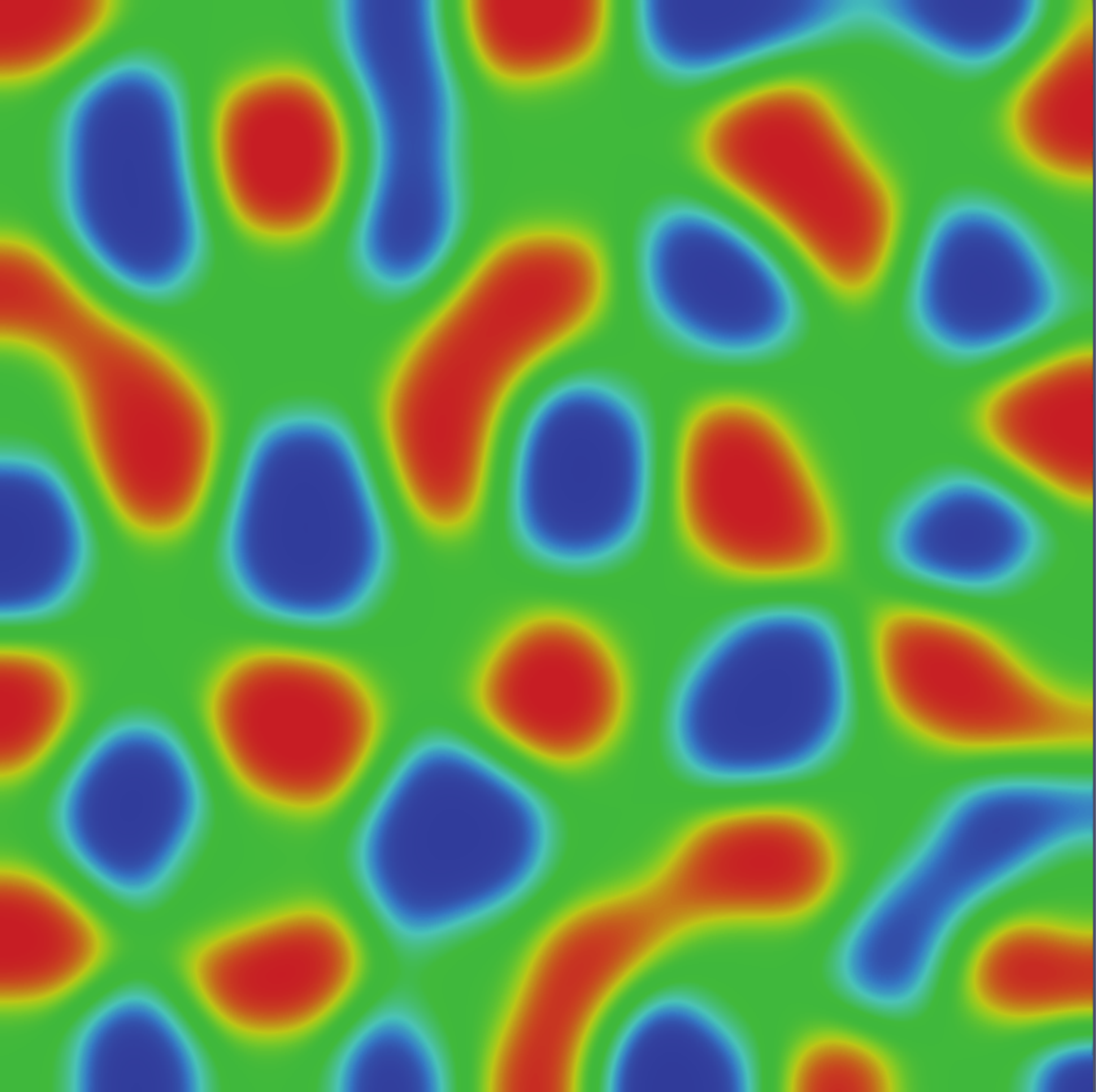}
\includegraphics[scale=0.12]{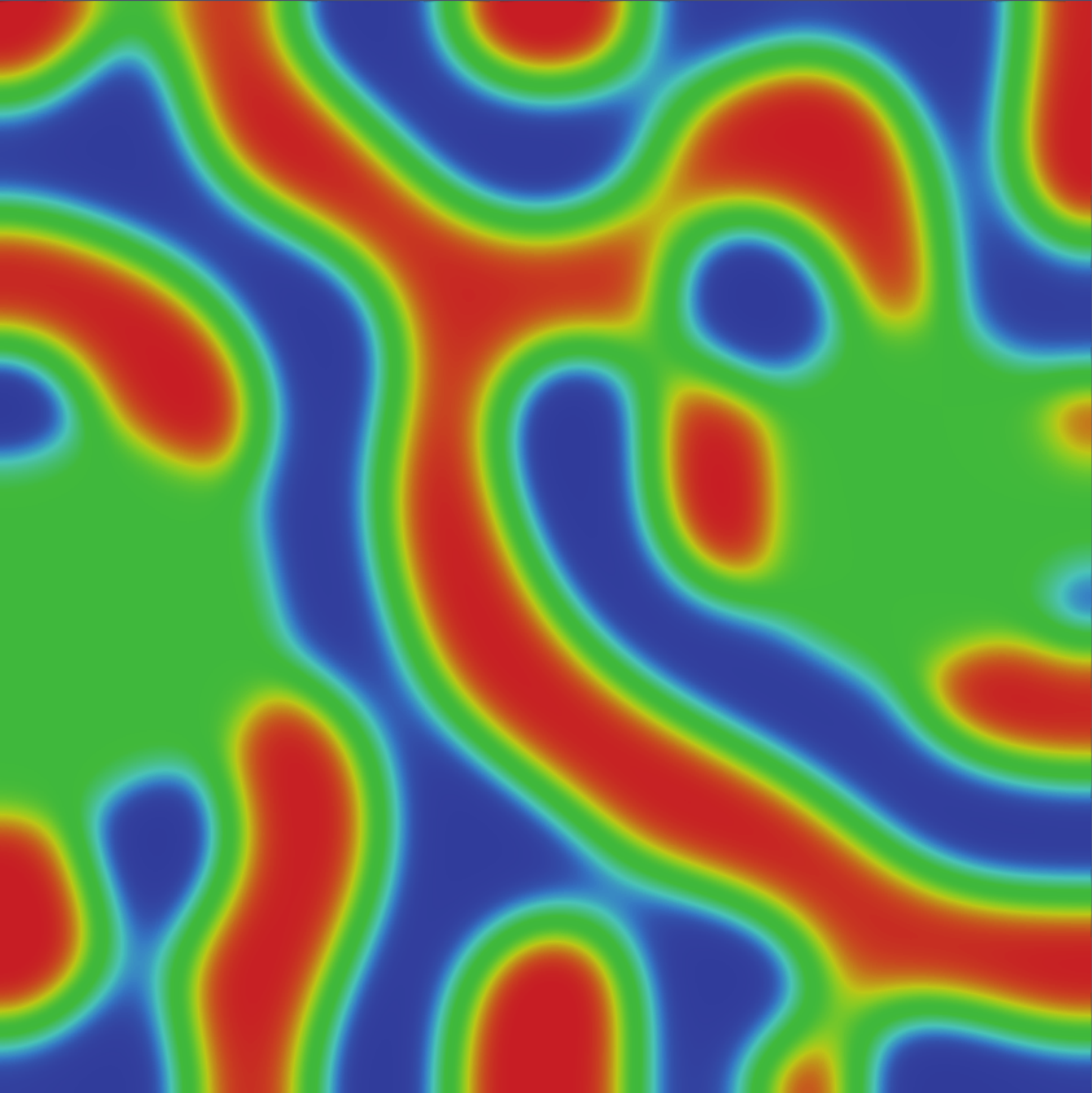}\\
$C_2$\\
\includegraphics[scale=0.12]{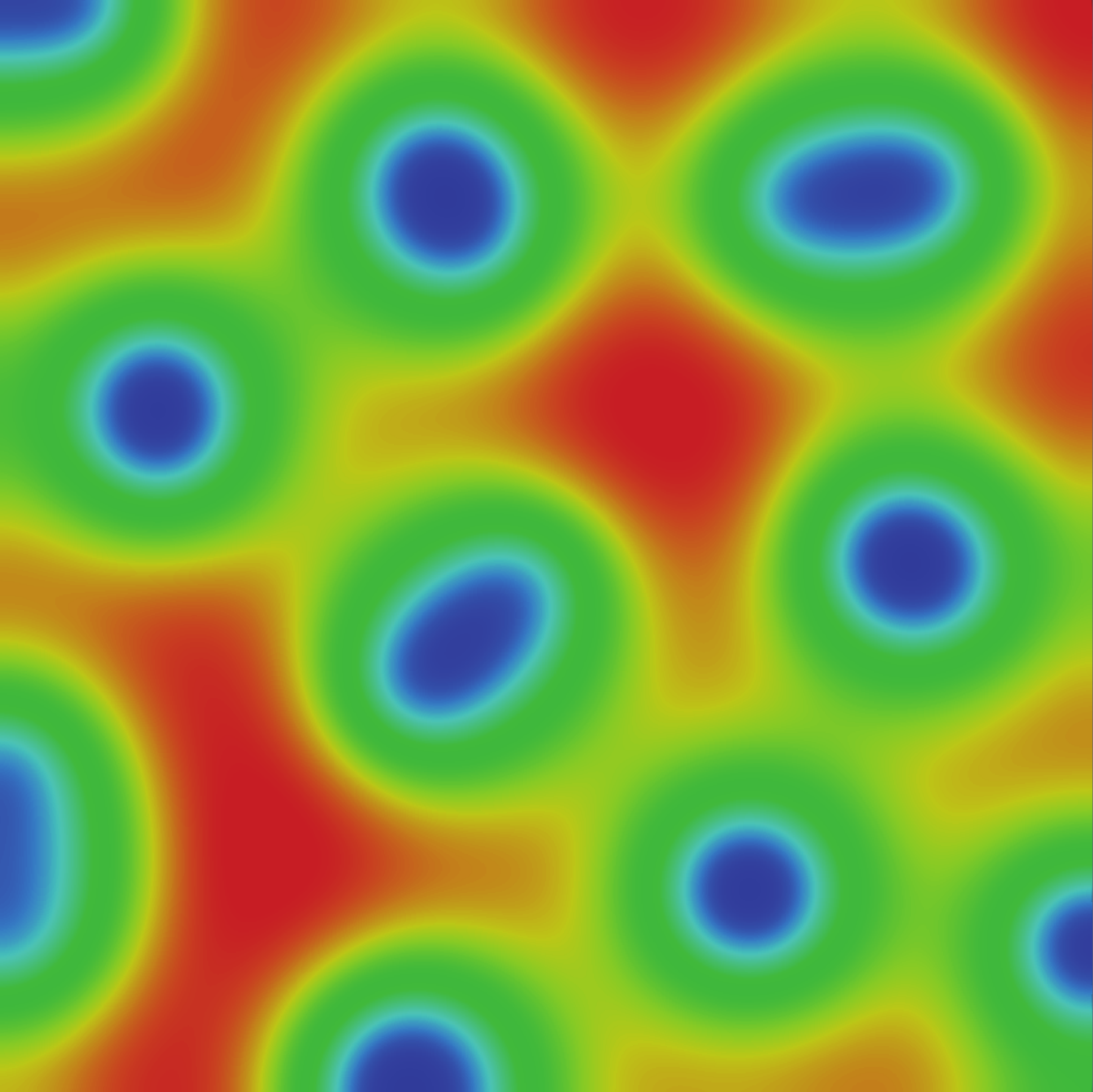}
\includegraphics[scale=0.12]{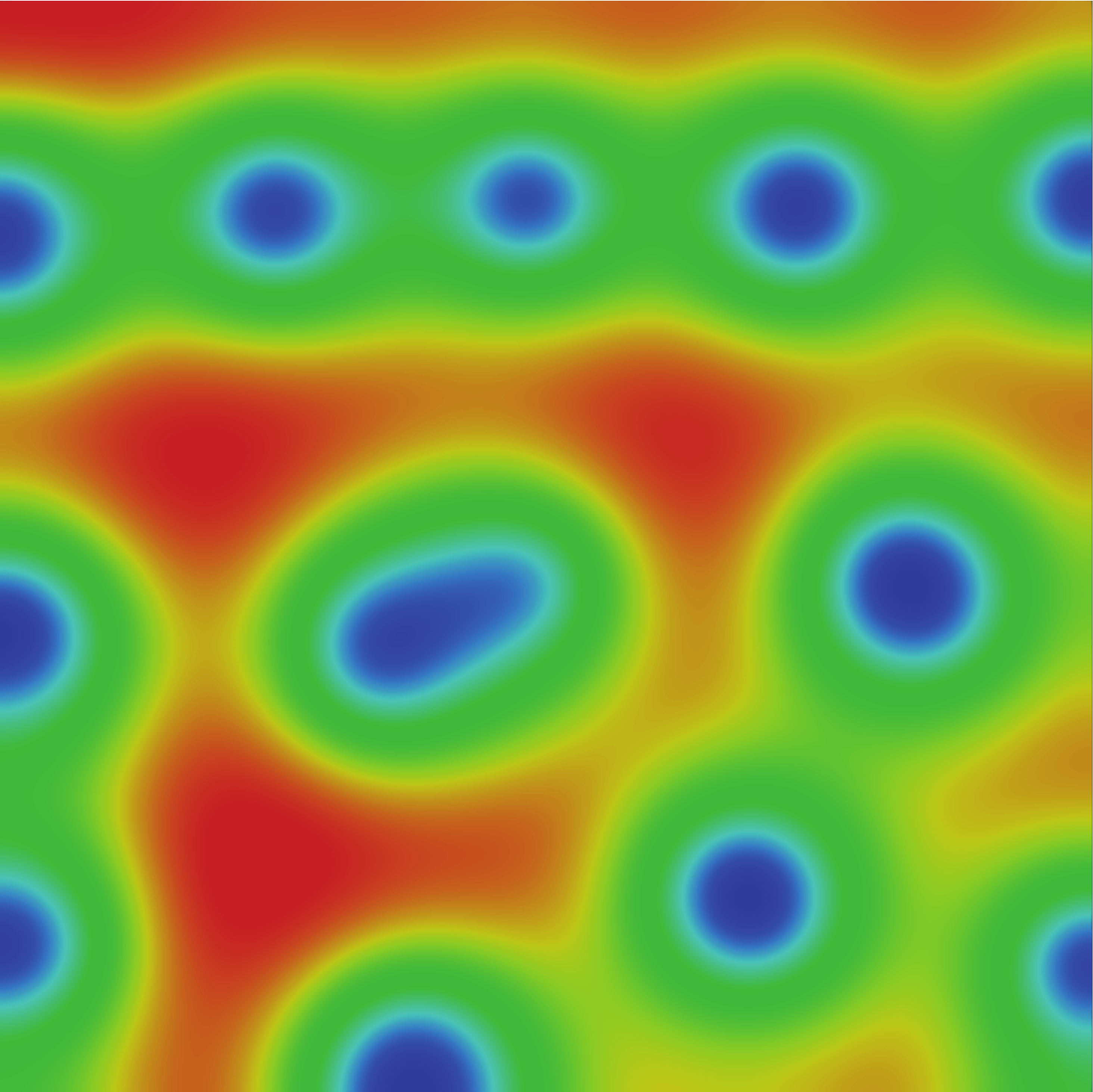}
\includegraphics[scale=0.12]{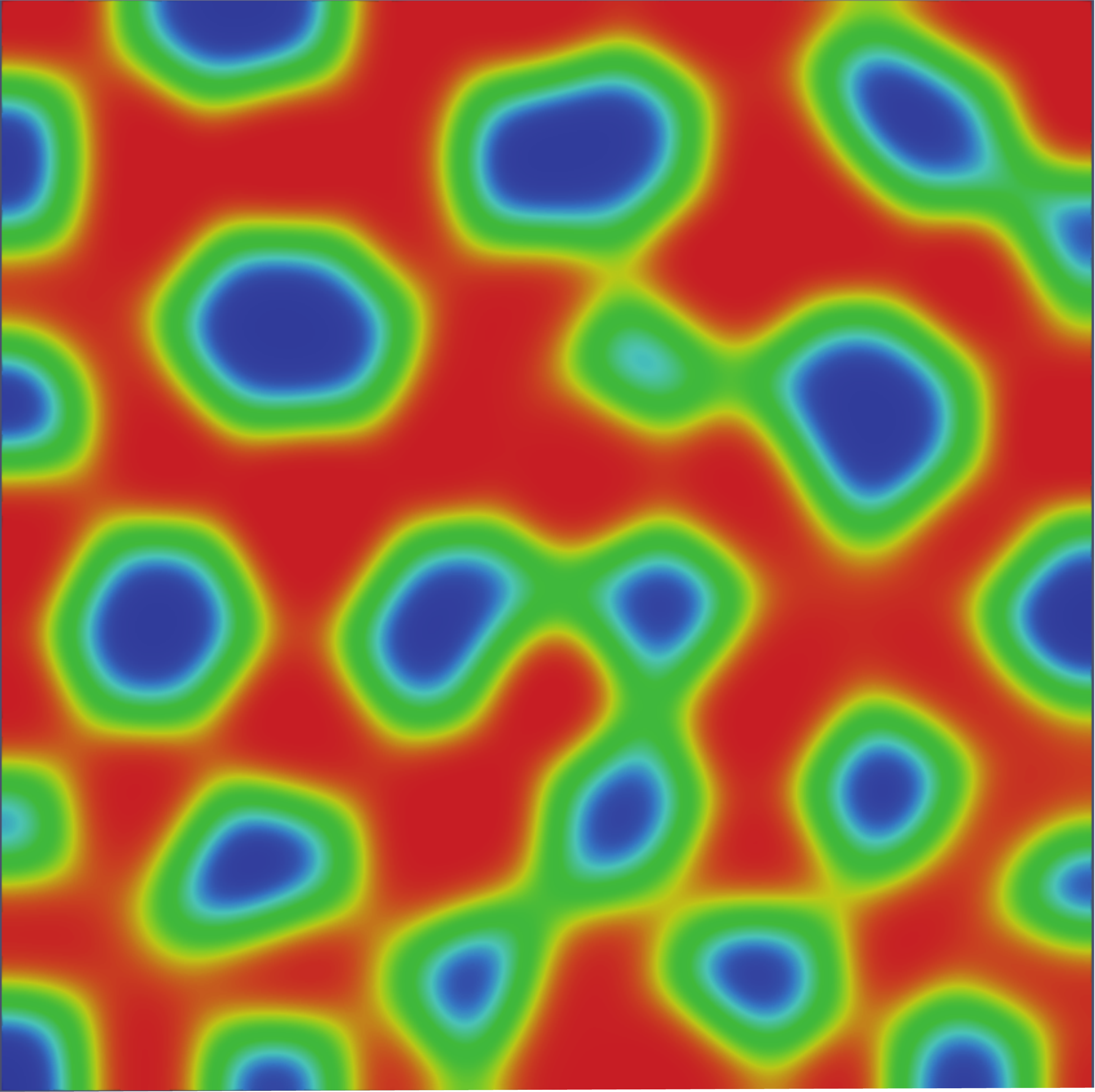}
\includegraphics[scale=0.12]{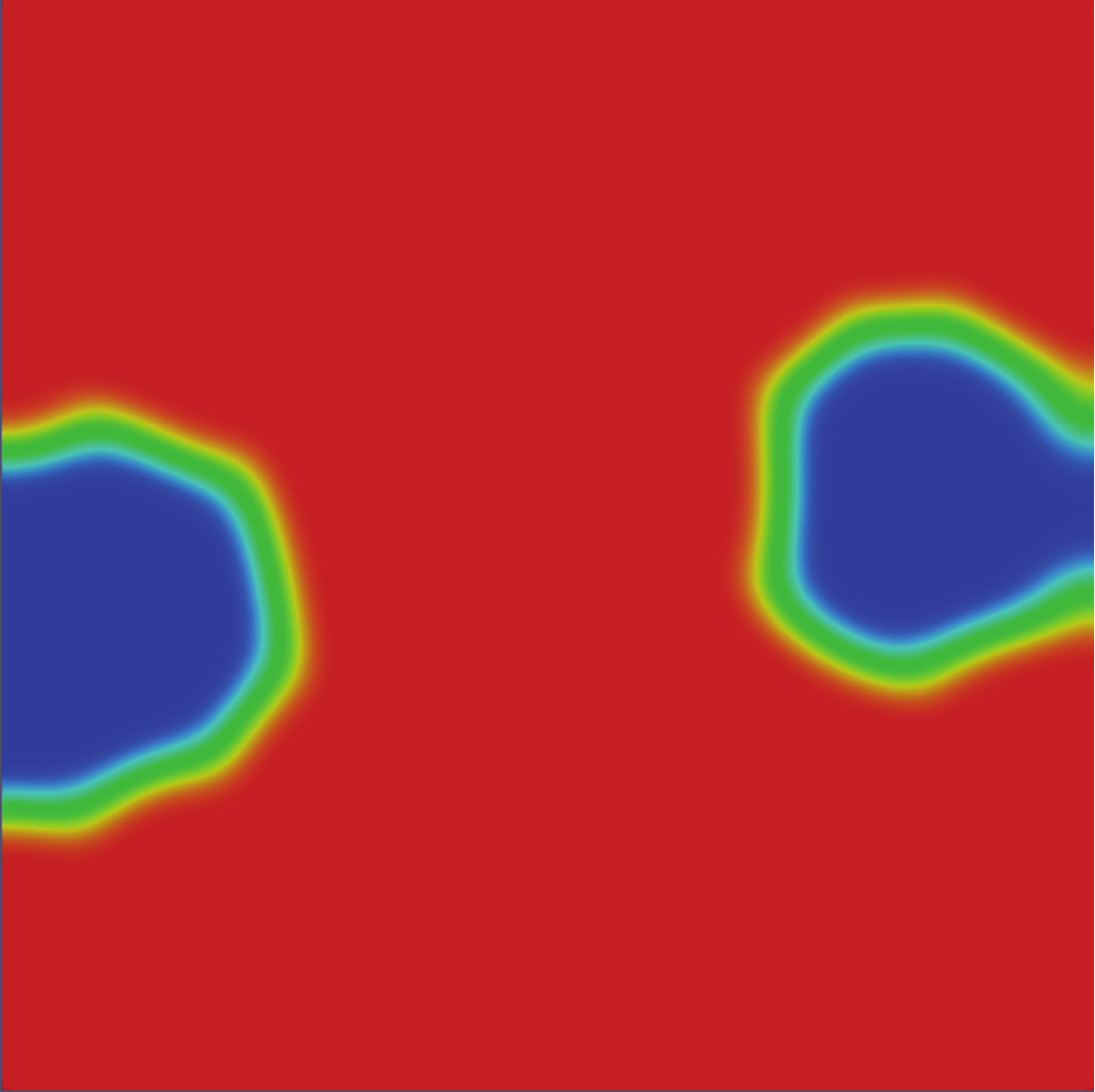}
\caption{$C_1$ and $C_2$ for Model 1 to 4 (left to right) at $t=32$. }
\label{fig:C_140}
\end{figure}

All computations have been implemented in the {\tt mechanoChemIGA} framework, a library for modeling mechano-chemical problems using isogeometric analytics, available at \url{https://github.com/mechanoChem/mechanoChem}. The IBVPs presented here are two-dimensional. Data generation by DNS was carried out on uniformly discretized $400\times400$ meshes. We start all Models from the same initial conditions obtained by running Model 1 for a very short time from a randomized state with small perturbations about $C_1, C_2 = 0.5$ (see Figure \ref{fig:ini_BC}). This state was relabelled as $t = 0$.
With this as initial conditions, the early stage dynamics of Models 1-4 cannot be distinguished by eye (see Figure \ref{fig:C_001}).
However driven by different boundary condition and/or governing equations, the dynamics of the four models soon evolve along distinct  trajectories, and form completely different patterns over a longer time as shown in Figure \ref{fig:C_140}.

The $400\times400$ mesh yields the DNS data for each model, in terms of the fields $(C_1, C_2)$, at the corresponding $401\times 401$ DOFs for each time step. The synthetic, high fidelity DNS data are regularized by having been obtained from PDEs with derivative constraints. Therefore, they are noise-free, which contributes to finding solutions $\Bomega$ that drive the residual equations \eqref{eq:residual} down to machine zero. However the data collected from physical experiments will not satisfy PDEs to the same precision, in general: The data could be at lower fidelity, having been collected at fewer spatial positions and time instants, and therefore may not approximate derivatives well. Background noise and collection error will further contribute to a poorer approximation by PDEs, as discussed in Section \ref{sec:Low_fidelity_noisy_data}. To mimic the experimental data, we take two approaches: i) As illustrated in Figure \ref{fig:mesh}, we use nested meshes for coarse-graining the data $(C_1,C_2)$ after generating them by DNS on the $400\times 400$ mesh. Doing so introduces a loss of information on derivatives. ii) We superpose noise with a Gaussian distribution having zero mean and standard deviation $\sigma$, on the collected $(C_1, C_2)$. We first collect the DNS data, from all four models, on DOFs corresponding to the $400\times400$ mesh, which was used for the forward computations. Data from the same forward computations were then collected on the DOFs corresponding only to the nested $200\times200$, $100\times100$ and $50\times50$ meshes. This yielded datasets with four different fidelities for each model, and 16 different datasets in total. We then superimpose noise with $\sigma=10^{-4}$ for data generated from Models 1 and 2, and $\sigma=10^{-5}$ for data generated from Models 3 and 4, yielding another 16 data sets. Having the clean and noisy data at different fidelity, we generate 38 candidate bases in addition to the time derivative terms, summarized in Table \ref{ta:basis}. Note that to construct the time derivative terms, we need data at two time steps, $t_{n-1}$ and $t_n$. Data at $t_{n-1}$ is only used in the time derivative terms, while other bases are constructed using data at $t_n$ (Backward Euler integration).
\begin{table}[h]
\centering
\caption{Candidate basis for model selection. Asterisks, ($\ast$) in the left column represent algebraic operators on $C_1$ and $C_2$.}
 \begin{tabular}{|c|c|}
 \hline
type of basis& basis in weak form\\\hline
$\by$&$\int_{\Omega}w\frac{\partial C_1}{\partial t}\text{d}v \quad \int_{\Omega}w\frac{\partial C_2}{\partial t}\text{d}v$\\
\hline
\multirow{3}{*}{$\nabla(*\nabla C_1)$}& $\int_{\Omega}\nabla w\nabla C_1\text{d}v \quad \int_{\Omega}\nabla wC_1\nabla C_1\text{d}v \quad \int_{\Omega}\nabla w\nabla C_2\text{d}v$\\
     &$\int_{\Omega}\nabla wC_1^2\nabla C_1\text{d}v \quad \int_{\Omega}\nabla wC_1C_2\nabla C_1\text{d}v \quad \int_{\Omega}\nabla wC_2^2\nabla C_1\text{d}v$ \\
     &$\int_{\Omega}\nabla wC_1^3\nabla C_1\text{d}v \quad \int_{\Omega}\nabla wC_1^2C_2\nabla C_1\text{d}v \quad \int_{\Omega}\nabla wC_1C_2^2\nabla C_1\text{d}v \quad \int_{\Omega}\nabla wC_2^3\nabla C_1\text{d}v$\\
     \hline
\multirow{3}{*}{$\nabla(*\nabla C_2)$}& $\int_{\Omega}\nabla w\nabla C_2\text{d}v \quad \int_{\Omega}\nabla wC_1\nabla C_2\text{d}v \quad \int_{\Omega}\nabla w\nabla C_2\text{d}v$\\
     &$\int_{\Omega}\nabla wC_1^2\nabla C_2\text{d}v \quad \int_{\Omega}\nabla wC_1C_2\nabla C_2\text{d}v \quad \int_{\Omega}\nabla wC_2^2\nabla C_2\text{d}v$ \\
     &$\int_{\Omega}\nabla wC_1^3\nabla C_2\text{d}v \quad \int_{\Omega}\nabla wC_1^2C_2\nabla C_2\text{d}v \quad \int_{\Omega}\nabla wC_1C_2^2\nabla C_2\text{d}v \quad \int_{\Omega}\nabla wC_2^3\nabla C_2\text{d}v$\\
     \hline  
$\nabla^2(*\nabla^2 C)$& $\int_{\Omega}\nabla^2 w\nabla^2 C_1\text{d}v \quad \int_{\Omega}\nabla^2 wC_1\nabla^2 C_1\text{d}v \quad \int_{\Omega}\nabla^2 w\nabla^2 C_2\text{d}v \quad \int_{\Omega}\nabla^2 wC_2\nabla^2 C_2\text{d}v $\\
\hline
\multirow{3}{*}{non-gradient}& $-\int_{\Omega} w1\text{d}v \quad -\int_{\Omega} wC_1 \text{d}v \quad -\int_{\Omega} w C_2\text{d}v$\\
     &$-\int_{\Omega} wC_1^2\text{d}v \quad -\int_{\Omega} wC_1C_2\text{d}v \quad -\int_{\Omega} wC_2^2\text{d}v$ \\
     &$-\int_{\Omega} wC_1^3\text{d}v \quad -\int_{\Omega} wC_1^2C_2\text{d}v \quad -\int_{\Omega} wC_1C_2^2\text{d}v \quad -\int_{\Omega} wC_2^3\text{d}v$\\
     \hline
boundary condition& $-\int_{\Gamma_1} w1\text{d}s \quad -\int_{\Gamma_4} w1\text{d}s \quad -\int_{\Gamma_3} w1\text{d}s \quad -\int_{\Gamma_4} w1\text{d}s$\\\hline
\end{tabular}
\label{ta:basis}
\end{table}
    
\subsection{System identification with data at different fidelity without noise}
\label{sec:results_fidelity}
\begin{figure}[hbtp]
\centering
\includegraphics[scale=0.2]{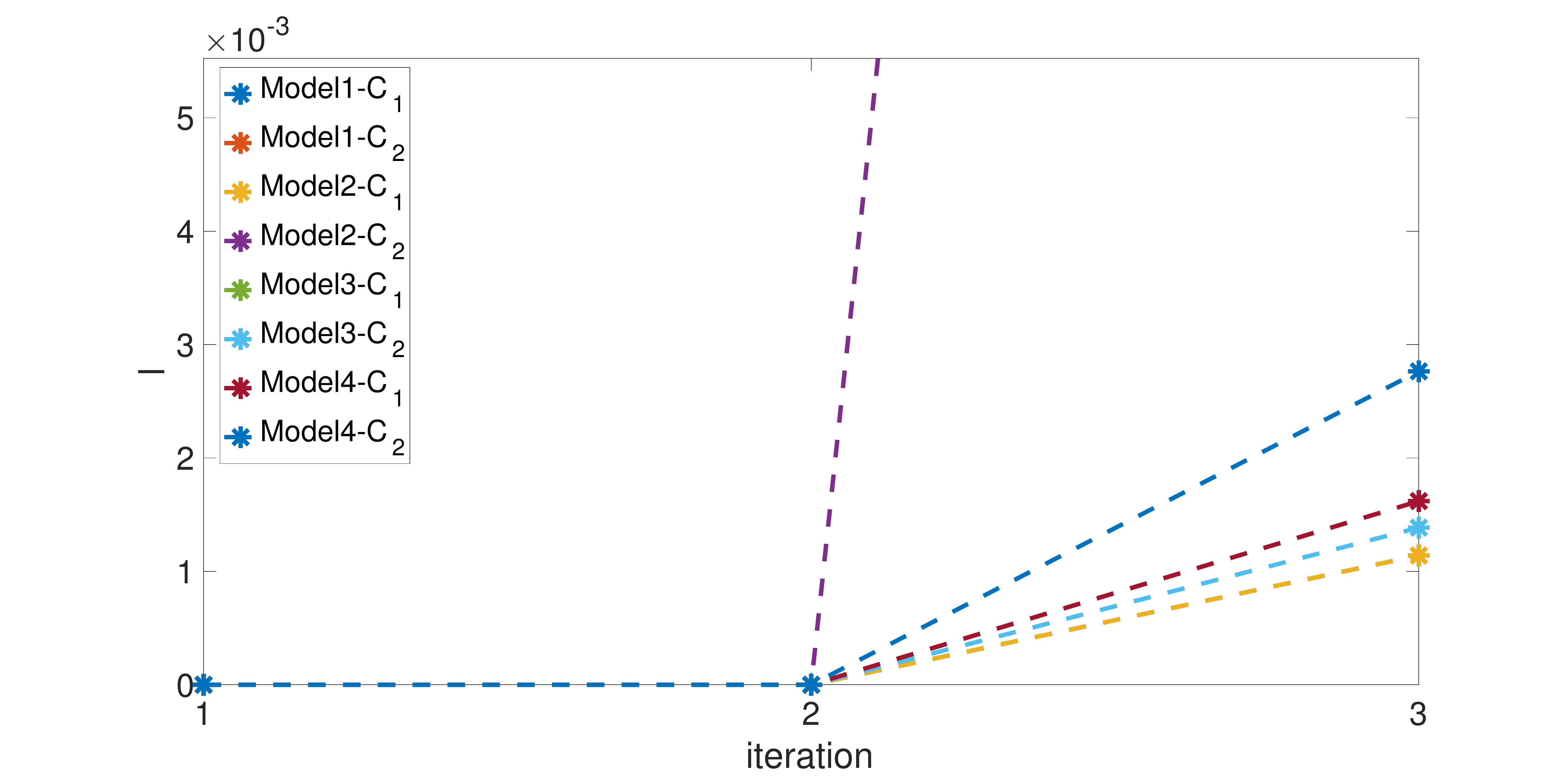}
\caption{Value of the loss function at each iteration during stepwise regression using the high fidelity data generated at $t=0$ and $t=0.01$. Note that the algorithm terminates at Iteration 2, and the value of the loss function increases dramatically if more bases are eliminated, as seen in Iteration 3. }
\label{fig:loss_clean}
\end{figure}

In order challenge our algorithms to distinguish the underlying PDEs from the simulation results, we first use only DNS results over certain time intervals that, to the eye, appear identical to each other. For this purpose, we first apply stepwise regression on the 16 data sets (four datasets of different fidelity from four models) without noise, generated at $t=0$ (Figure \ref{fig:ini_BC}) and $t=0.01$ (Figure \ref{fig:C_001}). For the four high fidelity data ($400\times400$ mesh), we found that in the first iteration, all the relevant bases have almost converged to the correct pre-factor, and irrelevant bases have trivial (approaching zero) pre-factors. By setting the tolerance $\epsilon=1.0\times10^{-5}$ in the threshold equation \eqref{eq:threshold}, all relevant bases are selected with exact pre-factors in the second iteration. The results are not shown here since the coefficients on operators are identical to those in Equations (\ref{eq:weak_value_model2-1} - \ref{eq:weak_value_model4-2}) for the four datasets respectively. In particular, our system identification algorithms correctly identify the zero flux boundary condition in Model 1, constant flux boundary condition on $\Gamma_4$ in Model 2, and higher order terms of the form $\nabla^2 (\ast\nabla^2 C)$ (see Table \ref{ta:basis}), in Models 3 and 4. The algorithm terminates after the second iteration since no more bases can be eliminated by the $F$ test. As shown in Figure \ref{fig:loss_clean}, for the first two iterations, the loss function remains machine zero as the synthetic data drive the residual equations \eqref{eq:residual} down to this limit. If any more bases are eliminated, the loss function increases dramatically in Iteration 3 due to underfitting.  

\begin{figure}[hbtp]
\centering
\includegraphics[scale=0.4]{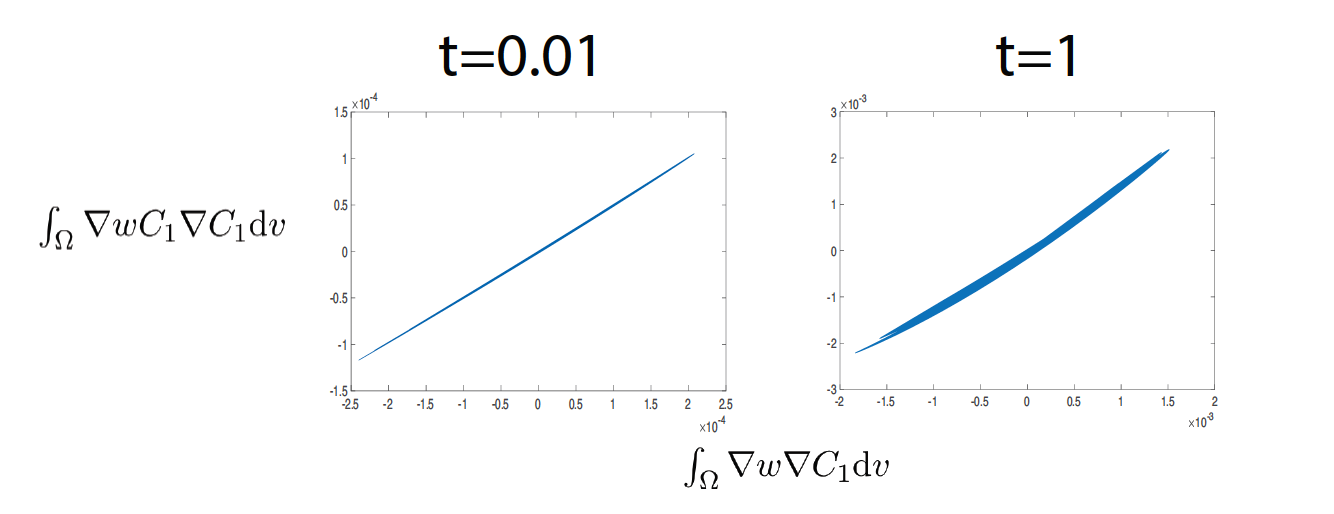}
\caption{The nearly uniform distribution, $C_1\approx0.5$, yields a slope of $\approx 1/2$ between $\int_{\Omega}\nabla wC_1\nabla C_1\text{d}v$ and $\int_{\Omega}\nabla w\nabla C_1\text{d}v$ at $t=0.01$, shown in the left plot. This spurious linearity disappears as the system evolves over longer times as shown in the right plot.}
\label{fig:bsis_1_2}
\end{figure}

Initially $C_1$ and $C_2$ are close to uniformly distributed with small perturbations. This uniformity yields bases, such as $\int_{\Omega}\nabla w\nabla C_1\text{d}v$ and $\int_{\Omega}\nabla wC_1\nabla C_1\text{d}v$ that appear to be linearly dependent (left plot in Figure in \ref{fig:bsis_1_2}). The high fidelity data ($400\times400$ mesh) drive the residual equations \eqref{eq:residual} down to machine zero, allowing the linear regression model Equation \eqref{eq:least-square} to converge. The algorithms can therefore select the relevant bases, distinguishing from other irrelevant bases that are close to being linearly dependent. However, low fidelidy data using NURBS basis functions yields poor representation of derivatives such as those in Equations (\ref{eq:back_euler}), (\ref{eq:nabla2_basis}) and (\ref{eq:nabla4_basis}). As a result, the linear regression model contains greater error using low fidelity data, which are created by directly sparsifying the high-fidelity data instead of being generated from a coarse grid. Therefore, an unresolvable discrepancy remains in the low-fidelity representation, and two terms that are nearly linearly dependent cannot be distinguished. Using low fidelity data at $t=0$ and $t=0.01$, the algorithms are unable to identify any of the four governing PDEs (Models 1-4). 

\begin{figure}[hbtp]
\centering
\includegraphics[scale=0.2]{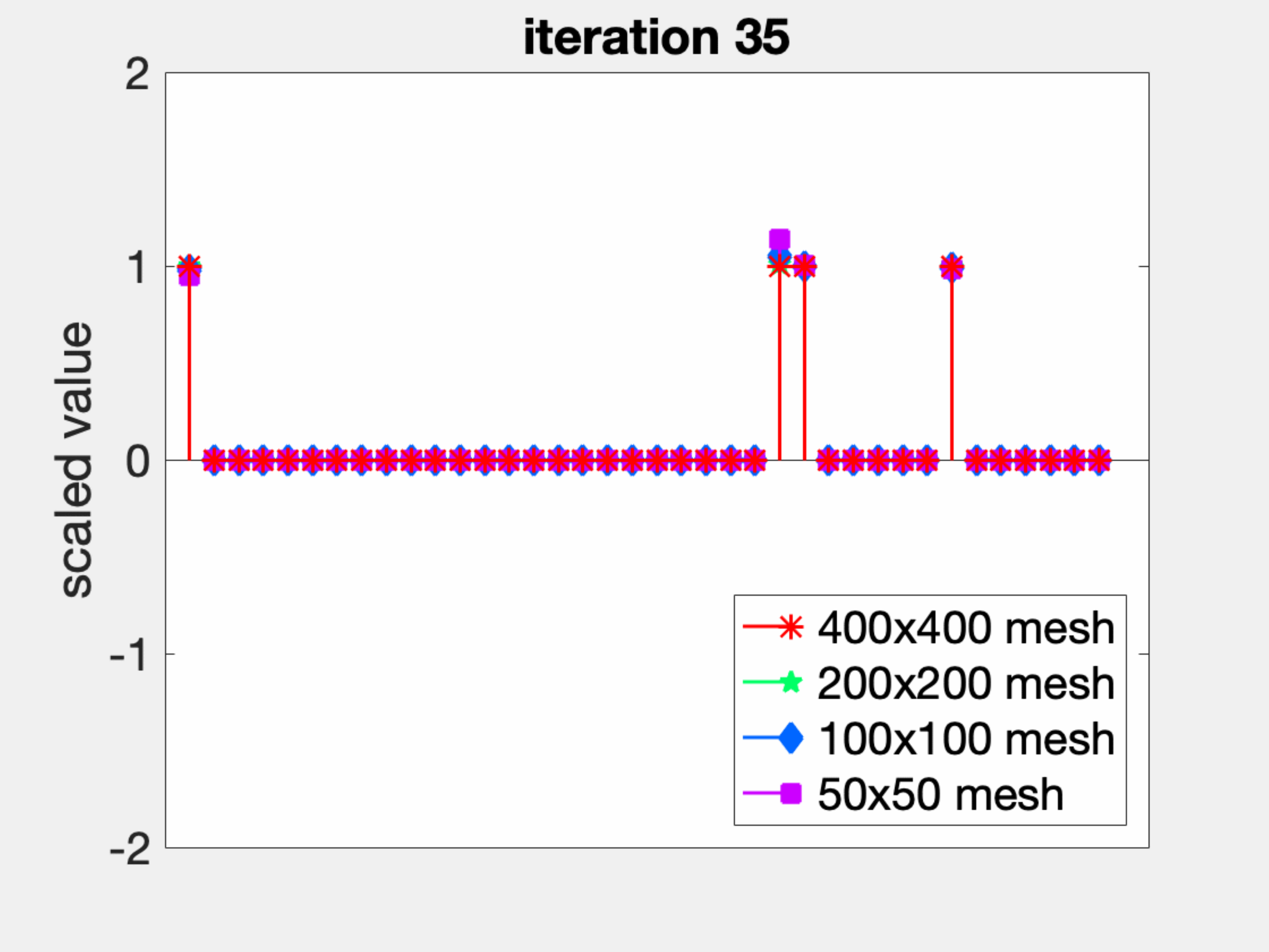}
\includegraphics[scale=0.2]{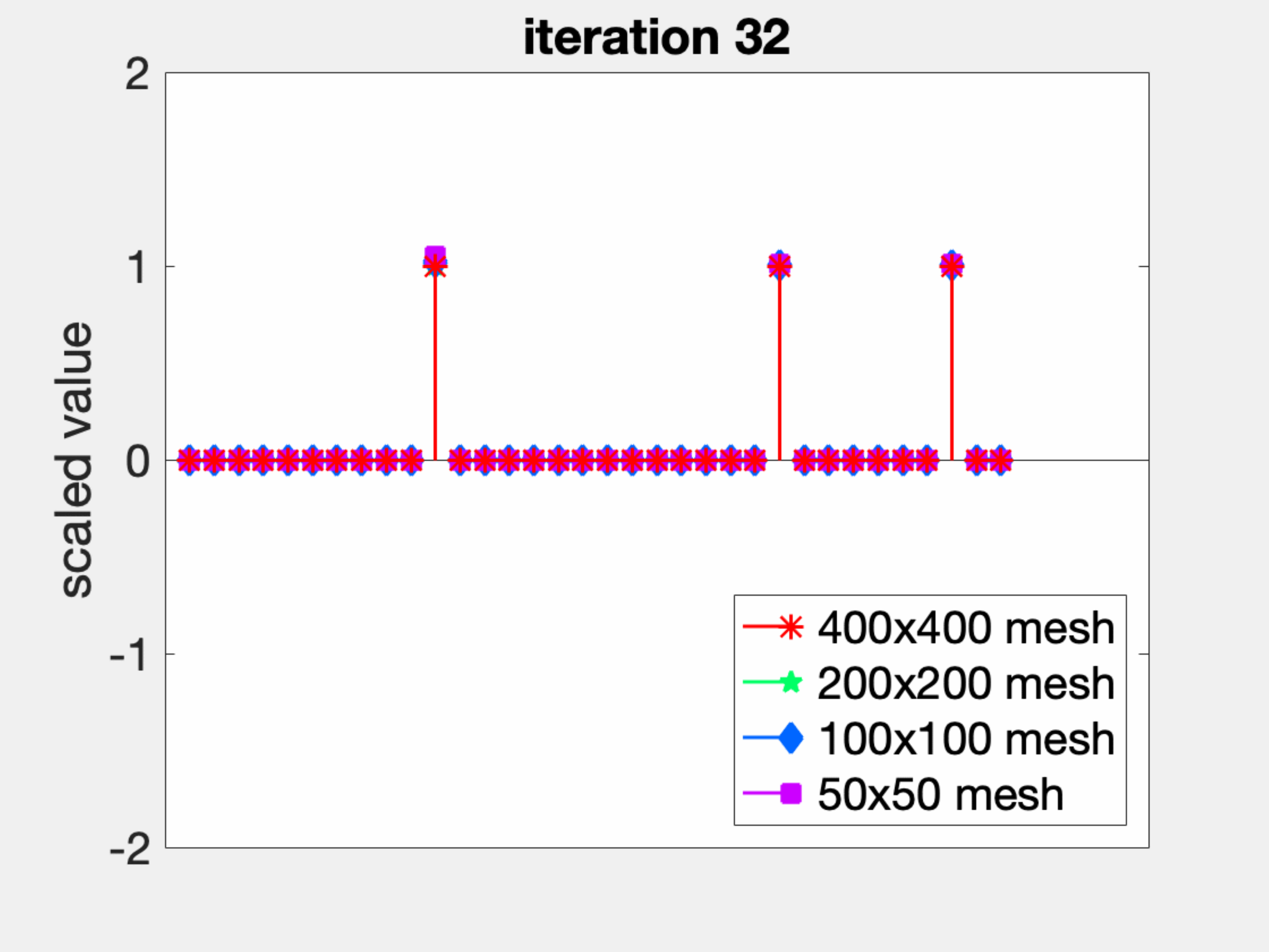}
\caption{Inferred operators for $C_1$ (left panel) and $C_2$ (right panel) using data generated from Model 1 using $\Delta t=1$, with 10 snapshots. The identified pre-factors of relevant terms are scaled by their true values: $\int_{\Omega}-1\nabla w_1 \cdot\nabla C_1\text{d}v
,\; 0.1\int_{\Omega}w_1\text{d}v,\;  -1\int_{\Omega}w_1C_1\text{d}v,\;  1\int_{\Omega}w_1C_1^2C_2\text{d}v$ in the left panel and $-40\int_{\Omega}\nabla w_2\cdot\nabla C_2\text{d}v,\;  \int_{\Omega}0.9\int_{\Omega}w_2\text{d}v,\;  -1\int_{\Omega}C_1^2C_2\text{d}v$ in right plot. They are also arranged from left to right in the same order as they appear in Equations \eqref{eq:weak_value_model1-1} and \eqref{eq:weak_value_model1-2}. The original results are summarized in Table \ref{ta:tab:results_noflux_noNoise_dt1}. The transient results have been included as Movies 1 and 2 in supplementary material.
}
\label{fig:results_noflux_noNoise_dt1}
\end{figure}

\begin{figure}[hbtp]
\centering
\includegraphics[scale=0.2]{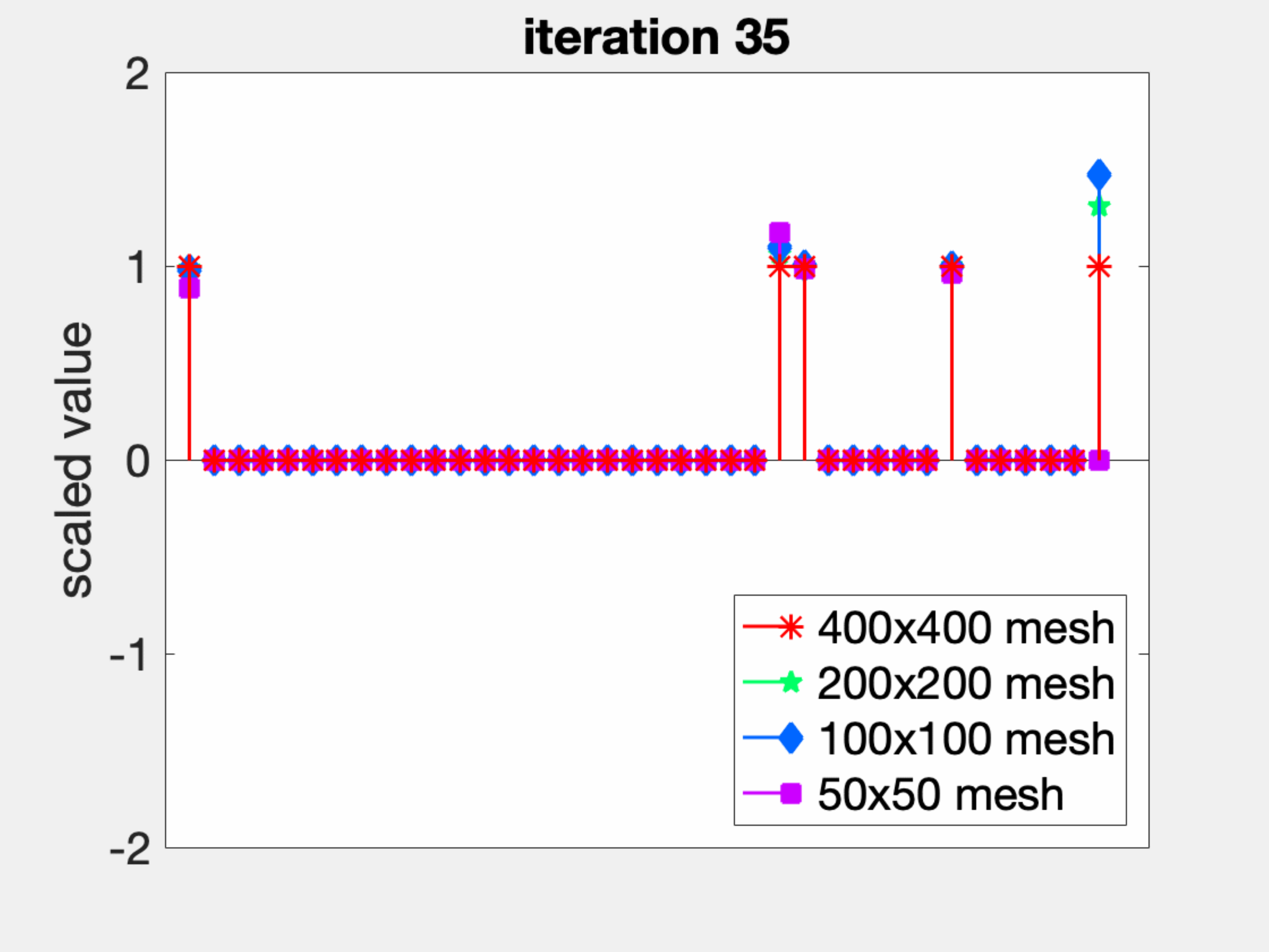}
\includegraphics[scale=0.2]{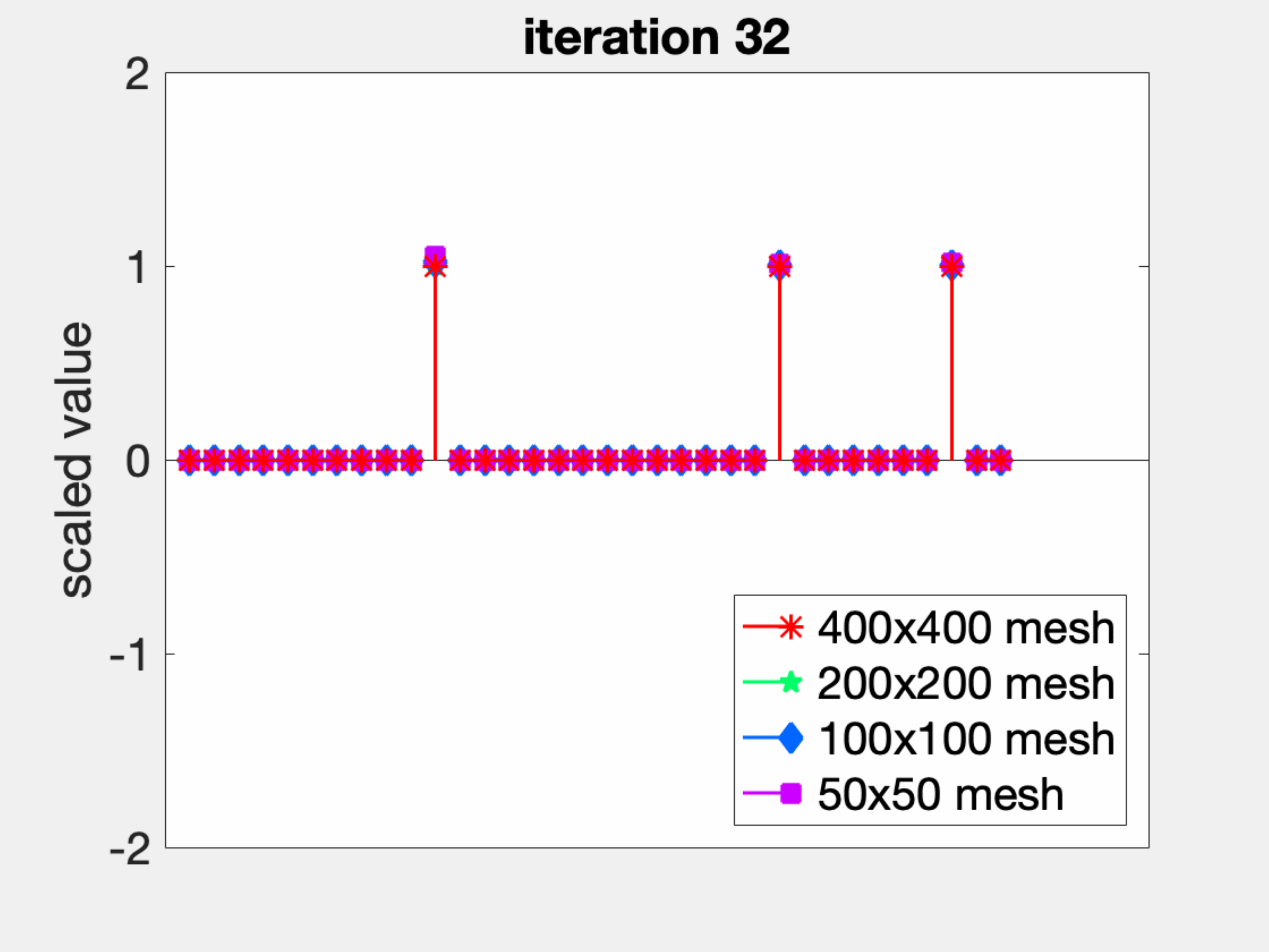}
\caption{Inferred operators for $C_1$ (left panel) and $C_2$ (right panel) using data generated from Model 2 using $\Delta t=1$, with 10 snapshots. The identified pre-factors of relevant terms are scaled by their true values: $\int_{\Omega}-1\nabla w_1 \cdot\nabla C_1\text{d}v
,\;0.1\int_{\Omega}w_1\text{d}v, \;-1\int_{\Omega}w_1C_1\text{d}v, \; 1\int_{\Omega}w_1C_1^2C_2\text{d}v, 0.1\int_{\Gamma_2}w_1\text{d}s$ in the left plot and $-40\int_{\Omega}\nabla w_2\cdot\nabla C_2\text{d}v,\; \int_{\Omega}0.9\int_{\Omega}w_2\text{d}v,\; -1\int_{\Omega}C_1^2C_2\text{d}v$ in right plot. They are also arranged from left to right in the same order as they appear in Equations \eqref{eq:weak_value_model2-1} and \eqref{eq:weak_value_model2-2}. The original results are summarized in Table \ref{ta:tab:results_influx_noNoise_dt1}.
As shown in the left panel, the Neumman boundary condition term (last leaf) is more poorly estimated than the other terms, and fails to be identified on the $50\times 50$ mesh. The transient results have been included as Movies 3 and 4 in supplementary material.
}
\label{fig:results_influx_noNoise_dt1}
\end{figure}

\begin{figure}[hbtp]
\centering
\includegraphics[scale=0.2]{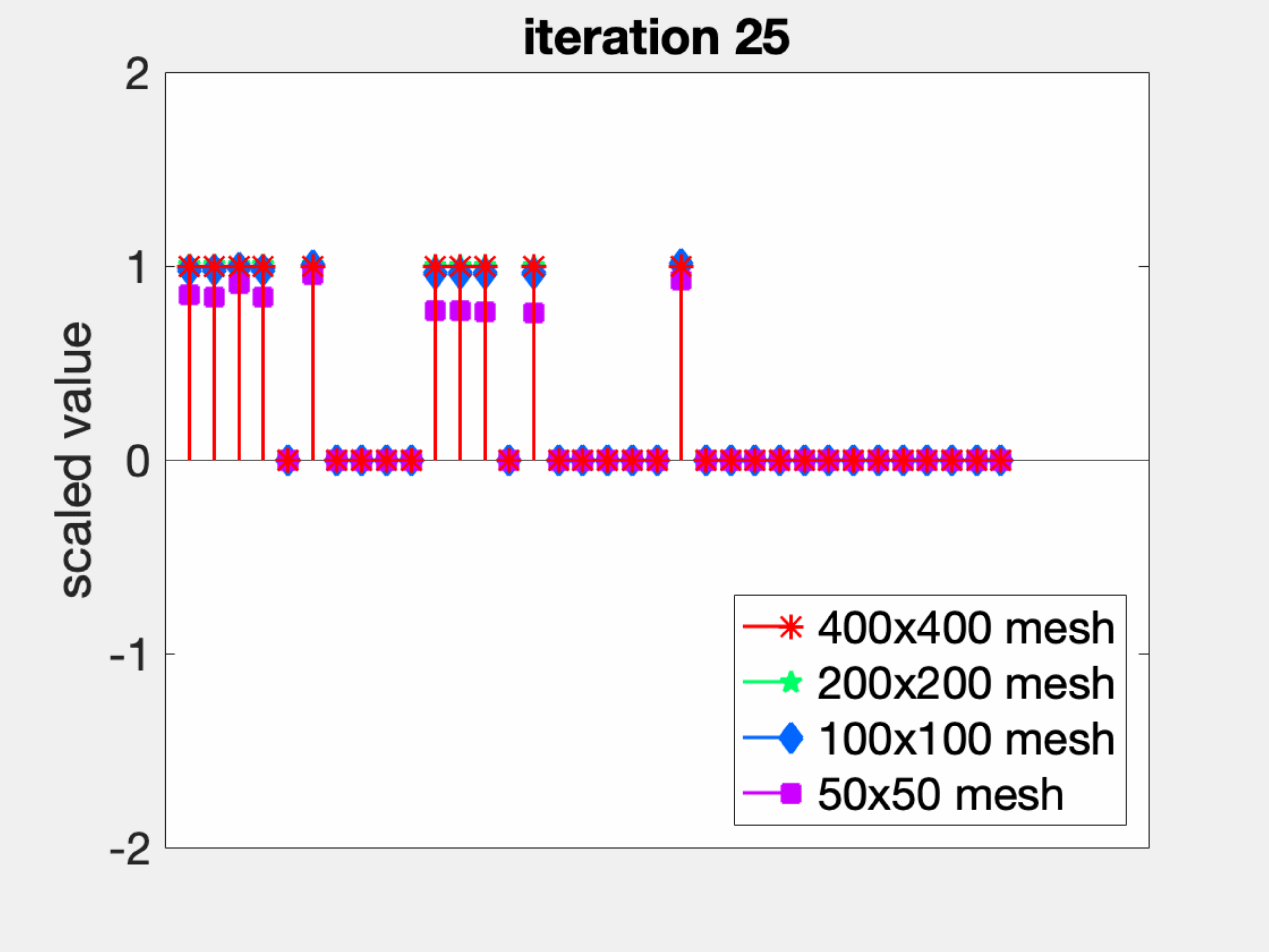}
\includegraphics[scale=0.2]{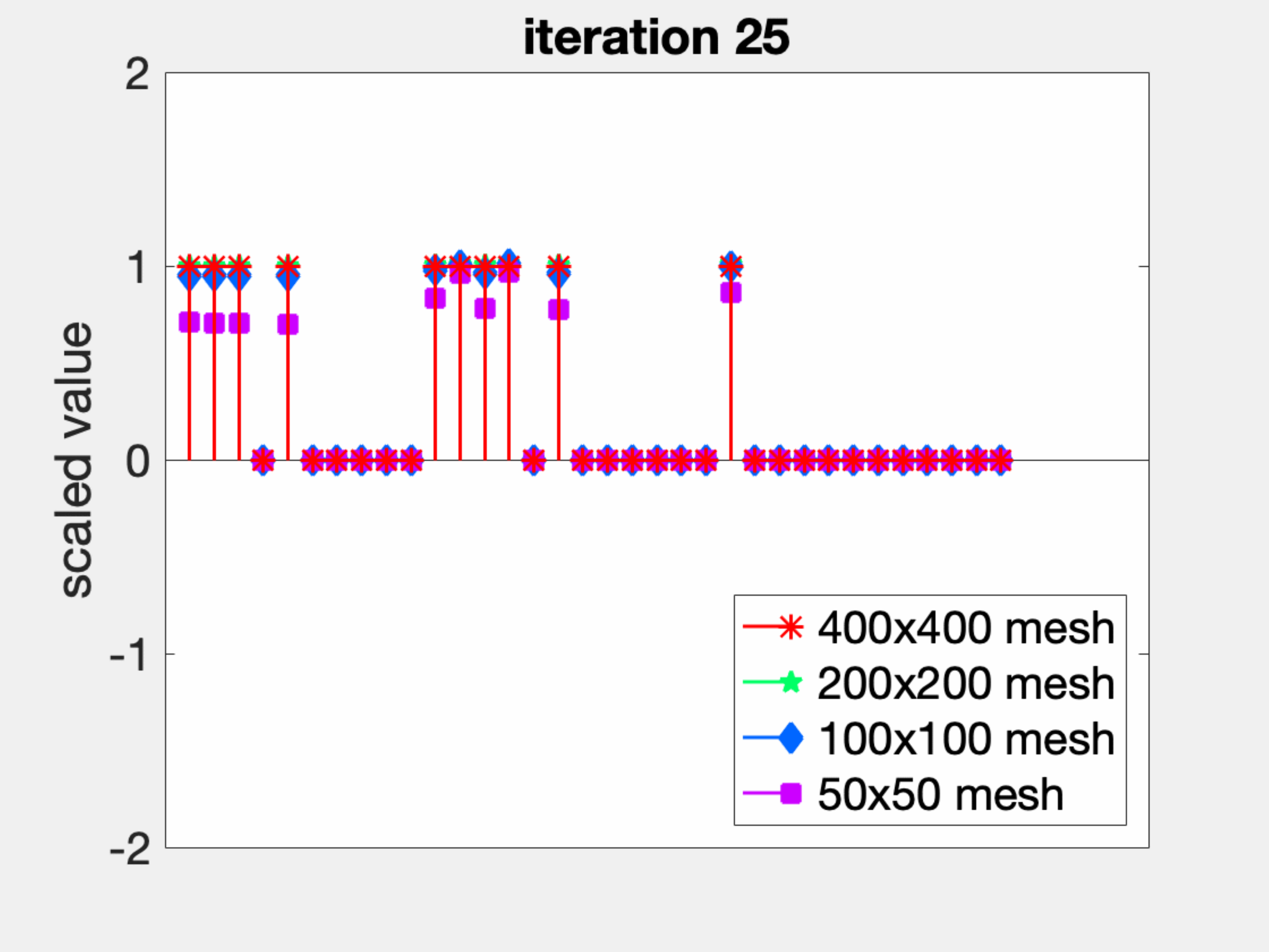}
\caption{Inferred operators for $C_1$ (left panel) and $C_2$ (right panel) using data generated from Model 3 using $\Delta t=1$, with 10 snapshots. The identified pre-factors of relevant terms are scaled by their true values, and arranged from left to right in the same order as they appear in Equations \eqref{eq:weak_value_model3-1} and \eqref{eq:weak_value_model3-2}. The original results are summarized in Table \ref{ta:tab:results_Cahn_noNoise_dt1}. When using the $50\times50$ mesh, a bias on $\int_{\Omega}\nabla wC_2^2\nabla C_1\text{d}v$ is needed to protect it from elimination in the first few iterations for inferring the governing PDE of $C_1$. The transient results have been included as Movies 5 and 6 in supplementary material.} 
\label{fig:results_Cahn_noNoise_dt1}
\end{figure}

\begin{figure}[hbtp]
\centering
\includegraphics[scale=0.2]{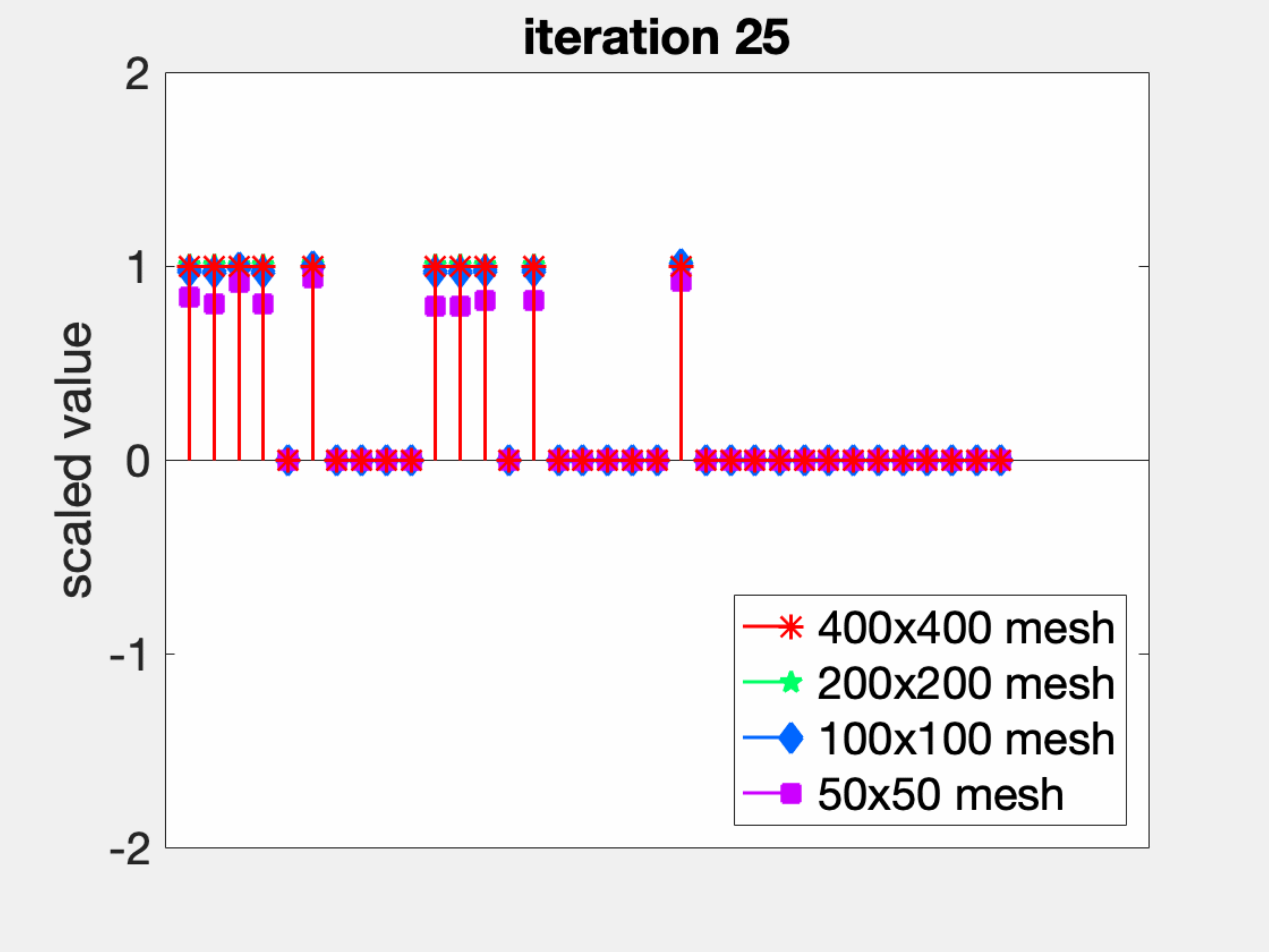}
\includegraphics[scale=0.2]{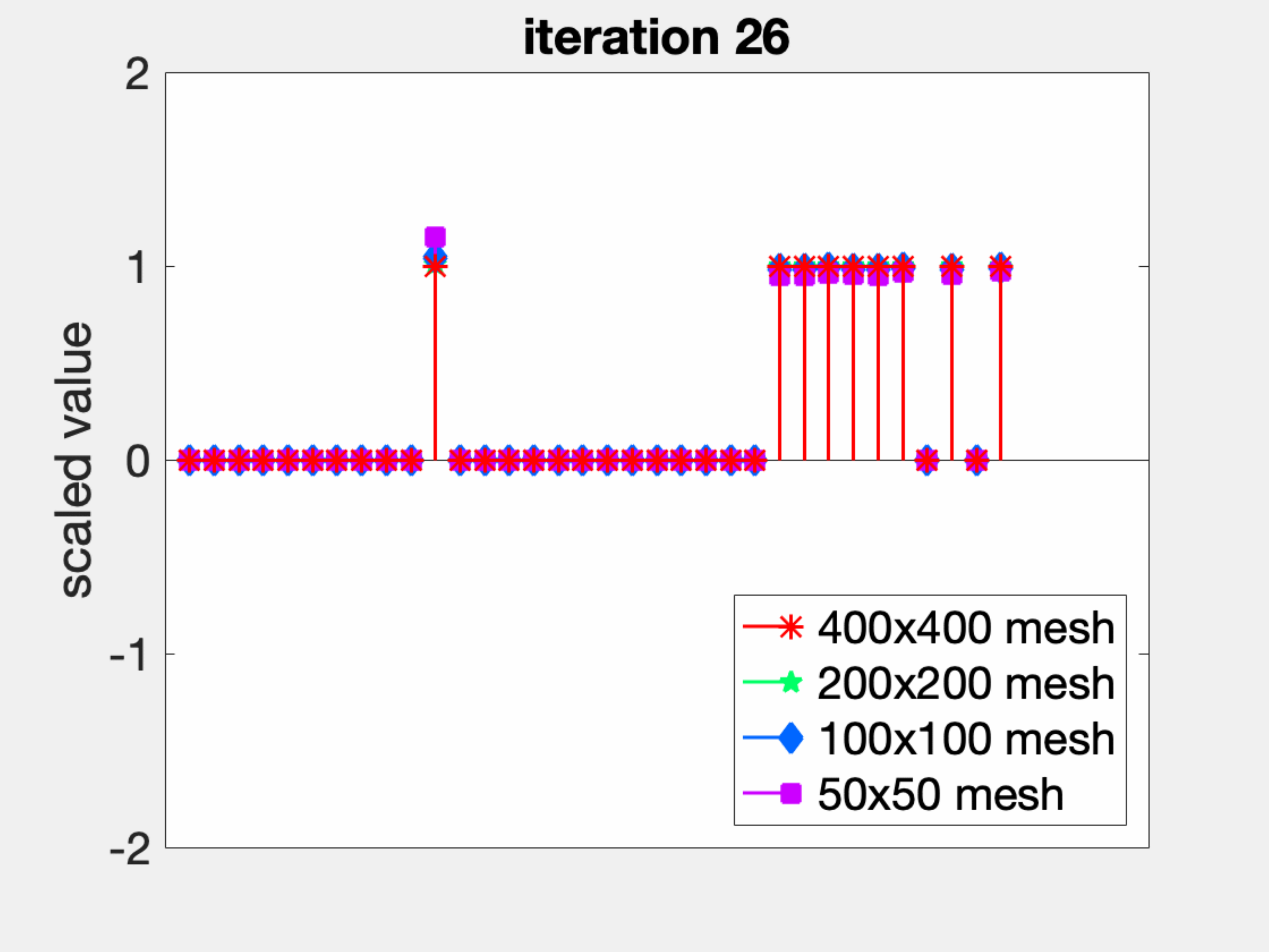}
\caption{Inferred operators for $C_1$ (left panel) and $C_2$ (right panel) using data generated from Model 4 using $\Delta t=1$, with 10 snapshots. The identified pre-factors of relevant terms are scaled by their true values, and arranged from left to right in the same order as they appear in Equations \eqref{eq:weak_value_model4-1} and \eqref{eq:weak_value_model4-2}. The original results are summarized in Table \ref{ta:tab:results_alen_noNoise_dt1}. A bias on $\int_{\Omega}\nabla wC_2^2\nabla C_1\text{d}v$ is needed to protect it from elimination in the first few iterations for inferring the governing PDE of $C_1$. The transient results have been included as Movies 7 and 8 in supplementary material.}
\label{fig:results_alen_noNoise_dt1}
\end{figure}

This spurious linearity however disappears as the system evolves (right plot in Figure \ref{fig:bsis_1_2}). 
By choosing data at 10 time steps: from $t=1$ with time step $\Delta t=1$, the algorithms are able to correctly identify all PDEs (Models 1-4) using low fidelity data. To quantify the algorithm's performance, we define the error of the estimated results to be:
\begin{align}
    \text{error}=\max \left|\frac{\widehat{\omega}_i-\omega_i}{\omega_i}\right|
\end{align}
where $\widehat{\omega}$ is the estimated pre-factor, and $\omega$ is the true pre-factor. Figures \ref{fig:results_noflux_noNoise_dt1} - \ref{fig:results_alen_noNoise_dt1} show the scaled results at the final iteration. The exact values and the error are summarized in Tables \ref{ta:tab:results_noflux_noNoise_dt1} - \ref{ta:tab:results_alen_noNoise_dt1}. 

Note that the Neumman boundary condition terms only have contributions from DOFs on the boundary, which are sparse relative to the total number. Consequently the basis for the Neumman boundary condition has a negligible contribution to the linear regression model compared to operators that appear via volume integrals in the weak forms of the PDEs. This results in large errors of the pre-factor for the Neumann terms. As shown in the left panel of Figure \ref{fig:results_influx_noNoise_dt1}, for the results using data generated from Model 2, the Neumman boundary condition operator is more poorly estimated than others. Using very low fidelity data ($50\times 50$ mesh), the Neumman boundary condition basis fails to be be identified. However, the errors in the other bases are similar to those obtained when using data generated from Model 1, for which the Neumann boundary condition and the corresponding operator are zero. 

The estimated results for data generated from Models 3 and 4, match the true values very well except for the lowest fidelity dataset using the $50\times50$ mesh (shown in Figures \ref{fig:results_Cahn_noNoise_dt1} and \ref{fig:results_alen_noNoise_dt1}).
This lowest fidelity data set poses even greater challenges, because, to identify the governing PDE for $C_1$, using the $50\times50$ mesh, we need to place a bias on the basis $\int_{\Omega}\nabla wC_2^2\nabla C_1\text{d}v$ to protect it from elimination in the first few iterations. Although we are able to select all relevant terms with this prior, the error increases significantly as the mesh cannot properly resolve the small variation of the basis as we see below in Figure \ref{fig:basis_7}.

Of the four datasets of different fidelity from each model, the high fidelity dataset ($400\times400$ mesh) yields exact pre-factors for all bases.
However for the lower fidelity datasets, variations in the data field over a small neighborhood of the domain may diminish as the decreasing mesh size resolves the variation poorly, i.e. the approximation in Equations (\ref{eq:nabla2_basis}) and (\ref{eq:nabla4_basis}) becomes poor. The variations may even completely disappear, as seen on the $50\times50$ mesh (see one example shown in Figure \ref{fig:basis_7}).
\begin{figure}[hbtp]
\centering
\includegraphics[scale=0.12]{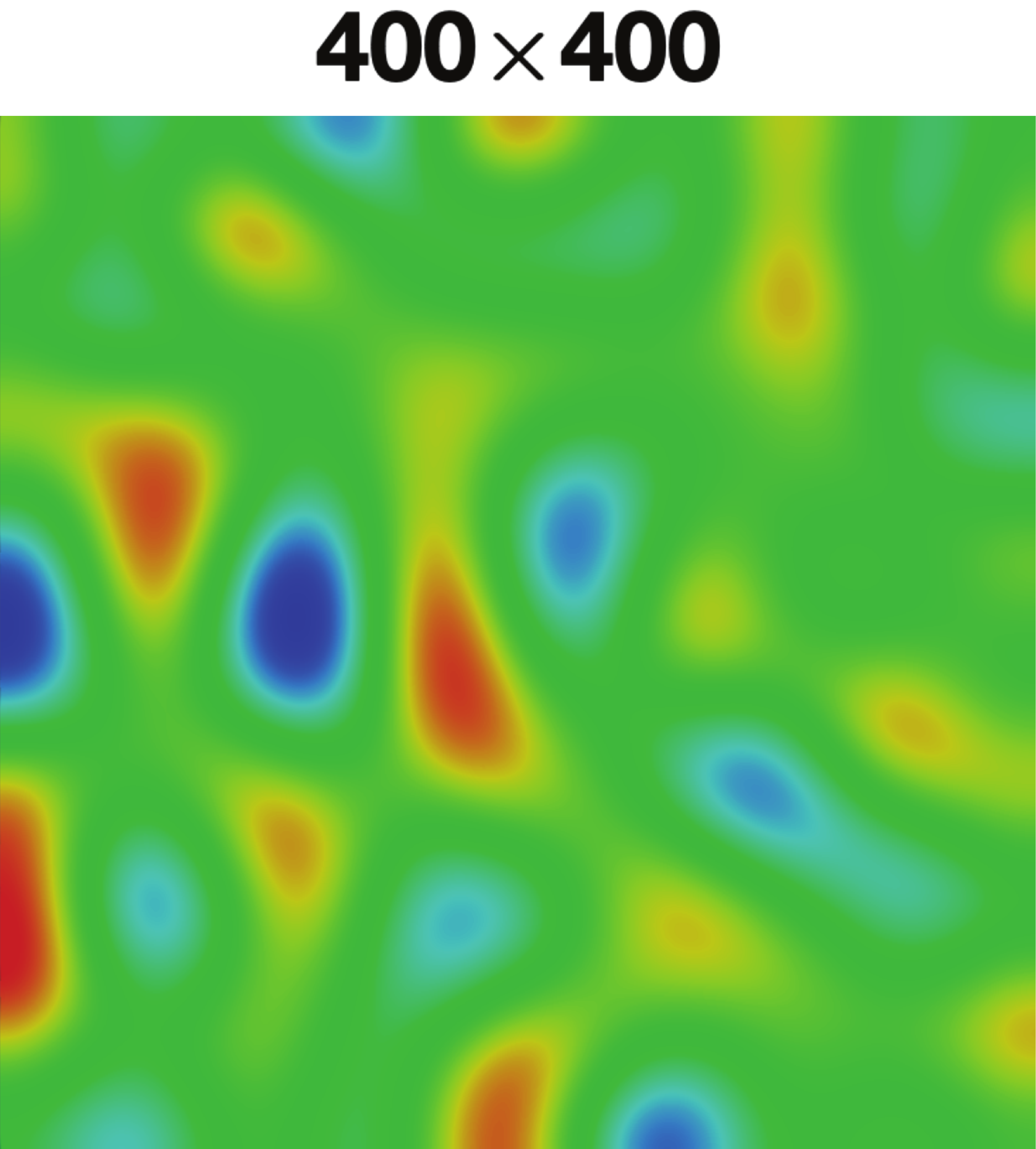}
\includegraphics[scale=0.12]{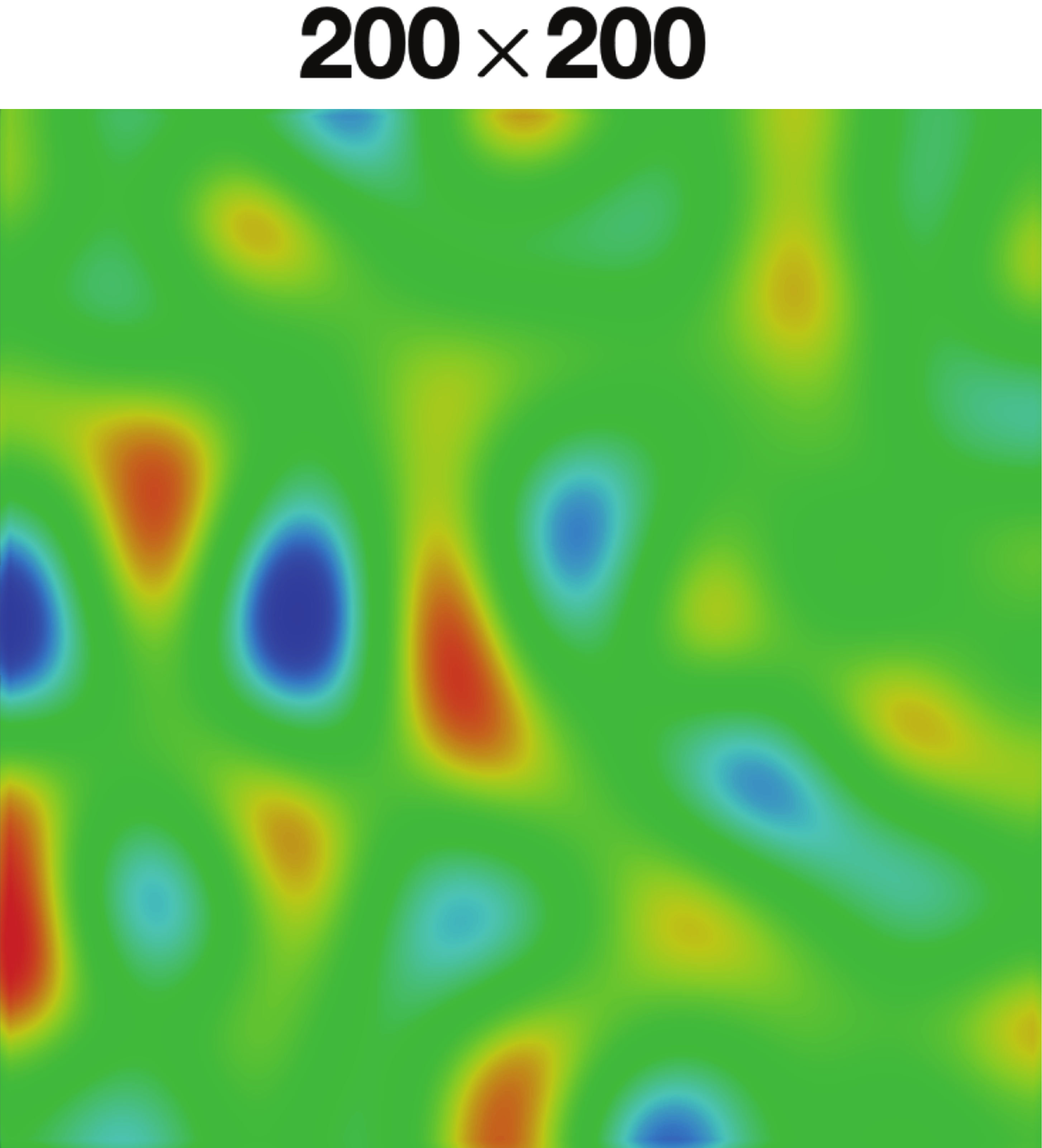}
\includegraphics[scale=0.12]{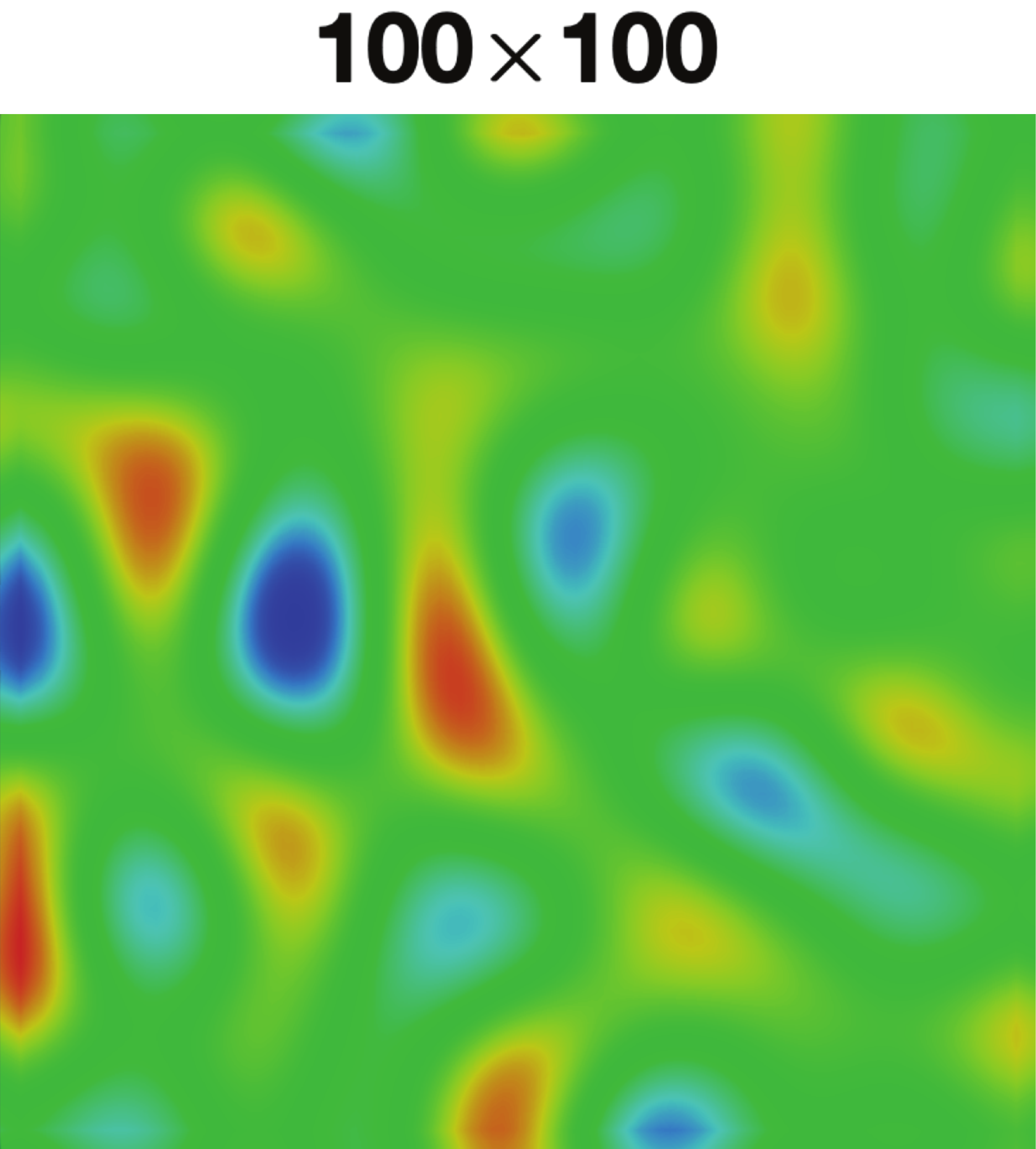}
\includegraphics[scale=0.12]{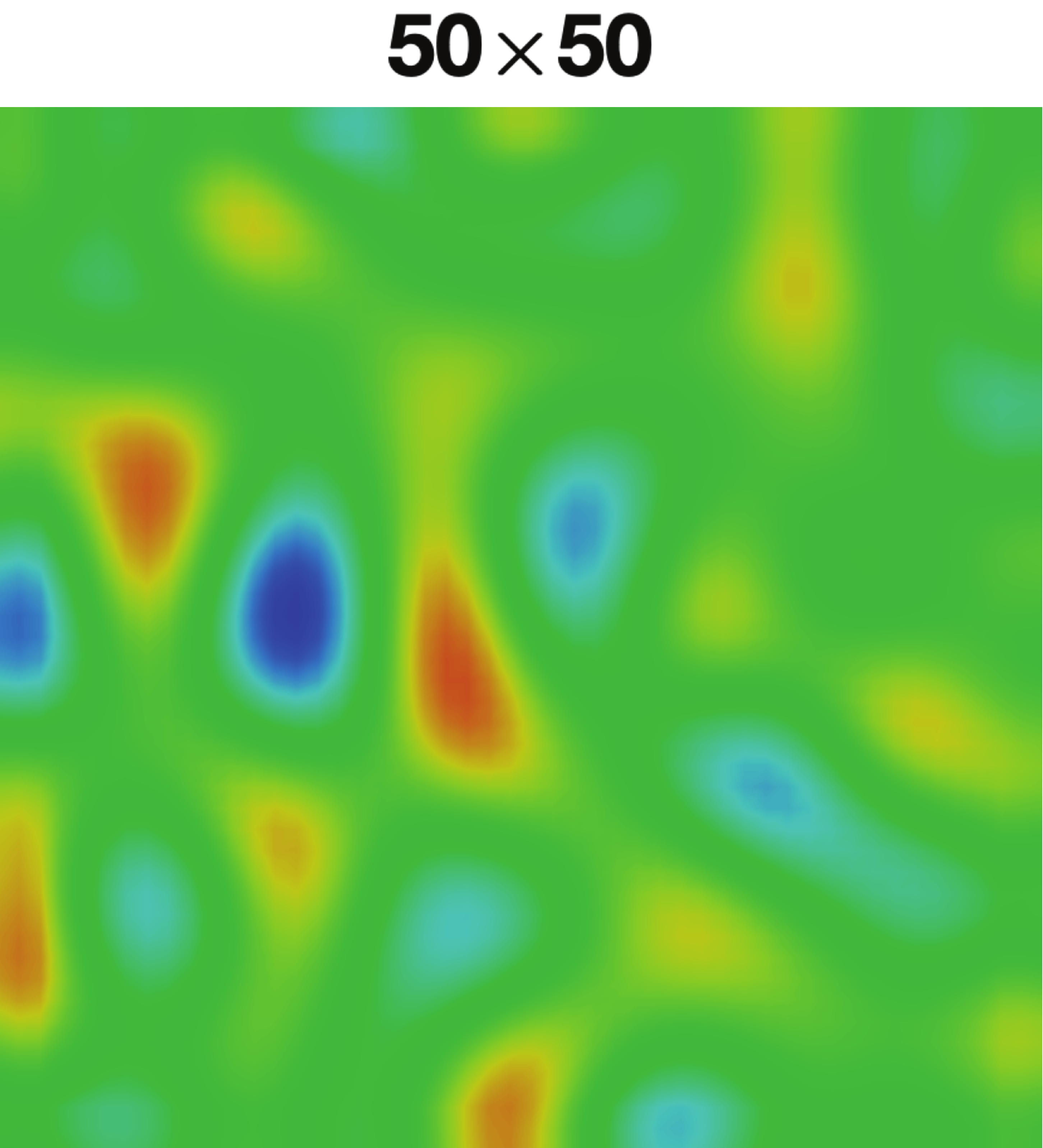}
\caption{The field of $\nabla^2 C_1$ using different fidelity datasets generated by Model 3, on the corresponding mesh. As the mesh size, $h$, decreases, the more diffuse features  diminish in sharpness, and even completely disappear in the $50\times50$ mesh.}
\label{fig:basis_7}
\end{figure}
Therefore, all the results deteriorate when using lower fidelity data. Specifically, for data generated from Model 1, the error gradually increases for decreasing mesh size. Due to the small pre-factor for the constant term, the error in the governing equation of $C_1$ is bigger than that in the governing equation of $C_2$.

\subsection{System identification under noisy data}
\label{sec:result_noise}

\begin{figure}[hbtp]
\centering
\includegraphics[scale=0.5]{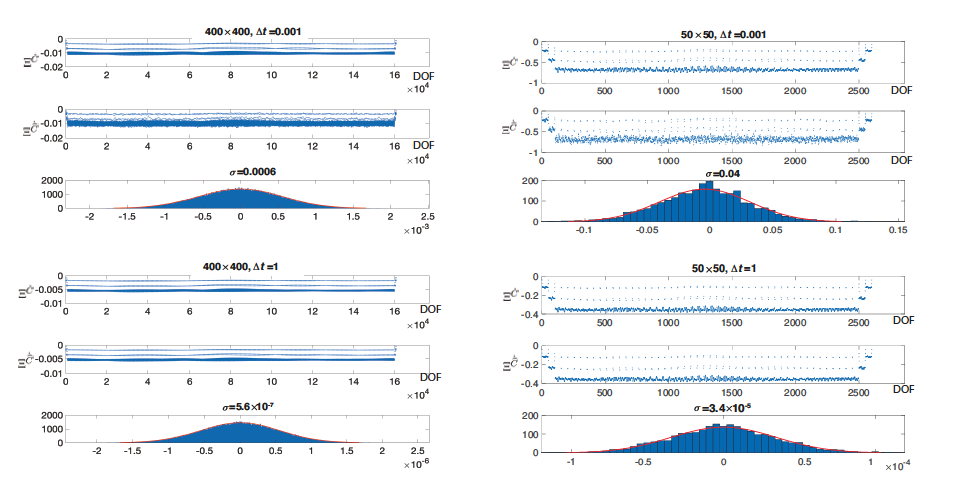}
\caption{Each plot shows the true value of the basis components $\Xi_i^{\dot{C}}$, $i = 1,\dots \text{Number(DOF)}$, generated from noise-free data, in the top subplot, basis generated from noisy data in the middle subplot, and the histogram of error between them in the bottom subplot. The error follows a Gaussian distribution with standard deviation denoted by $\sigma$. The ratio of the error to the maximum true value, $\frac{\sigma}{\max_i\vert\Xi_i^{\dot{C}}\vert} \sim \mathcal{O}(10^{-2})$ for the top two plots ($\Delta t = 0.001$), and decreases with increase in $\Delta t$ to $\sim\mathcal{O}(10^{-4})$ for the bottom two plots ($\Delta t = 1$).}
\label{fig:basis_y}
\end{figure}

Using meshes of different fidelity, we superimpose noise with $\sigma=10^{-4}$ for data generated from Models 1 and 2, and $\sigma=10^{-5}$ for data generated from Models 3 and 4. The noise on $C$ will be amplified in the time derivative and spatial gradients as discussed in Section \ref{sec:Low_fidelity_noisy_data}. Figure \ref{fig:basis_y} shows the target vector, $\Xi^{\dot{\widehat{C} }}$ with two different time steps, $\Delta t=0.001$ and $\Delta t = 1$, on different meshes. At the smaller time step, $\Delta t=0.001$, the noise, scaled by $\Delta t$, washes out the true value. Decreasing the mesh size cannot alleviate the noise, as the ratio of error, denoted by $\sigma$, to its true value largely remains fixed. At the larger time step, $\Delta t=1$, the noise is suppressed, as expressed in Equation (\ref{eq:basis_back_euler_error}). For the following analysis, We choose data at the same 10 time steps starting from $t=10$, with time step $\Delta t=1$.

\begin{figure}[hbtp]
\centering
\includegraphics[scale=0.5]{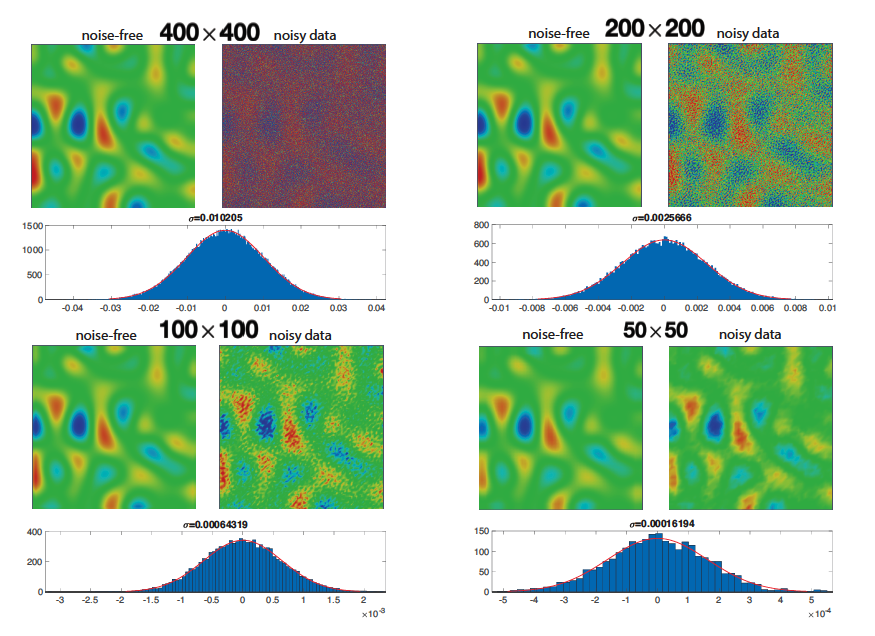}
\caption{Each panel shows the field $\nabla^2 C_1$ extracted from datasets with varying fidelity and generated by Model 1 with and without noise. The histogram of error between the noise-free and noisy data appears in the bottom plot. The error appears to follow a Gaussian distribution with standard deviation denoted by $\sigma$. All the noise-free data produce similar results, while the noise increasingly dominates as the mesh size decreases. (The standard deviations of the error distributions are scaled by $h^{-2}$). The true value begins to emerge in the basis generated from noisy data as the mesh size decreases.}
\label{fig:basis_11}
\end{figure}

\begin{figure}[hbtp]
\centering
\includegraphics[scale=0.2]{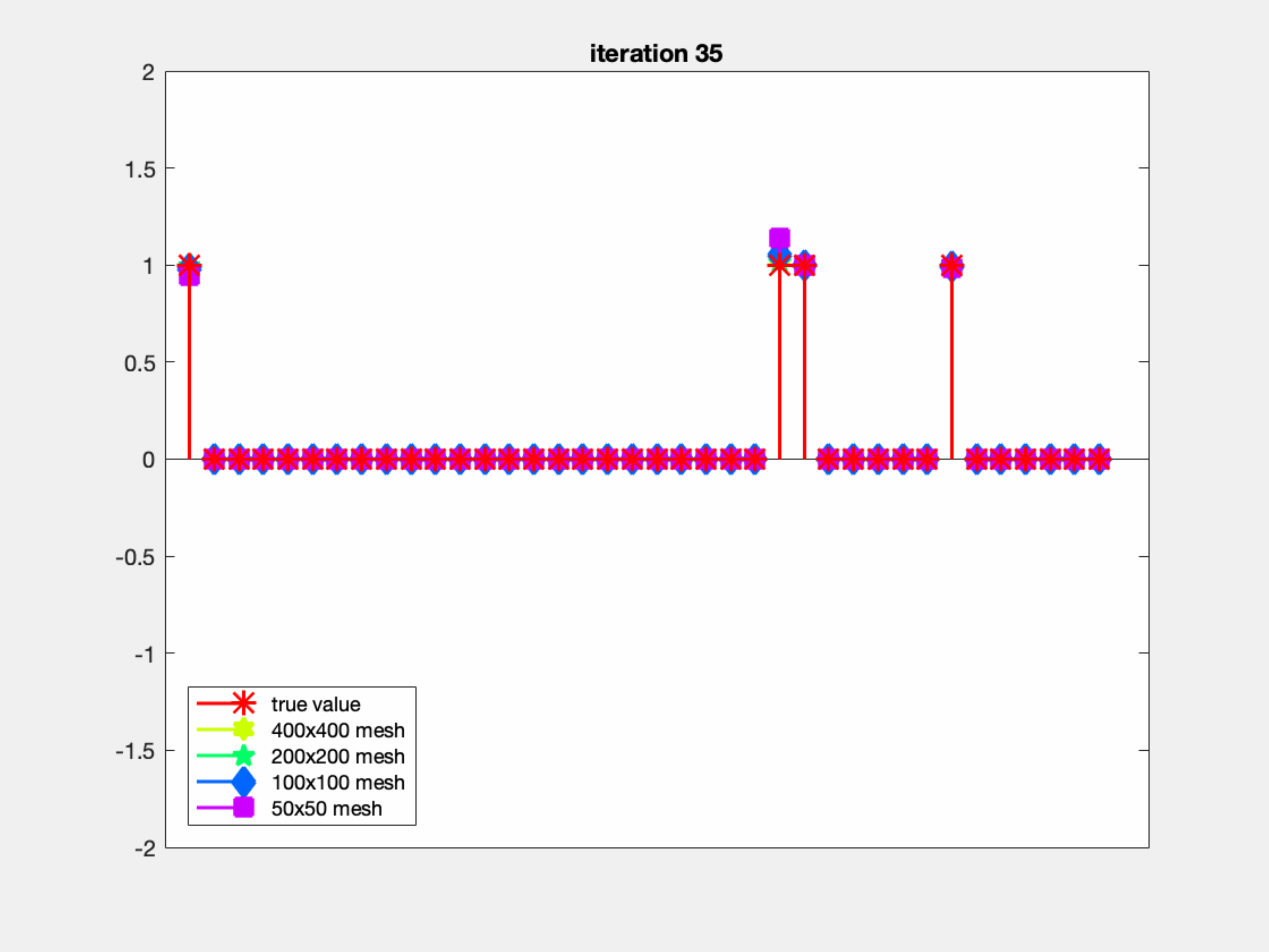}
\includegraphics[scale=0.2]{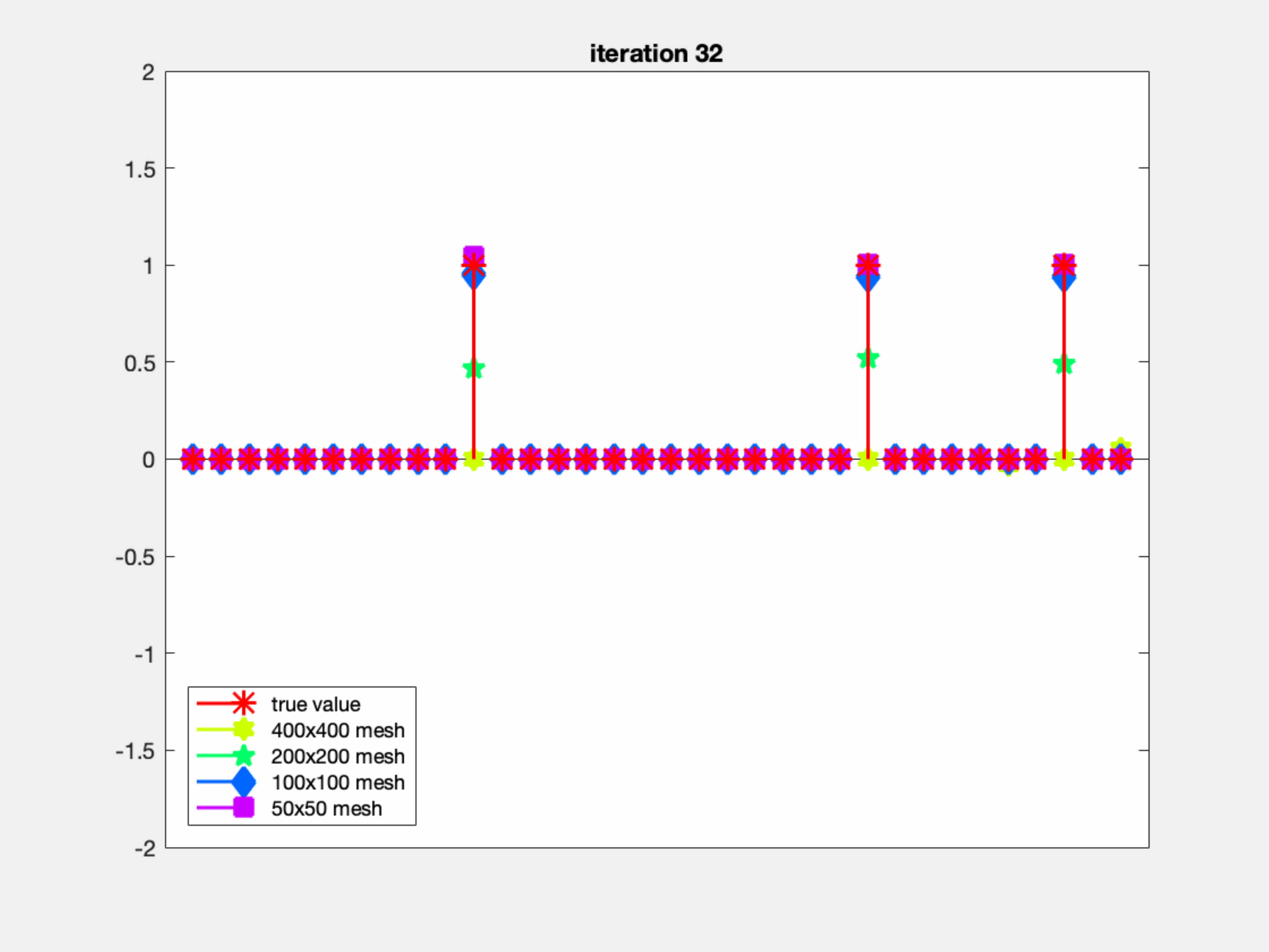}
\caption{Inferred operators for $C_1$  (left panel) and $C_2$ (right panel) using data generated from Model 1 using $\Delta t =1$, with 10 snapshots and $\sigma=10^{-4}$. The identified pre-factors of relevant terms are scaled by their true values: $\int_{\Omega}-1\nabla w_1 \cdot\nabla C_1\text{d}v
,\; 0.1\int_{\Omega}w_1\text{d}v, -1\int_{\Omega}w_1C_1\text{d}v,\; 1\int_{\Omega}w_1C_1^2C_2\text{d}v$ in the left plot and $-40\int_{\Omega}\nabla w_2\cdot\nabla C_2\text{d}v,\; \int_{\Omega}0.9\int_{\Omega}w_2\text{d}v,\; -1\int_{\Omega}C_1^2C_2\text{d}v$ in right plot. They also are arranged from left to right in the same order as they appear in Equations \eqref{eq:weak_value_model1-1} and \eqref{eq:weak_value_model1-2}. The original results are summarized in Table \ref{ta:results_noflux_Noise_dt1}. Using data from the $400\times400$ mesh , the governing equation for $C_2$ is identified incorrectly, as seen in the right panel. 
The transient results have been included as Movies 9 and 10 in supplementary material.
}
\label{fig:results_noflux_Noise_dt1}
\end{figure}

\begin{figure}[hbtp]
\centering
\includegraphics[scale=0.2]{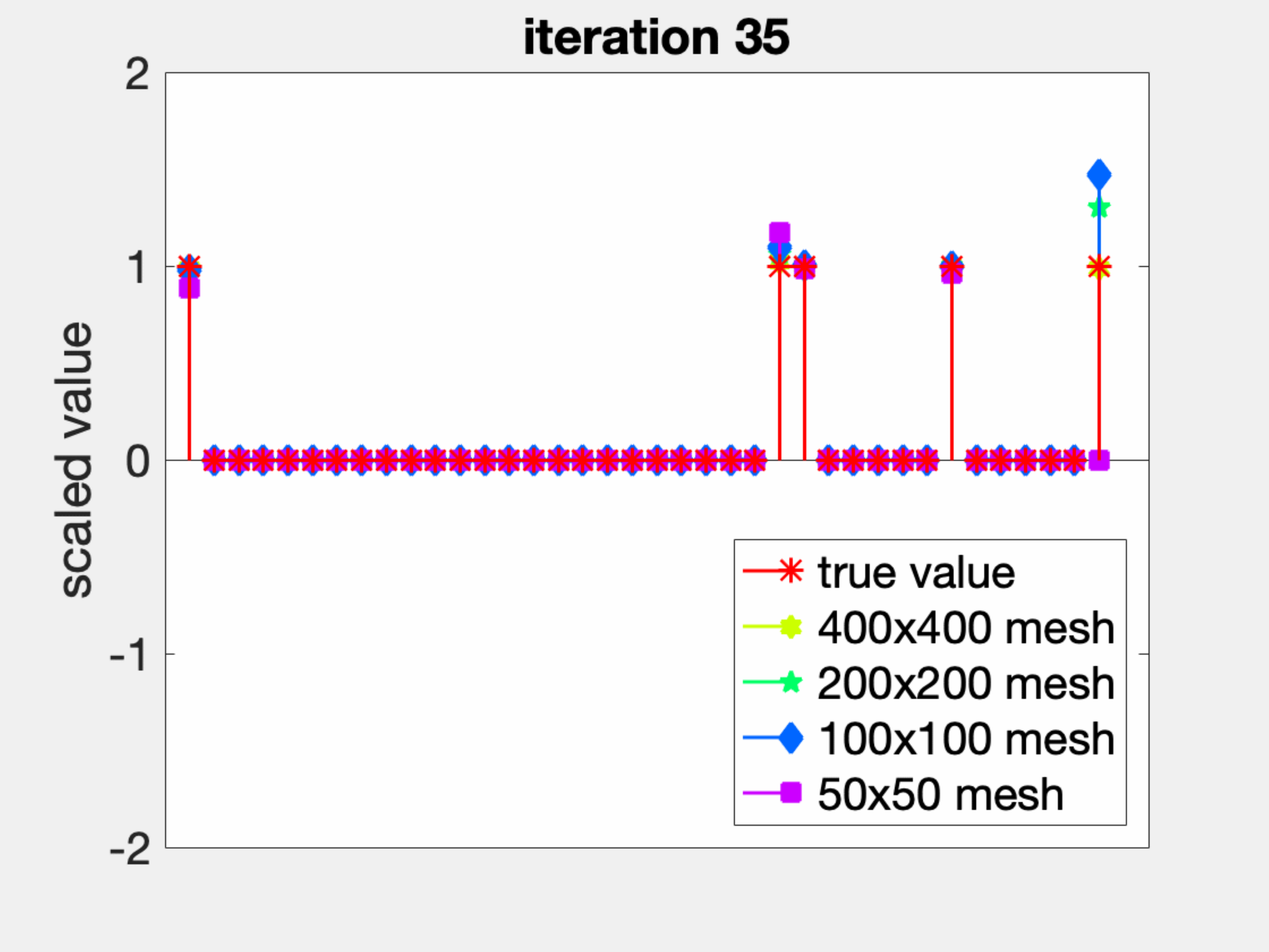}
\includegraphics[scale=0.2]{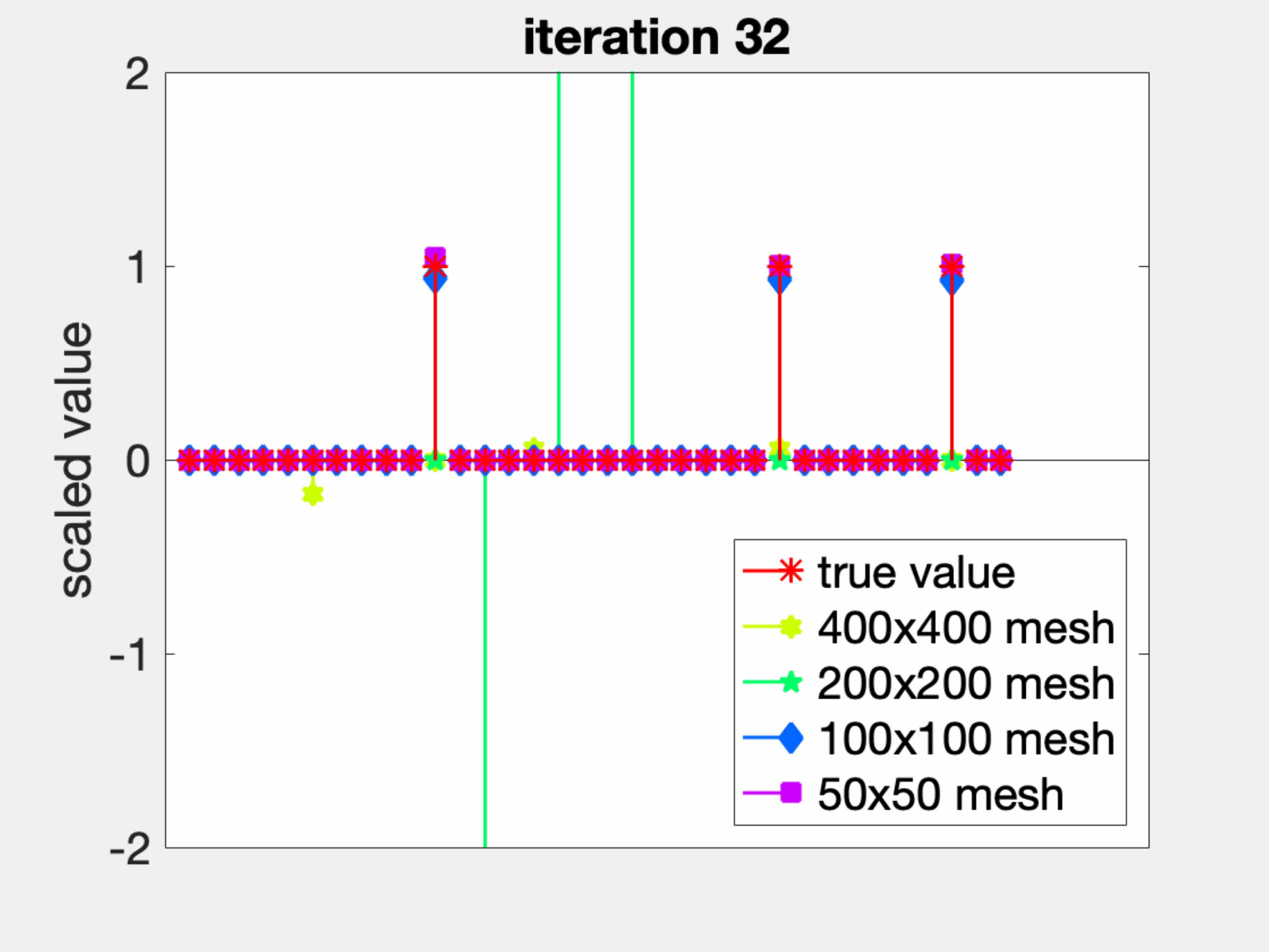}
\caption{Inferred operators for $C_1$ (left panel) and $C_2$ (right panel) with data generated from Model 2 using $\Delta t =1$, with 10 snapshots and $\sigma=10^{-4}$. The identified pre-factors of relevant terms are scaled by their true values: $\int_{\Omega}-1\nabla w_1 \cdot\nabla C_1\text{d}v
,\; 0.1\int_{\Omega}w_1\text{d}v, -1\int_{\Omega}w_1C_1\text{d}v, \; 1\int_{\Omega}w_1C_1^2C_2\text{d}v,\; 0.1\int_{\Gamma_2}w_1\text{d}s$ in the left plot and $-40\int_{\Omega}\nabla w_2\cdot\nabla C_2\text{d}v,\; \int_{\Omega}0.9\int_{\Omega}w_2\text{d}v,\; -1\int_{\Omega}C_1^2C_2\text{d}v$ in right plot. They are also arranged from left to right in the same order as they appear in  Equations \eqref{eq:weak_value_model2-1} and \eqref{eq:weak_value_model2-2}. The original results are summarized in Table \ref{ta:results_influx_noise_dt1}. Using the data from the $400\times400$ or $200\times200$ mesh, the governing equation for $C_2$ is identified incorrectly, as seen in the right panel. The transient results have been included as Movies 11 and 12 in supplementary material.
}
\label{fig:results_influx_Noise_dt1}
\end{figure}

Figure \ref{fig:basis_11} shows the Laplacian basis operator on $C_2$, i.e., $\Xi^{\nabla^2 \widehat{C_2}}$, with data generated by Model 1 using different meshes.
For the fine mesh, the noise remains at a high level over an element, and washes out the true value. By decreasing the mesh size, the ratio of final error to true value decreases by a factor of $h^{-2}$ as shown in Equation (\ref{eq:basis_laplace_error}), and the true value begins to emerge in the basis generated from noisy data. Consequently, as shown in Figure \ref{fig:results_noflux_Noise_dt1}, using high fidelity data (the $400\times400$ mesh) we are unable to correctly identify the governing equation for $C_2$ using data generated by Model 1. Using lower fidelity data, however, we are able to identify all the relevant terms. The error also is smaller when using lower fidelity data. For the same reason, the
governing equation for $C_2$ using data generated by Model 2 also cannot be identified using the $400\times400$ and $200\times200$ meshes, but becomes feasible for lower fidelity data (see Figure \ref{fig:results_influx_Noise_dt1}). On the other hand the error is small for the basis operator $C_1$ in both models. As a result, the estimation of the governing PDE for $C_1$ is comparable with that by using same data of the same fidelity, but without noise. Again, as seen in the previous section, the Neumman boundary condition  cannot able be identified by using very low fidelity data on the $50\times50$ mesh.

\begin{figure}[hbtp]
\centering
\includegraphics[scale=0.5]{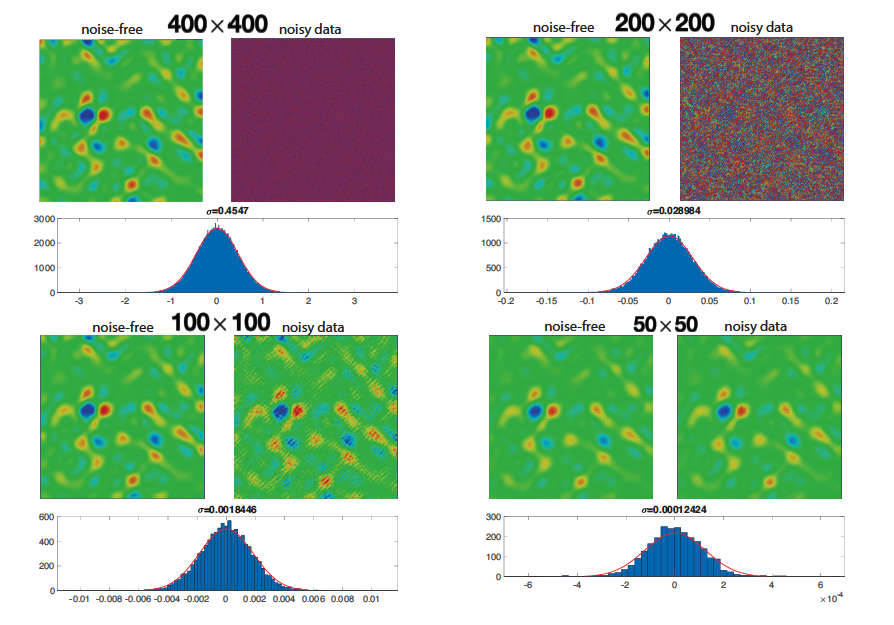}
\caption{Each plot shows the field $\nabla^4 C_1$ using different fidelity datasets with and without noise, generated by Model 3, and the histogram of error between them in the bottom plot. The error appears to follow a Gaussian distribution with standard deviation denoted by $\sigma$. All the noise-free data produce similar results, while the noise washes out the fields very quickly as mesh size increases (The standard deviations of errors are scaled by $h^{-4}$). The true value begins to emerge in the basis generated from noisy data with decreasing mesh size. } 
\label{fig:basis_21}
\end{figure}

\begin{figure}[hbtp]
\centering
\includegraphics[scale=0.2]{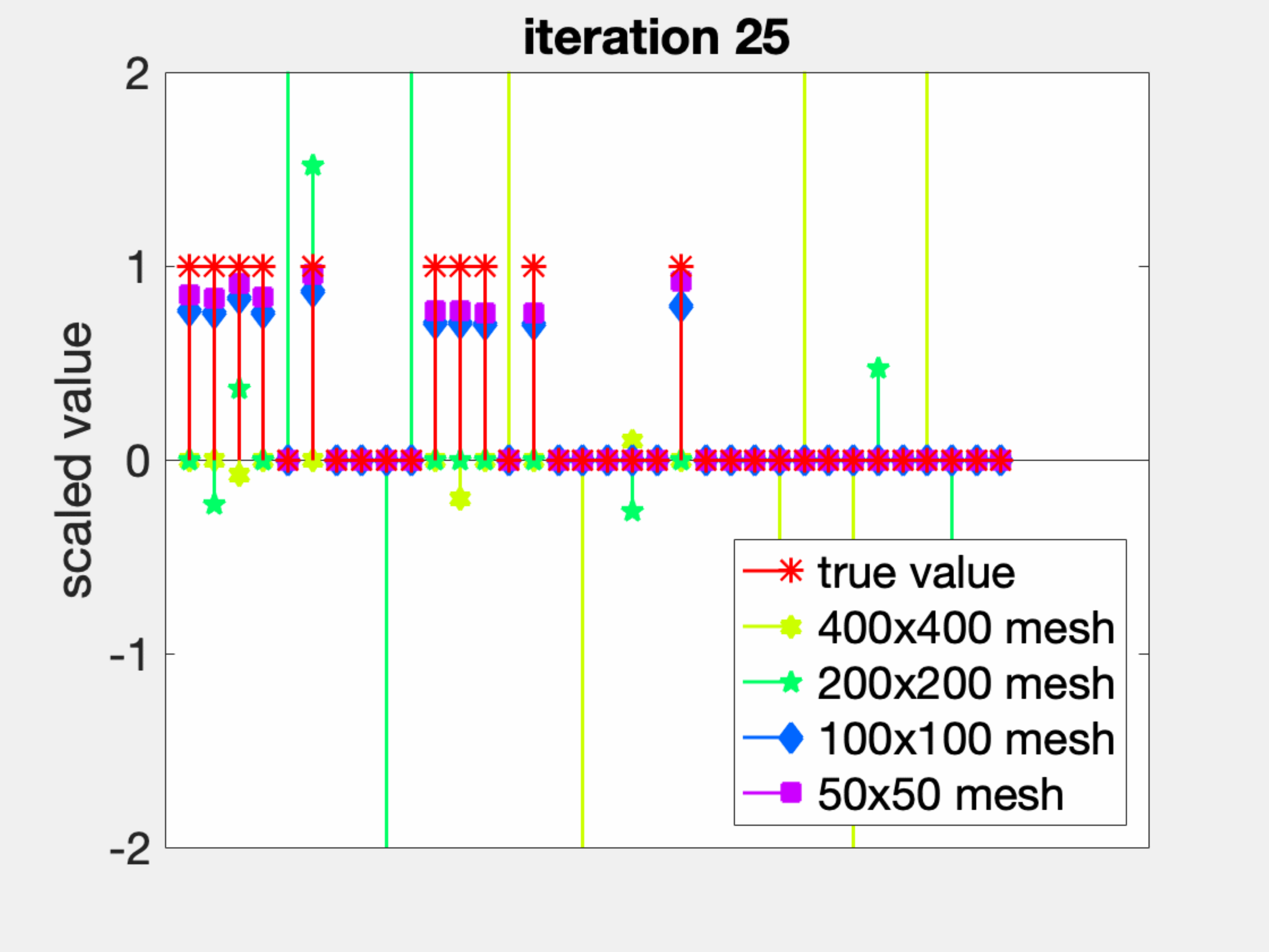}
\includegraphics[scale=0.2]{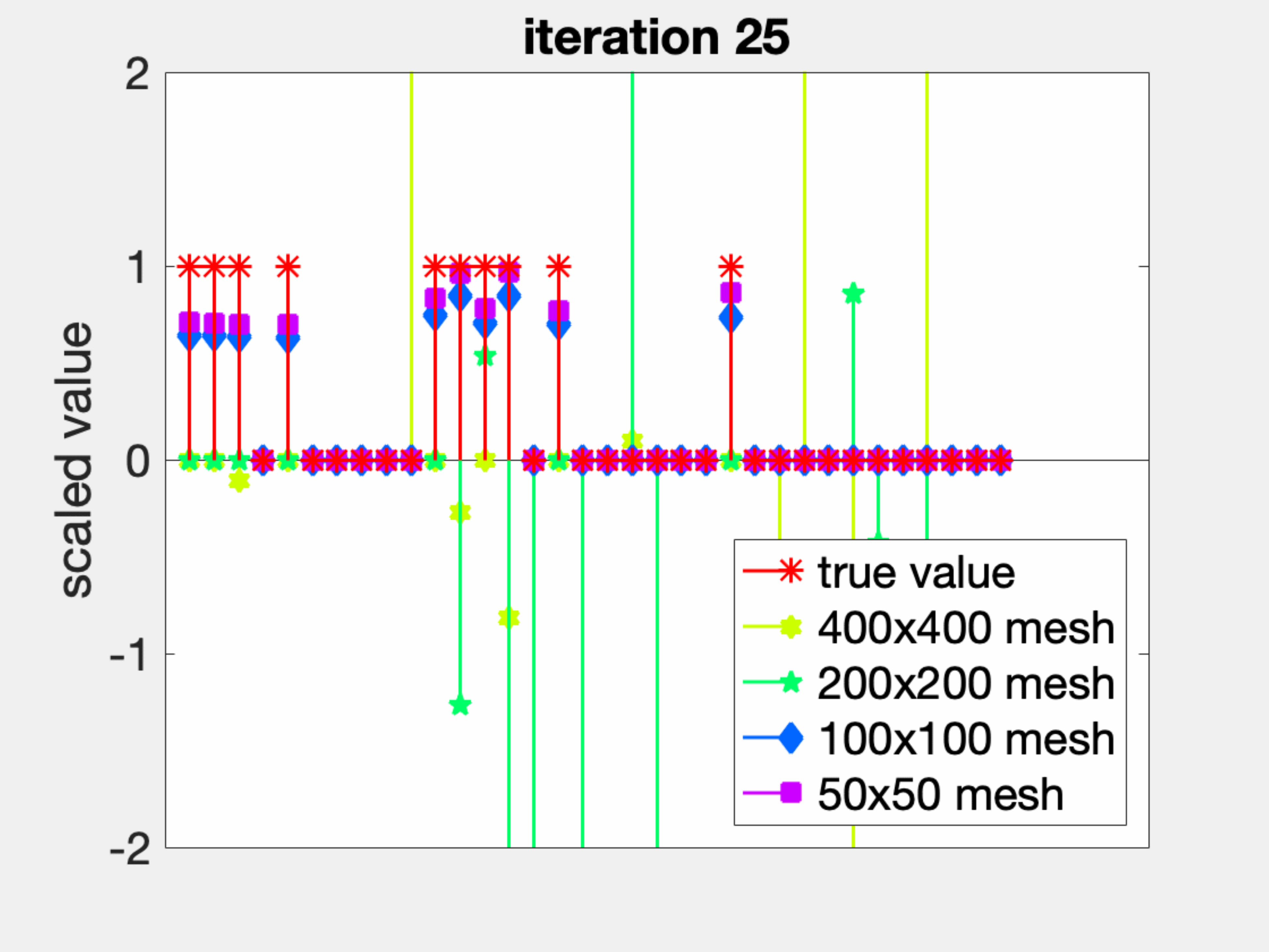}
\caption{Inferred governing equations for $C_1$( left plot) and $C_2$( right plot) using data generating from Model 3 using dt=1, 10 snapshot. $\sigma=10^{-5}$. The estimated pre-factors of relevant terms are scaled by their true values, and arranged from left to right in the same order as they appear in Equation \eqref{eq:weak_value_model3-1} and \eqref{eq:weak_value_model3-2}. The original results are summarized in Table \ref{ta:results_cahn_Noise_dt1}. Using data from the $400\times400$ or $200\times200$ meshes, neither governing equation can be identified correctly. The transient results have been included as Movies 13 and 14 in supplementary material.}
\label{fig:results_cahn_Noise_dt1}
\end{figure}

\begin{figure}[hbtp]
\centering
\includegraphics[scale=0.2]{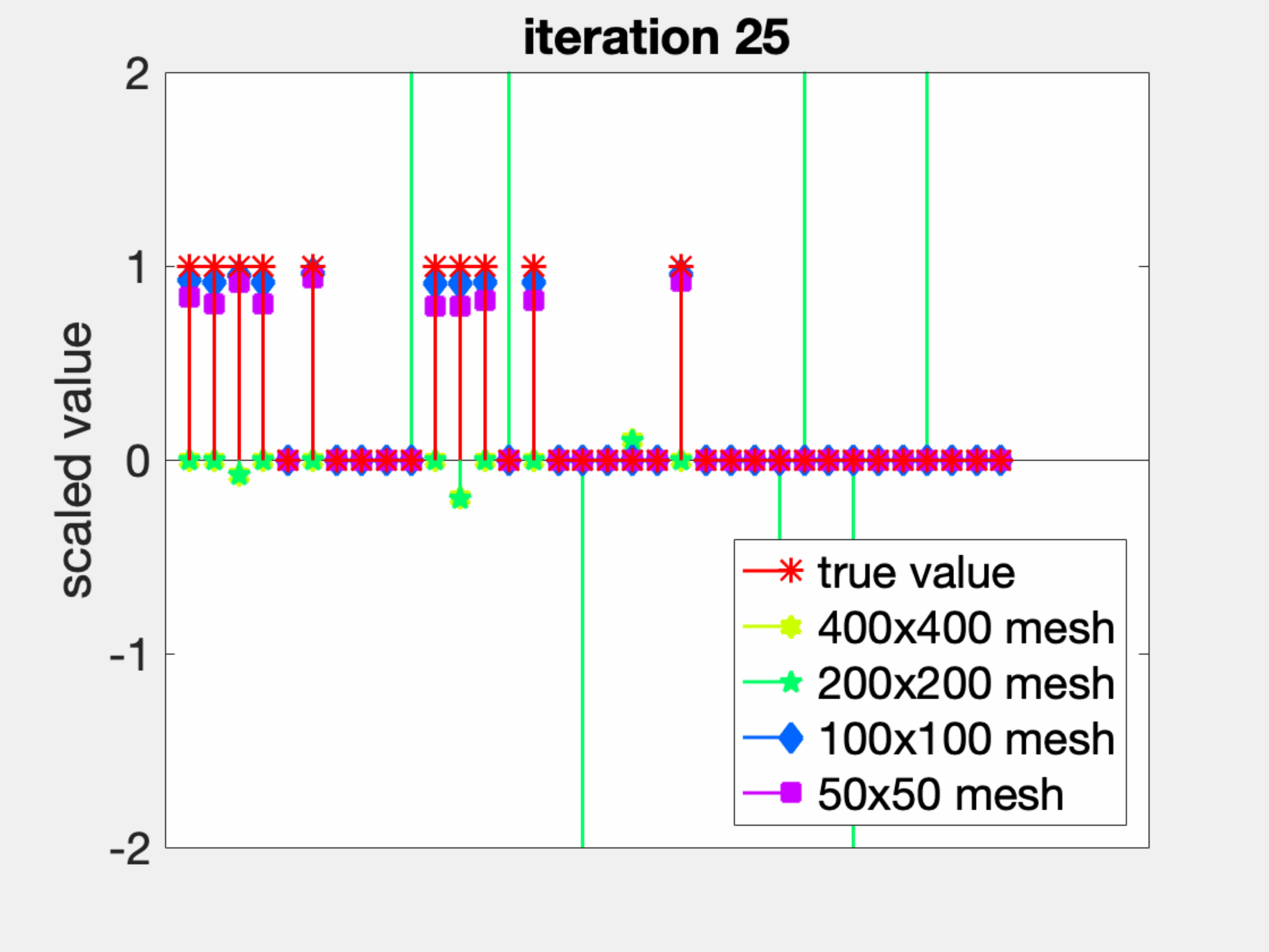}
\includegraphics[scale=0.2]{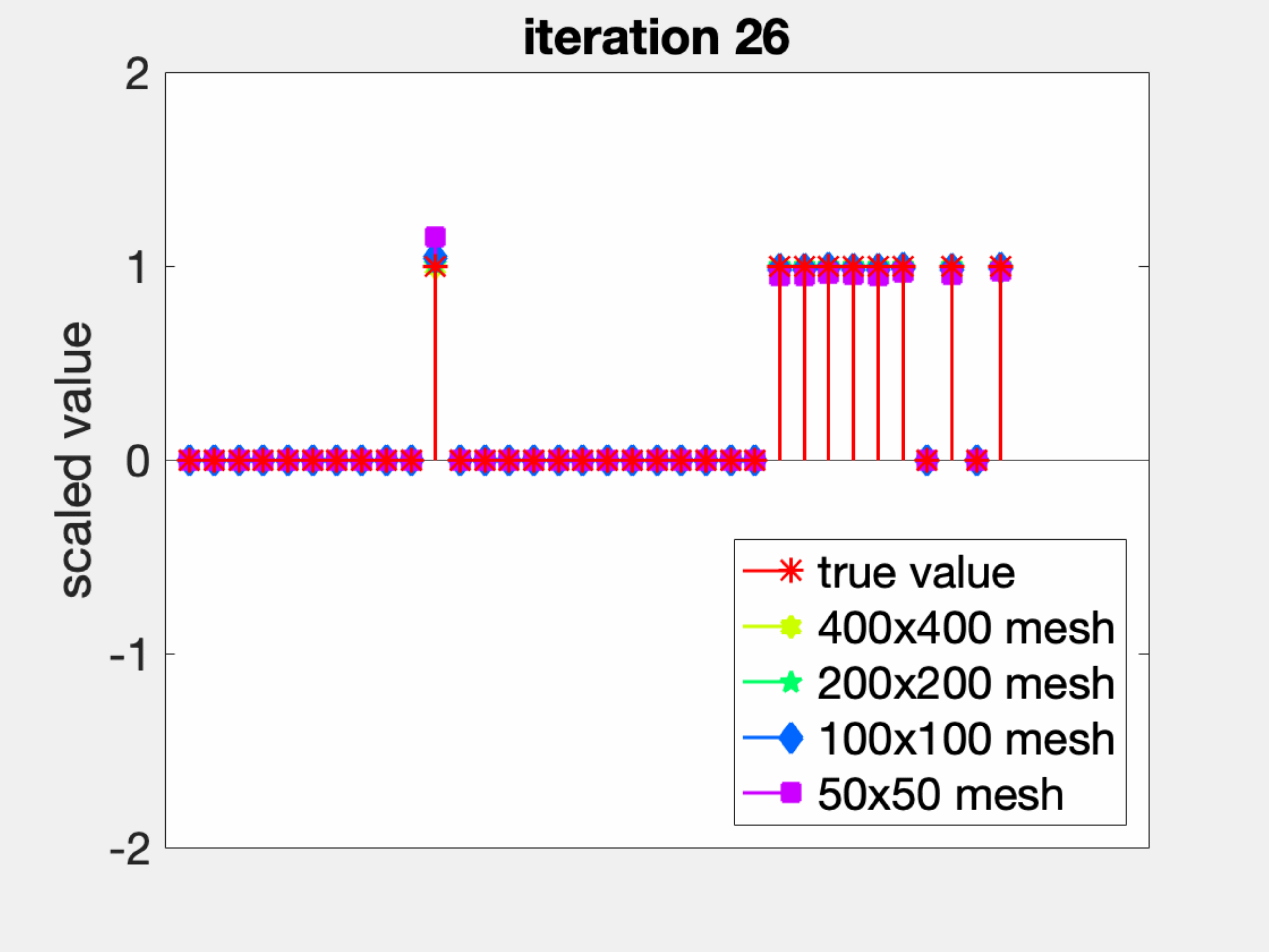}
\caption{Inferred operators for $C_1$ (left plot) and $C_2$ (right plot) using data generating from Model 4 using $\Delta t =1$, 10 snapshots and $\sigma=10^{-5}$. The identified pre-factors of relevant terms are scaled by their true values, and arranged from left to right in the same order as they appear in Equation \eqref{eq:weak_value_model4-1} and \eqref{eq:weak_value_model4-2}. The original results are summarized in Table \ref{ta:results_cahn_Noise_dt1}. Using data from the $400\times400$ or $200\times200$ meshes, the governing equation for $C_1$ is identified incorrectly seen in the left panel. The transient results have been included as Movies 15 and 16 in supplementary material.}
\label{fig:results_alen_noise_dt1}
\end{figure}

Figure \ref{fig:basis_21} shows the biharmonic operator, $\Xi^{\nabla^4 \widehat{C_1}}$ using data generated by Model 3 with different meshes.\footnote{This field was computed by projecting the data on to a fourth-order $C^3$ NURBS basis.} Similar to the Laplace operator, the true value of the basis is washed out as the noise overwhelms the high fidelity data on the fine meshes. By using low fidelity data, we are able to correctly identify the governing equations for $C_1$ and $C_2$ using data generated from Model 3 as shown in Figure \ref{fig:results_cahn_Noise_dt1}. The results for data generated from Model 4 are mixed. The governing equation for $C_1$ is similar to that in Model 3 with the fourth-order biharmonic operator in the concentration. Similar to that case, and for the same reason of noise overwhelming the true data, Model 4 cannot be identified by using high fidelity data, but can be identified by using low fidelity data (see  Figure \ref{fig:results_alen_noise_dt1}). However, the governing equation for $C_2$ does not have the fourth-order biharmonic operator. Given the small error on $(C_1, C_2)$, the identification  of the governing PDE for $C_2$ is comparable with that by using same fidelity data, but without noise. Again, as seen in the last section, for the $50\times50$ mesh, we need to place a bias on $\int_{\Omega}\nabla wC_2^2\nabla C_1\text{d}v$ to protect it from elimination in the first few iterations for data generated from Models 3 and 4. Also note that the error increases significantly to $20.7\%$ with decreasing fidelity as shown in Table \ref{ta:results_alen_noise_dt1}.

Finally for all results using noisy data, the errors are, in general, higher than those obtained by using same fidelity data, but without noise.

\section{Discussion and conclusions}
\label{sec:conclusions}
The development of patterns in biophysics and material physics are governed by a range of spatio-temporal PDEs. It is compelling to attempt to discover the analytic forms of these PDEs from data, because doing so immediately provides insight to the governing physics. System identification has been explored using the strong form of the PDEs \cite{SchmidtSCI2009, SchmidtPB2011,KutzPNAS2015, KutzIEEE2016, KutzSCIADV2017}, as discussed in the Introduction. However the equivalent weak forms, which can offer capabilities not achievable by strong forms, have remained under-exploited for system identification. The weak form transfers derivatives to the weighting function, and allows smooth evaluation of the remaining, possibly higher-order, derivatives when NURBS basis functions are employed. In our numerical experiments, NURBS basis functions allow the fourth-order terms in the Cahn-Hilliard and Allen-Cahn equations to be correctly identified with higher accuracy estimation of the pre-factors using data at different fidelity, with and without noise (results in Tables \ref{ta:tab:results_Cahn_noNoise_dt1}, \ref{ta:tab:results_alen_noNoise_dt1}, \ref{ta:results_cahn_Noise_dt1} and \ref{ta:results_alen_noise_dt1}). Another advantage of writing the PDEs in weak form is that Neumann boundary conditions are explicitly included as surface integrals. During system identification, this class of boundary conditions can be, therefore, constructed along with other operators in the PDEs. By incorporating these terms in the regression model, Equation \eqref{eq:least-square}, identification of boundary conditions becomes feasible (results in Tables \ref{ta:tab:results_influx_noNoise_dt1} and \ref{ta:results_influx_noise_dt1}). 

Intuitively high fidelity data is favored as it minimizes the residual equations, and can accurately evaluate the time derivative and spatial gradient. For data without noise, the results enjoy these advantages (results in Tables
\ref{ta:tab:results_noflux_noNoise_dt1},
\ref{ta:tab:results_influx_noNoise_dt1}, \ref{ta:tab:results_Cahn_noNoise_dt1} and  \ref{ta:tab:results_alen_noNoise_dt1} and Figures \ref{fig:results_noflux_noNoise_dt1} - \ref{fig:results_alen_noNoise_dt1}). However, the error embedded in noisy data is exaggerated relative to the true signal by small time steps and element lengths, as discussed in the context of Equations (\ref{eq:back_euler}), (\ref{eq:basis_laplace_noise}) and (\ref{eq:basis_Biharmonic_noise}). Low fidelity data in time and space may allow greater variations in the true signal relative to the noise, as shown in Figures \ref{fig:basis_y}, \ref{fig:basis_11} and \ref{fig:basis_21}, and thus improve the results of system identification (results in Tables \ref{ta:results_noflux_Noise_dt1}, \ref{ta:results_influx_noise_dt1}, \ref{ta:results_cahn_Noise_dt1} and \ref{ta:results_alen_noise_dt1}). On the other hand, low fidelity data may yield inaccurate estimation of the pre-factors due to poor evaluation of the operators (Figure \ref{fig:basis_7}). {\color{black} Data of very low fidelity may be incapable of capturing the high-frequency content of solution fields, thus leading to entirely incorrect identification of the operators. For example, without an injection of prior knowledge, we were unable to identify the high frequency operators such as $\int_{\Omega}\nabla wC_2^2\nabla C_1\text{d}v$ in Model 3 and 4 using low fidelity data on the 50$\times$ 50 mesh.  }  The careful selection of data could improve these results, but that direction for study lies beyond the scope of this first communication.

We have demonstrated our algorithm on data at different fidelity with and without noise, generated from three different types of PDEs. In this first communication, we do not aim to systematically study the uncertainty in the models induced by the lack of fidelity and noise, nor to optimize the data extraction from simulations/experiments. However we point out several factors that could influence the results:

\begin{enumerate}
\item Two analytically independent bases may appear to be linearly dependent over certain time intervals (Figure \ref{fig:bsis_1_2}). This spurious linearity could disappear as the system evolves.

\item Low fidelity data on coarse meshes resolves the variation of fields poorly (see Figure \ref{fig:basis_7}). However it helps by allowing larger variation of the true signal relative to the noise. It thus can  effectively smooth out the error in bases constructed from noisy data (Figures \ref{fig:basis_11} and \ref{fig:basis_21}).  

\item Fields may evolve at different rates, and be distributed with different gradients (Figures \ref{fig:basis_7} and \ref{fig:basis_y} ). As a result a chosen basis may be more prominent at certain DOFs  than at others, and contribute more in the linear regression model. Using DOFs over subdomains may improve the results of system identification. Using partial data in system identification has been seen to reduce the computational expense \cite{KutzSCIADV2017}, but atempting to systematically improve the results by choosing partial data at certain times has not been widely explored. 

\item We found that results improve when the Neumman boundary condition bases are not included. We could correctly identify the governing equation generated from Model 1 with larger noise by not including the boundary condition bases. We suspect that since the Neumman boundary condition basis is zero at all inner DOFs it renders the matrix of bases, $\BXi$, close to singular. 

\item Any prior knowledge of the system would help to select the correct basis from many candidates. For example we placed a bias on certain bases when identifying PDEs using data generated from Model 3 with low fidelity data. Similarly, knowing that Model 3 is a conserved system, all bases contributing to the reaction (source) terms can be pre-eliminated.  
\end{enumerate}

We note that our treatment is finite-dimensional by being discretization-based, similar to previous work that has used the finite-difference method for representing operators \cite{KutzPNAS2015, KutzIEEE2016, KutzSCIADV2017, KutzChaos2018,KutzHybrid2018, KutzSIAM2019}. 
A different, neural network-based approach \cite{Raissi2019} enforced PDE constraints at random collocation points in one or two spatial dimensions and in time, thus effectively introducing a finite-dimensional representation with the global basis induced by activation functions. 
We also note that the neural network approach, while illustrated for learning coefficients in a predefined PDE, remains to be tested for discovery of PDE structures by downselection from a large set of operators, which is central to the treatment in this work and other recent approaches \cite{KutzPNAS2015, KutzIEEE2016, KutzSCIADV2017, KutzChaos2018,KutzHybrid2018, KutzSIAM2019}.

Finally, we have presented system identification of parabolic PDEs in two dimensions. This is not a very serious restriction for the targeted physics of pattern formation in materials and biophysics. The preponderance of data in these fields is also restricted to two dimensions. Apart from an increase in computational cost, as the set of differential operators increases under the curse of dimensionality, there are no fundamental limitations. Even in this regard, the physics itself restricts the set of three-dimensional operators actually admissible in PDEs. Such classical knowledge could be leveraged to sparsify the operator basis set in three dimensions.

In this work, we currently use standard linear regression models for each iteration of the step-wise regression algorithm. Bayesian frameworks present a natural path to systematically {\color{black}inject prior knowledge and domain expertise, and to} quantify and propagate the uncertainty induced by lack of model fidelity and data noise. Furthermore, they enable techniques for comparing competing models that are rooted in a rigorous probabilistic paradigm, such as with the Bayesian model selection and model averaging methods. 
Bayesian analysis of our methods will appear in future communications. 

\section*{Acknowledgements}
\label{sec:acknowledgements}
We acknowledge the support of Toyota Research Institute, Award \#849910, ``Computational framework for data-driven, predictive, multi-scale and multi-physics modeling of battery materials" (ZW and KG). Additional support: This material is based upon work supported by the Defense Advanced Research Projects Agency (DARPA) under Agreement No. HR0011199002, ``Artificial Intelligence guided multi-scale multi-physics framework for discovering complex emergent materials phenomena'' (ZW, XH and KG).  Simulations in this work were performed using the Extreme Science and Engineering Discovery Environment (XSEDE) Stampede2 at the Texas Advance Computing Center through allocations TG-MSS160003 and TG-DMR180072. XSEDE is supported by National Science Foundation grant number ACI-1548562.


\section*{Appendix}
\label{sec:Appendix}
\begin{table}[h]
\centering
\footnotesize
 \begin{tabular}{|c|c|c|}
 \hline
 mesh size& results&error\\
 \hline
\multirow{2}{*}{$400\times 400$} &$\int_{\Omega}w_1\frac{\partial C_1}{\partial t}\text{d}v=\int_{\Omega}-1\nabla w_1 \cdot\nabla C_1\text{d}v
+\int_{\Omega}w_1 (0.1-C_1+1C_1^2C_2)\text{d}v$ &0\%\\
 & $\int_{\Omega}w_2\frac{\partial C_2}{\partial t}\text{d}v=\int_{\Omega}-40\nabla w_2\cdot\nabla C_2+\int_{\Omega}w_2(0.9-1C_1^2C_2)\text{d}v$  &0\% \\
\hline
\multirow{2}{*}{$200\times 200$} &$\int_{\Omega}w_1\frac{\partial C_1}{\partial t}\text{d}v=\int_{\Omega}-0.9971\nabla w_1 \cdot\nabla C_1\text{d}v
+\int_{\Omega}w_1 (0.10116-0.9998C_1+0.9985C_1^2C_2)\text{d}v$ & 1.15\% \\ 
  & $\int_{\Omega}w_2\frac{\partial C_2}{\partial t}\text{d}v=\int_{\Omega}-40.2205\nabla w_2\cdot\nabla C_2+\int_{\Omega}w_2(0.89982-0.9998C_1^2C_2)\text{d}v$& 0.56\%  \\ \hline
\multirow{2}{*}{$100\times 100$} & $\int_{\Omega}w_1\frac{\partial C_1}{\partial t}\text{d}v=\int_{\Omega}-0.9805\nabla w_1 \cdot\nabla C_1\text{d}v
+\int_{\Omega}w_1 (0.1051-0.9995C_1+0.9936C_1^2C_2)\text{d}v$& 5.14\%  \\ 
&$\int_{\Omega}w_2\frac{\partial C_2}{\partial t}\text{d}v=\int_{\Omega}-40.73\nabla w_2\cdot\nabla C_2+\int_{\Omega}w_2(-0.9019-1.0026C_1^2C_2)\text{d}v$ &1.83\%\\
\hline
\multirow{2}{*}{$50\times 50$}  & $\int_{\Omega}w_1\frac{\partial C_1}{\partial t}\text{d}v=\int_{\Omega}-0.9489\nabla w_1 \cdot\nabla C_1\text{d}v
+\int_{\Omega}w_1 (0.1138-1.0007C_1+0.9858C_1^2C_2)\text{d}v$ &13.9\%  \\ 
&$\int_{\Omega}w_2\frac{\partial C_2}{\partial t}\text{d}v\int_{\Omega}-41.9339\nabla w_2\cdot\nabla C_2+\int_{\Omega}w_2(-0.9075-1.01C_1^2C_2)\text{d}v$ &4.83\%  \\ \hline
\end{tabular}
\caption{Results using data generated from Model 1 using $\Delta t=1$, with 10 snapshots.}
\label{ta:tab:results_noflux_noNoise_dt1}
\end{table}

\begin{table}[h]
\centering
\footnotesize
 \begin{tabular}{|c|c|c|}
 \hline
 mesh size& results&error\\
 \hline
\multirow{2}{*}{$400\times 400$} &$\int_{\Omega}w_1\frac{\partial C_1}{\partial t}\text{d}v=\int_{\Omega}-1\nabla w_1 \cdot\nabla C_1\text{d}v
+\int_{\Omega}w_1 (0.1-C_1+1C_1^2C_2)\text{d}v+\int_{\Gamma_2}w_10.1\text{d}s$&0\% \\
 & $\int_{\Omega}w_2\frac{\partial C_2}{\partial t}\text{d}v=\int_{\Omega}-40\nabla w_2\cdot\nabla C_2+\int_{\Omega}w_2(0.9-1C_1^2C_2)\text{d}v$&0\%   \\
\hline
\multirow{2}{*}{$200\times 200$} &$\int_{\Omega}w_1\frac{\partial C_1}{\partial t}\text{d}v=\int_{\Omega}-0.982\nabla w_1 \cdot\nabla C_1\text{d}v+\int_{\Omega}w_1 (0.106-1.0027C_1+0.997C_1^2C_2)\text{d}v+\int_{\Gamma_2}w_10.1305\text{d}s$& 30\%\\
 & $\int_{\Omega}w_2\frac{\partial C_2}{\partial t}\text{d}v=\int_{\Omega}-40.23\nabla w_2\cdot\nabla C_2+\int_{\Omega}w_2(0.90043-1.0006C_1^2C_2)\text{d}v$&0.57\%   \\
\hline
\multirow{2}{*}{$100\times 100$} &$\int_{\Omega}w_1\frac{\partial C_1}{\partial t}\text{d}v=\int_{\Omega}-0.9793\nabla w_1 \cdot\nabla C_1\text{d}v+\int_{\Omega}w_1 (0.1095-1.0036C_1+0.9949C_1^2C_2)\text{d}v+\int_{\Gamma_2}w_10.1471\text{d}s$&47.1\% \\
 & $\int_{\Omega}w_2\frac{\partial C_2}{\partial t}\text{d}v=\int_{\Omega}-40.751\nabla w_2\cdot\nabla C_2+\int_{\Omega}w_2(0.9028-1.0031C_1^2C_2)\text{d}v$&1.87\%   \\
\hline
\multirow{2}{*}{$50\times 50$} & $\int_{\Omega}w_1\frac{\partial C_1}{\partial t}\text{d}v=\int_{\Omega}-0.884\nabla w_1 \cdot\nabla C_1\text{d}v+\int_{\Omega}w_1 (0.1172-0.98576C_1+0.9643C_1^2C_2)\text{d}v$  &17\% \\
 &  $\int_{\Omega}w_2\frac{\partial C_2}{\partial t}\text{d}v=\int_{\Omega}-41.985\nabla w_2\cdot\nabla C_2+\int_{\Omega}w_2(0.9076-1.0118C_1^2C_2)\text{d}v$&5\% \\
\hline
\end{tabular}
\caption{Results using data generating from Model 2 with $\Delta t=1$, with 10 snapshots. The high error in the governing equation of $C_1$ is induced by the Neumman boundary condition term, which cannot be identified on the $50\times 50$ mesh.}
\label{ta:tab:results_influx_noNoise_dt1}
\end{table}

\begin{table}[h]
\centering
\scriptsize
 \begin{tabular}{|c|c|c|}
 \hline
 mesh size& results&error\\
 \hline
\multirow{4}{*}{$400\times 400$} &$\int_{\Omega}w_1\frac{\partial C_1}{\partial t}\text{d}v=\int_{\Omega}\nabla w_1\cdot\left(-17.8126+47.98C_1+21.591C_2-47.98C_1^2-15.9933C_2^2 \right)\nabla C_1\text{d}v$&\multirow{2}{*}{0\%} \\
& $+ \int_{\Omega}\nabla w_2\nabla\cdot\left(-10.7955+21.591C_1+15.9933C_2-31.9867C_1C_2 \right)\nabla C_2\text{d}v+\int_{\Omega}-1\nabla^2w_1\nabla^2C_1 $& \\
& $\int_{\Omega}w_2\frac{\partial C_2}{\partial t}\text{d}v=\int_{\Omega}\nabla w_2\cdot\left(-10.7955+21.591C_1+15.9933C_2-31.9867C_1C_2 \right)\nabla C_1\text{d}v$  &\multirow{2}{*}{0\%}\\
& $+ \int_{\Omega}\nabla w_2\nabla\cdot\left(-12.2149+15.9933C_1+42.3823C_2-15.9933C_1^2-47.98C_2^2 \right)\nabla^2C_2\text{d}v+\int_{\Omega}-1\nabla^2w_2\nabla^2C_2$ & \\
\hline
\multirow{4}{*}{$200\times 200$} &$\int_{\Omega}w_1\frac{\partial C_1}{\partial t}\text{d}v=\int_{\Omega}\nabla w_1\cdot\left(-17.8498+48.07C_1+21.674C_2-48.07C_1^2-16.078C_2^2 \right)\nabla C_1\text{d}v$ &\multirow{2}{*}{1.01\%} \\
& $+ \int_{\Omega}\nabla w_2\nabla\cdot\left(-10.789+21.5766C_1+15.9762C_2-31.9503C_1C_2 \right)\nabla C_2\text{d}v+\int_{\Omega}-1.01\nabla^2w_1\nabla^2C_1 $ &\\
& $\int_{\Omega}w_2\frac{\partial C_2}{\partial t}\text{d}v=\int_{\Omega}\nabla w_2\cdot\left(-10.7845+21.567C_1+15.9752C_2-31.9493C_1C_2 \right)\nabla C_1\text{d}v$  &\multirow{2}{*}{0.92\%} \\
& $+ \int_{\Omega}\nabla w_2\nabla\cdot\left(-12.2452+16.1062C_1+42.4017C_2-16.1108C_1^2-47.985C_2^2 \right)\nabla^2C_2\text{d}v+\int_{\Omega}-1.0093\nabla^2w_2\nabla^2C_2$  &\\
\hline
\multirow{4}{*}{$100\times 100$} &$\int_{\Omega}w_1\frac{\partial C_1}{\partial t}\text{d}v=\int_{\Omega}\nabla w_1\cdot\left(-17.4653+46.9488C_1+21.445C_2-46.959C_1^2-16.0377C_2^2 \right)\nabla C_1\text{d}v$ &\multirow{2}{*}{3.9\%} \\
& $+ \int_{\Omega}\nabla w_2\nabla\cdot\left(-10.4004+20.7943C_1+15.367C_2-30.7248C_1C_2 \right)\nabla C_2\text{d}v+\int_{\Omega}-1.0103\nabla^2w_1\nabla^2C_1 $& \\
& $\int_{\Omega}w_2\frac{\partial C_2}{\partial t}\text{d}v=\int_{\Omega}\nabla w_2\cdot\left(-10.2816+20.551C_1+15.2133C_2-30.4078C_1C_2 \right)\nabla C_1\text{d}v$  &\multirow{2}{*}{4.9\%} \\
& $+ \int_{\Omega}\nabla w_2\nabla\cdot\left(-11.9343+16.0866C_1+40.8853C_2-16.1143C_1^2-46.1839C_2^2 \right)\nabla^2C_2\text{d}v+\int_{\Omega}-0.998\nabla^2w_2\nabla^2C_2$ &\\
\hline
\multirow{4}{*}{$50\times 50$} &$\int_{\Omega}w_1\frac{\partial C_1}{\partial t}\text{d}v=\int_{\Omega}\nabla w_1\cdot\left(-15.1267+40.1515C_1+19.6646C_2-40.1784C_1^2\underline{-15.2573}C_2^2 \right)\nabla C_1\text{d}v$  &\multirow{2}{*}{24\%}\\
& $+ \int_{\Omega}\nabla w_2\nabla\cdot\left(-8.3232+16.6306C_1+12.16C_2-24.3068C_1C_2 \right)\nabla C_2\text{d}v+\int_{\Omega}-0.9249\nabla^2w_1\nabla^2C_1 $& \\
& $\int_{\Omega}w_2\frac{\partial C_2}{\partial t}\text{d}v=\int_{\Omega}\nabla w_2\cdot\left(-7.6813+15.2688C_1+11.2396C_2-22.3164C_1C_2 \right)\nabla C_1\text{d}v$   &\multirow{2}{*}{30\%}\\
& $+ \int_{\Omega}\nabla w_2\nabla\cdot\left(-10.187+15.361C_1+33.0694C_2-15.4857C_1^2-37.0552C_2^2 \right)\nabla^2C_2\text{d}v+\int_{\Omega}-0.8645\nabla^2w_2\nabla^2C_2$ & \\
\hline
\end{tabular}
\caption{Results using data generated from Model 3 using $\Delta t=1$, with 10 snapshots. A bias on $\int_{\Omega}\nabla wC_2^2\nabla C_1\text{d}v$ is needed to protect it from elimination in the first few iterations.}
\label{ta:tab:results_Cahn_noNoise_dt1}
\end{table}
\begin{table}[h]
\centering
\footnotesize
 \begin{tabular}{|c|c|c|}
 \hline
 mesh size& results&error\\
 \hline
\multirow{4}{*}{$400\times 400$} & $\int_{\Omega}w_1\frac{\partial C_1}{\partial t}\text{d}v=\int_{\Omega}\nabla w_1\cdot\left(-17.8126+47.98C_1+21.591C_2-47.98C_1^2-15.9933C_2^2 \right)\nabla C_1\text{d}v$ &\multirow{2}{*}{0\%}\\
& $+ \int_{\Omega}\nabla w_2\nabla\cdot\left(-10.7955+21.591C_1+15.9933C_2-31.9867C_1C_2 \right)\nabla C_2\text{d}v+\int_{\Omega}-1\nabla^2w_1\nabla^2C_1$ &\\
& $\int_{\Omega}w_2\frac{\partial C_2}{\partial t}\text{d}v=\int_{\Omega}-1\nabla w_2\cdot\nabla C_2+\int_{\Omega}+w_2(3.5085-10.7955C_1-12.2149C_2$ &\multirow{2}{*}{0\%}\\
& $+10.7955c_1^2+15.9933C_1C_2+21.1912C_2^2-15.9933C_1^2C_2-15.9933C_2^3)\text{d}v$  &\\
\hline
\multirow{4}{*}{$200\times 200$} &$\int_{\Omega}w_1\frac{\partial C_1}{\partial t}\text{d}v=\int_{\Omega}\nabla w_1\cdot\left(-17.8589+48.06C_1+21.7044C_2-48.0635C_1^2-16.09C_2^2 \right)\nabla C_1\text{d}v$ &\multirow{2}{*}{1.2\%}\\
& $+ \int_{\Omega}\nabla w_2\nabla\cdot\left(-10.8207+21.64C_1+16.0478C_2-32.089C_1C_2 \right)\nabla C_2\text{d}v+\int_{\Omega}-1.0123\nabla^2w_1\nabla^2C_1 $ &\\
& $\int_{\Omega}w_2\frac{\partial C_2}{\partial t}\text{d}v=\int_{\Omega}-1.0114\nabla w_2\cdot\nabla C_2+\int_{\Omega}+w_2(3.498-10.7636C_1-12.191C_2$ &\multirow{2}{*}{1.14\%}\\
& $+10.7663C_1^2+15.9459C_1C_2+21.1655C_2^2-15.9499C_1^2C_2-15.9752C_2^3)\text{d}v$  &\\
\hline
\multirow{4}{*}{$100\times 100$} &$\int_{\Omega}w_1\frac{\partial C_1}{\partial t}\text{d}v=\int_{\Omega}\nabla w_1\cdot\left(-17.3435+46.4025C_1+21.4151C_2-46.4119C_1^2-15.9544C_2^2 \right)\nabla C_1\text{d}v$&\multirow{2}{*}{3.28\%}\\
& $+ \int_{\Omega}\nabla w_2\nabla\cdot\left(-10.4451+20.8839C_1+15.5847C_2-31.1354C_1C_2 \right)\nabla C_2\text{d}v+\int_{\Omega}-1.0092\nabla^2w_1\nabla^2C_1 $& \\
& $\int_{\Omega}w_2\frac{\partial C_2}{\partial t}\text{d}v=\int_{\Omega}-1.0458\nabla w_2\cdot\nabla C_2+\int_{\Omega}+w_2(3.46-10.6416C_1-12.096C_2$&\multirow{2}{*}{4.58\%} \\
& $+10.6545C_1^2+15.7652C_1C_2+21.056C_2^2-15.7843C_1^2C_2-15.8966C_2^3)\text{d}v$ & \\
\hline
\multirow{4}{*}{$50\times 50$} &$\int_{\Omega}w_1\frac{\partial C_1}{\partial t}\text{d}v=\int_{\Omega}\nabla w_1\cdot\left(-14.9026+38.6896C_1+19.735C_2-38.689C_1^2-\underline{14.989}C_2^2 \right)\nabla C_1\text{d}v$&\multirow{2}{*}{20.6\%} \\
& $+ \int_{\Omega}\nabla w_2\nabla\cdot\left(-8.5697+17.1244C_1+13.1685C_2-26.231C_1C_2 \right)\nabla C_2\text{d}v+\int_{\Omega}-0.9216\nabla^2w_1\nabla^2C_1 $ &\\
& $\int_{\Omega}w_2\frac{\partial C_2}{\partial t}\text{d}v=\int_{\Omega}-1.1476\nabla w_2\cdot\nabla C_2+\int_{\Omega}+w_2(3.3323-10.2516C_1-11.7393C_2$ &\multirow{2}{*}{14.7\%}\\
& $+10.3c_1^2+15.191C_1C_2+20.5589C_2^2-15.2618C_1^2C_2-15.5277C_2^3)\text{d}v$ & \\
\hline
\end{tabular}
\caption{Results using data generated from Model 4 using $\Delta t=1$, with 10 snapshots. A bias on $\int_{\Omega}\nabla wC_2^2\nabla C_1\text{d}v$ is needed to protect it from elimination in the first few iterations.}
\label{ta:tab:results_alen_noNoise_dt1}
\end{table}

\begin{table}[h]
\centering
\small
 \begin{tabular}{|c|c|c|}
 \hline
 mesh size& results&error\\
 \hline
\multirow{2}{*}{$400\times 400$} &$\int_{\Omega}w_1\frac{\partial C_1}{\partial t}\text{d}v=\int_{\Omega}-0.9693\nabla w_1 \cdot\nabla C_1\text{d}v
+\int_{\Omega}w_1 (0.1041-0.9991C_1+0.9941C_1^2C_2)\text{d}v$&4.05\%\\
 & $\int_{\Omega}w_2\frac{\partial C_2}{\partial t}\text{d}v$ identify wrongly& N/A  \\
\hline
\multirow{2}{*}{$200\times 200$} &$\int_{\Omega}w_1\frac{\partial C_1}{\partial t}\text{d}v=\int_{\Omega}-0.9950\nabla w_1 \cdot\nabla C_1\text{d}v
+\int_{\Omega}w_1 (0.1014-0.9998C_1+0.9981C_1^2C_2)\text{d}v$&1.42\% \\ 
  & $\int_{\Omega}w_2\frac{\partial C_2}{\partial t}\text{d}v=\int_{\Omega}-18.6482\nabla w_2\cdot\nabla C_2+\int_{\Omega}w_2(0.5309-0.4921C_1^2C_2)\text{d}v$&53\%  \\ \hline
\multirow{2}{*}{$100\times 100$} & $\int_{\Omega}w_1\frac{\partial C_1}{\partial t}\text{d}v=\int_{\Omega}-0.9804\nabla w_1 \cdot\nabla C_1\text{d}v
+\int_{\Omega}w_1 (0.1052-0.9995C_1+0.9936C_1^2C_2)\text{d}v$ &5.15\% \\ 
&$\int_{\Omega}w_2\frac{\partial C_2}{\partial t}\text{d}v=\int_{\Omega}-37.9655\nabla w_2\cdot\nabla C_2+\int_{\Omega}w_2(0.8473-0.9381C_1^2C_2)\text{d}v$&6.19\% \\
\hline
\multirow{2}{*}{$50\times 50$}  & $\int_{\Omega}w_1\frac{\partial C_1}{\partial t}\text{d}v=\int_{\Omega}-0.9488\nabla w_1 \cdot\nabla C_1\text{d}v
+\int_{\Omega}w_1 (0.1139-1.0007C_1+0.9858C_1^2C_2)\text{d}v$&13.8\%   \\ 
&$\int_{\Omega}w_2\frac{\partial C_2}{\partial t}\text{d}v=\int_{\Omega}-41.7368\nabla w_2\cdot\nabla C_2+\int_{\Omega}w_2(0.90375-1.0057C_1^2C_2)\text{d}v$ &4.3\% \\ \hline
\end{tabular}
\caption{Results using data generated from Model 1 using $\Delta t =1$, with 10 snapshots and $\sigma=10^{-4}$}
\label{ta:results_noflux_Noise_dt1}
\end{table}
\begin{table}[h]
\centering
\scriptsize
 \begin{tabular}{|c|c|c|}
 \hline
 mesh size& results&error\\
 \hline
\multirow{2}{*}{$400\times 400$} &$\int_{\Omega}w_1\frac{\partial C_1}{\partial t}\text{d}v=\int_{\Omega}-0.9831\nabla w_1 \cdot\nabla C_1\text{d}v
+\int_{\Omega}w_1 (0.1018-0.9983C_1+0.9958C_1^2C_2)\text{d}v+\int_{\Gamma_2}w_10.0984\text{d}s$ &1.8\%\\
 & $\int_{\Omega}w_2\frac{\partial C_2}{\partial t}\text{d}v$ identify wrongly &N/A \\
\hline
\multirow{2}{*}{$200\times 200$} &$\int_{\Omega}w_1\frac{\partial C_1}{\partial t}\text{d}v=\int_{\Omega}-0.983\nabla w_1 \cdot\nabla C_1\text{d}v+\int_{\Omega}w_1 (0.1059-1.0028C_1+0.9973C_1^2C_2)\text{d}v+\int_{\Gamma_2}w_10.1305\text{d}s$&30.3\% \\
 &  $\int_{\Omega}w_2\frac{\partial C_2}{\partial t}\text{d}v$ identify wrongly &N/A\\
\hline
\multirow{2}{*}{$100\times 100$} &$\int_{\Omega}w_1\frac{\partial C_1}{\partial t}\text{d}v=\int_{\Omega}-0.9794\nabla w_1 \cdot\nabla C_1\text{d}v+\int_{\Omega}w_1 (0.1094-1.0036C_1+0.9949C_1^2C_2)\text{d}v+\int_{\Gamma_2}w_10.1472\text{d}s$&47\% \\
 & $\int_{\Omega}w_2\frac{\partial C_2}{\partial t}\text{d}v=\int_{\Omega}-47.5561\nabla w_2\cdot\nabla C_2+\int_{\Omega}w_2(0.8375-0.9269C_1^2C_2)\text{d}v$&7.3\%   \\
\hline
\multirow{2}{*}{$50\times 50$} & $\int_{\Omega}w_1\frac{\partial C_1}{\partial t}\text{d}v=\int_{\Omega}-0.884\nabla w_1 \cdot\nabla C_1\text{d}v+\int_{\Omega}w_1 (0.1172-1.0036C_1+0.9856C_1^2C_2)\text{d}v$&17.22\%  \\
 & $\int_{\Omega}w_2\frac{\partial C_2}{\partial t}\text{d}v=\int_{\Omega}-41.755\nabla w_2\cdot\nabla C_2+\int_{\Omega}w_2(0.903-1.0064C_1^2C_2)\text{d}v$&4.38\%   \\
\hline
\end{tabular}
\caption{Results using data generated from Model 2 using $\Delta =1$, with 10 snapshots and $\sigma=10^{-4}$.}
\label{ta:results_influx_noise_dt1}
\end{table}
\begin{table}[h]
\centering
\tiny
 \begin{tabular}{|c|c|c|}
 \hline
 mesh size& results&error\\
   \hline
 \multirow{2}{*}{$400\times 400$} & $\int_{\Omega}w_1\frac{\partial C_1}{\partial t}\text{d}v$ identify wrongly&N/A\\
 & $\int_{\Omega}w_2\frac{\partial C_2}{\partial t}\text{d}v$ identify wrongly&N/A\\
  \hline
 \multirow{2}{*}{$200\times 200$} & $\int_{\Omega}w_1\frac{\partial C_1}{\partial t}\text{d}v$ identify wrongly&N/A\\
 & $\int_{\Omega}w_2\frac{\partial C_2}{\partial t}\text{d}v$ identify wrongly&N/A\\
   \hline
\multirow{4}{*}{$100\times 100$} & $\int_{\Omega}w_1\frac{\partial C_1}{\partial t}\text{d}v=\int_{\Omega}\nabla w_1\cdot\left(-13.7499+36.3723C_1+17.9356C_2-36.3702C_1^2-13.9C_2^2 \right)\nabla C_1\text{d}v$& \multirow{2}{*}{30.2\%} \\
& $+ \int_{\Omega}\nabla w_2\nabla\cdot\left(-7.6263+15.2494C_1+11.158C_2-22.32C_1C_2 \right)\nabla C_2\text{d}v+\int_{\Omega}-0.7919\nabla^2w_1\nabla^2C_1$&  \\
& $\int_{\Omega}w_2\frac{\partial C_2}{\partial t}\text{d}v=\int_{\Omega}\nabla w_2\cdot\left(-6.9381+13.8269C_1+10.1406C_2-20.2006C_1C_2 \right)\nabla C_1\text{d}v$& \multirow{2}{*}{30.3\%}\\
& $+ \int_{\Omega}\nabla w_2\nabla\cdot\left(-9.1114+13.4938C_1+29.8767C_2-13.5301C_1^2-33.5644C_2^2 \right)\nabla^2C_2\text{d}v+\int_{\Omega}-0.735\nabla^2w_2\nabla^2C_2$& \\
\hline
\multirow{4}{*}{$50\times 50$} & $\int_{\Omega}w_1\frac{\partial C_1}{\partial t}\text{d}v=\int_{\Omega}\nabla w_1\cdot\left(-15.1116+40.105C_1+19.6542C_2-40.132C_1^2-\underline{15.2351}C_2^2 \right)\nabla C_1\text{d}v$& \multirow{2}{*}{24.1\%} \\
& $+ \int_{\Omega}\nabla w_2\nabla\cdot\left(-8.3106+16.6052C_1+12.1407C_2-24.2681C_1C_2 \right)\nabla C_2\text{d}v+\int_{\Omega}-0.9238\nabla^2w_1\nabla^2C_1$ & \\
& $\int_{\Omega}w_2\frac{\partial C_2}{\partial t}\text{d}v=\int_{\Omega}\nabla w_2\cdot\left(-7.6666+15.2396C_1+11.2172C_2-22.272C_1C_2 \right)\nabla C_1\text{d}v$& \multirow{2}{*}{30.3\%}\\
& $+ \int_{\Omega}\nabla w_2\nabla\cdot\left(-10.1748+15.3511C_1+33.0199C_2-15.4756C_1^2-36.9985C_2^2 \right)\nabla^2C_2\text{d}v+\int_{\Omega}-0.8631\nabla^2w_2\nabla^2C_2$ &\\
\hline
\end{tabular}
\caption{results using data generating from Model 3 using dt=1, 10 snapshot. $\sigma=10^{-5}$}
\label{ta:results_cahn_Noise_dt1}
\end{table}
\begin{table}[h]
\centering
\scriptsize
 \begin{tabular}{|c|c|c|}
 \hline
 mesh size& results&error\\
 \hline
\multirow{3}{*}{$400\times 400$} & $\int_{\Omega}w_1\frac{\partial C_1}{\partial t}\text{d}v$ identify wrongly&N/A \\
& $\int_{\Omega}w_2\frac{\partial C_2}{\partial t}\text{d}v=\int_{\Omega}-0.9894\nabla w_2\cdot\nabla C_2+\int_{\Omega}+w_2(3.4733-10.6805C_1-12.1079C_2$ &\multirow{2}{*}{1.065\%}\\
& $+10.6816c_1^2+15.8249C_1C_2+21.0192C_2^2-15.8266C_1^2C_2-15.8597C_2^3)\text{d}v$&  \\
\hline
\multirow{3}{*}{$200\times 200$} &$\int_{\Omega}w_1\frac{\partial C_1}{\partial t}\text{d}v$ identify wrongly&N/A \\
& $\int_{\Omega}w_2\frac{\partial C_2}{\partial t}\text{d}v=\int_{\Omega}-1.0107\nabla w_2\cdot\nabla C_2+\int_{\Omega}+w_2(3.497-10.756C_1-12.183C_2$ &\multirow{2}{*}{1.066\%}\\
& $+10.7587C_1^2+15.9347C_1C_2+21.1536C_2^2-15.9389C_1^2C_2-15.966C_2^3)\text{d}v$&  \\
\hline
\multirow{4}{*}{$100\times 100$} &$\int_{\Omega}w_1\frac{\partial C_1}{\partial t}\text{d}v=\int_{\Omega}\nabla w_1\cdot\left(-16.4594+43.8416C_1+20.5547C_2-43.8509C_1^2-15.3705C_2^2 \right)\nabla C_1\text{d}v$ &\multirow{2}{*}{9.17\%}\\
& $+ \int_{\Omega}\nabla w_2\nabla\cdot\left(-9.8091+19.6113C_1+14.6763C_2-29.3139C_1C_2 \right)\nabla C_2\text{d}v+\int_{\Omega}-0.9544\nabla^2w_1\nabla^2C_1 $ &\\
& $\int_{\Omega}w_2\frac{\partial C_2}{\partial t}\text{d}v=\int_{\Omega}-1.0457\nabla w_2\cdot\nabla C_2+\int_{\Omega}+w_2(3.46-10.641C_1-12.096C_2$ &\multirow{2}{*}{4.57\%}\\
& $+10.6537C_1^2+15.7641C_1C_2+21.055C_2^2-15.7832C_1^2C_2-15.8958C_2^3)\text{d}v$ & \\
\hline
\multirow{4}{*}{$50\times 50$} &$\int_{\Omega}w_1\frac{\partial C_1}{\partial t}\text{d}v=\int_{\Omega}\nabla w_1\cdot\left(-14.898+38.6644C_1+19.7295C_2-38.678C_1^2-\underline{14.984}C_2^2 \right)\nabla C_1\text{d}v$&\multirow{2}{*}{20.7\%} \\
& $+ \int_{\Omega}\nabla w_2\nabla\cdot\left(-8.5672+17.119C_1+13.1645C_2-26.22C_1C_2 \right)\nabla C_2\text{d}v+\int_{\Omega}-0.9213\nabla^2w_1\nabla^2C_1 $ &\\
& $\int_{\Omega}w_2\frac{\partial C_2}{\partial t}\text{d}v=\int_{\Omega}-1.148\nabla w_2\cdot\nabla C_2+\int_{\Omega}+w_2(3.33-10.2485C_1-11.739C_2$&\multirow{2}{*}{14.7\%} \\
& $+10.2984c_1^2+15.19C_1C_2+20.5588C_2^2-15.26C_1^2C_2-15.5276C_2^3)\text{d}v$ & \\
\hline
\end{tabular}
\caption{results using data generating from Model 4 using dt=1, 10 snapshot, $\sigma=10^{-5}$.}
\label{ta:results_alen_noise_dt1}
\end{table}

\clearpage
\bibliography{reference}
\bibliographystyle{unsrtnat}
\end{document}